\documentclass{aa}  
\usepackage[english]{babel}
\usepackage{amsmath}
\usepackage{graphicx}
\usepackage{natbib}
\usepackage[colorlinks=true, allcolors=blue]{hyperref}
\DeclareUnicodeCharacter{2061}{}

\usepackage{txfonts}
\begin{document}

 \title{Application of the FRADO model of BLR formation to the Seyfert galaxy NGC 5548 and the first step toward determining the Hubble constant 
 }
 \titlerunning{FRADO model and H$_0$ determination for NGC 5548}


   \author{Vikram Kumar Jaiswal\inst{1}
          \and
         Amit Kumar Mandal \inst{1}
          \and
          Raj Prince \inst{1,2}
          \and 
          Ashwani Pandey 
          \inst{1,3}
          \and
          Mohammad Hassan Naddaf\inst{4}
          \and 
          Bo\.zena Czerny \inst{1}
          \and
          Swayamtrupta Panda\inst{5,\thanks{Gemini Science Fellow}}
          \and
          Francisco Pozo Nu\~nez \inst{6}
          }

   \institute{Center for Theoretical Physics, Polish Academy of Sciences, Al. Lotnik\'ow 32/46, 02-668 Warsaw, Poland
    \and
    Department of Physics, Institute of Science, Banaras Hindu University, Varanasi-221005, India
    \and 
    Department of Physics and Astronomy, University of Utah, Salt Lake City, UT 84112, USA
    \and
    Institut d'Astrophysique et de Géophysique, Université de Liège Allée du six août 19c, B-4000 Liège (Sart-Tilman), Belgium
    \and
    International Gemini Observatory/NSF NOIRLab, Casilla 603, La Serena, Chile
    \and
    Astroinformatics, Heidelberg Institute for Theoretical Studies, Schloss-Wolfsbrunnenweg 35, 69118 Heidelberg, Germany
    }

   \date{}

 
  \abstract
   {The dynamical and geometric structures of the Broad Line Region (BLR), along with the origins of continuum time delays in active galaxies, remain topics of ongoing debate.}
   {In this study, we aim to reproduce the observed broadband spectrum, the H$\beta$ line delay, and the continuum time delays using our newly developed model for the source NGC 5548.}
   {We adopt the standard accretion disk model, with the option of an inner hot flow, and employ the lamp-post model to account for disk irradiation. Additionally, we model the BLR structure based on radiation pressure acting on dust. The model is parameterized by the black hole mass, $M_{\text{BH}}$ (which is fixed), the accretion rate, the viewing angle, the height of the lamp-post, the cloud density, and the cloud covering factor. The resulting continuum time delays arise from a combination of disk reprocessing and the reprocessing of a fraction of radiation by the BLR.}
   {Our model reasonably reproduces the observed broad-band continuum, the H$\beta$ time delay, and the continuum inter-band time delays measured during the observational campaign. When the accretion rate is not constrained by the known distance to the source, our approach allows for a direct estimation of the distance. The resulting Hubble constant, $H_0$ = $66.9^{+10.6}_{-2.1}$ km s$^{-1}$ Mpc$^{-1}$, represents a significant improvement over previously reported values derived from continuum time delays in the literature.}
  {This pilot study demonstrates that, with sufficient data coverage, it is possible to disentangle the time delays originating from the accretion disk and the BLR. This paves the way for effectively using inter-band continuum time delays as a method for determining the Hubble constant. Additionally, the findings provide strong support for the adopted model for the formation of the H$\beta$ line.}


   \maketitle
%


\section{Introduction}

The central components of active galactic nuclei (AGNs) include a massive black hole, a compact X-ray emitting region, a relatively cold Keplerian accretion disk, and the broad line region (BLR) \citep[see, e.g.,][for a basic compendium]{krolik1999}. The Keplerian disk serves as the primary source of continuum emission, while the BLR predominantly produces broad emission lines, along with a minor contribution to the continuum. In terms of spatial extent, the innermost regions of the accretion disk span approximately 10$-$100 gravitational radii ($r_g$), whereas the BLR extends outward at distances of order of $\sim$ 1000 $r_g$. Beyond the BLR, a dusty molecular torus is located at even larger radii ($\sim 10^4$ to $10^5 r_g$). Since the early studies of active galaxies by \citet{Seyfert1943} and the discovery of quasars by \citet{schmidt1963}, there has been tremendous progress in our understanding of the structure of AGNs \citep[see e.g.][for a review]{1999agnc.book.....K}.

The innermost regions of an AGN are too compact ($\sim$ a few microarcseconds in the case of BLR) to be spatially resolved directly. However, their structure can be probed using intrinsic flux variability. Among the various wavelengths, X-ray variability is the fastest, originating from the most compact region near the black hole. The reprocessing of this X-ray emission at longer wavelengths, observed through light echo studies, provides insights into the structure and dynamics of the surrounding material, a technique known as reverberation mapping \citep[RM;][]{1982ApJ...255..419B, 1993PASP..105..247P}.
RM has been widely applied to AGNs \citep{2005ApJ...622..129S, 2014MNRAS.444.1469M, 2018ApJ...862..123M, 2019ApJ...880..126H, 2020ApJS..246...16Y, 2022ApJ...940...20G}, particularly in studies of continuum variability, which is believed to arise from the irradiated accretion disk. 

Although our understanding of AGN structure is still evolving, significant progress has enabled the exploration of AGNs as potential tools for cosmology. Unlike standard candles, AGNs exhibit luminosities that span several orders of magnitude. Nevertheless, the luminosity distance to individual AGNs can be estimated either through geometric methods or by standardizing them in a statistical sense \citep[see e.g.][for reviews]{czerny_distances2018}. Early successful approaches relied on the non-linear relation between UV and X-ray luminosities \citep{risaliti2015,lusso2025}, or the correlation between the BLR radius and the source luminosity \citep{martinezAldama2019,cao2022,cao2024, 2025PhRvD.111h3545C}. However, these methods require external scaling and, without it, can only constrain the curvature of the Hubble diagram. As such, they are not directly suitable for determining the Hubble constant.

The Hubble constant can instead be derived using strong gravitational lensing techniques \citep[see][for a recent review]{Suyu2024}, or through a combination of RM of the BLR with angular size measurements obtained from high-resolution interferometric observations \citep{LiYanRong2022, 2025arXiv250218856L}. While both methods are promising, they demand extensive, dedicated observational campaigns and are currently feasible only for a limited number of AGNs, specifically, nearby, bright sources at low redshift where sufficient spatial resolution can be achieved.

The primary goal of this work is to develop a practical method for measuring the Hubble constant using a single mean spectrum of an AGN combined with photometric RM results at multiple wavelengths. High-quality photometric RM measurements are already available for a number of sources \citep{fausnaugh2016, 2018ApJ...857...53C, 2024ApJ...973..152E, 2025arXiv250606731P}, and a substantial influx of data is expected from large-scale time-domain surveys, such as the Zwicky Transient Facility (ZTF) and the Vera C. Rubin Observatory’s Legacy Survey of Space and Time (LSST).

The idea of using photometric RM to measure the Hubble constant is not new; it was first proposed by \citet{collier1999}, but has yet to produce successful results. The method is based on the standard accretion disk model of \citet{SS1973}, which predicts a direct relationship between the time delay at a given wavelength and the monochromatic luminosity: intrinsically brighter sources are physically larger and thus produce longer time delays. More specifically, the model anticipates that the time delays ($\tau$) should scale with wavelength ($\lambda$) as $\tau \propto \lambda^{4/3}$. While all coefficients in this relation are determined by the model, the viewing angle ($i$) to the source remains a free parameter and must be estimated independently.

 While observational results have roughly confirmed this trend, they have not matched the predicted normalization, as observed delays continue to exceed theoretical expectations \citep[e.g.][]{cackett2007,shappee2014,kokubo2018, 2022ApJ...940...20G, 2025ApJ...985...30M}. This discrepancy, known as the accretion disk size problem, was first recognized in optical microlensing studies of gravitationally lensed quasars \citep{2010ApJ...712.1129M} and till recently remained an ongoing challenge in our understanding of AGN accretion disk structure.

 Recent intensive RM campaigns \citep{edelson_2015ApJ, derosa_2015ApJ, fausnaugh2016, pei2017, horne2021, Lu2022, 2025arXiv250606731P} have revealed that the observed optical continuum is not solely emitted by the accretion disk. \citet{korista2001} first noted that radiation reprocessed by the BLR contributes not only to the formation of emission lines but also to the diffuse continuum emission. This finding was later confirmed by \citet{fausnaugh2016} and further emphasized by \citet{cackett_2021iSci, netzer2022}, who demonstrated that the diffuse BLR contribution can significantly dominate the measured continuum time delays.  This effect poses challenges for using continuum time delays as a method to determine the Hubble constant, $H_0$, as originally proposed by \citet{collier1999}. For instance, \citet{cackett2007} derived an $H_0$ value of $15 \pm 3$ km s$^{-1}$ Mpc$^{-1}$; however, at the time, not all aspects of the physical mechanisms driving continuum time delays were fully understood.

The contribution of the BLR to the continuum time delay implies that, in order to use photometric continuum RM of AGNs for measuring the Hubble constant, we require not only a reliable accretion disk model but also a robust model of the BLR structure. Importantly, such a BLR model must be fully determined by the absolute luminosity of the source. However, fully parametric BLR models \citep[e.g.][]{baldwin1995,li2013,pancoast2014,grier2017} may not be adequate in this context, as the ability to recover the absolute luminosity of the source can be compromised unless wavelength-resolved spectroscopic RM data are available. Without such spectral resolution, key constraints on the BLR geometry and its luminosity dependence may be lost, limiting the accuracy of distance measurements based on photometric continuum RM.

We thus propose the use of the Failed Radiatively Accelerated Dusty Outflow (FRADO) model of the BLR \citep{czhr2011, naddaf2021, naddaf2022}. The model assumes    BLR formation based on radiation pressure acting on dust in disk regions where the disk atmosphere temperature falls below the sublimation threshold. It was proposed to explain the vertical rise of material responsible for low-ionization lines, such as H$\beta$.  In this study, we thus investigate the reprocessing of central radiation through a combination of two reprocessors: the Keplerian disk, which we modify to include the warm corona in the innermost part, and the BLR, using the BLR response derived from the FRADO model. This approach provides a three-dimensional distribution of BLR clouds, which we then combine with a spectral shape model of the BLR generated using the CLOUDY code (version 23.00; \citealt{2023RMxAA..59..327C}). We fit this model to the observed spectrum and time delays of NGC 5548 and evaluate its potential for determining the distance to the source and the Hubble constant. We test the model for a specific source, and we treat the results as the pilot study: a base for further model developments and the later use for other objects.

NGC 5548 ($z$ = 0.017175), a Seyfert 1 galaxy, is among the most extensively studied AGNs. Its brightness, relatively close distance, and significant variability made it an early candidate for studying the inner, unresolved structure of an active nucleus \citep[e.g.,][]{clavel1991, peterson1991, krolik1991, rokaki1993}.  The recent high-quality photometric monitoring \citep{fausnaugh2016}, broad-band spectral analysis \citep{mehdipour2015}, measured H$\beta$ time delay from spectroscopic RM \citep{pei2017}, and wavelength-resolved RM that captured the detailed response of the BLR \citep{horne2021} collectively make NGC 5548 an ideal laboratory for testing our methodology.

The structure of this paper is as follows. In Section \ref{sec:spectral}, we describe the data used for fitting. Section \ref{sec:method} outlines our methodology, which involves generating the response function of the accretion disk using the lamp-post model and incorporating the BLR contribution via FRADO and CLOUDY to estimate the combined time delays. Section \ref{sec:result} presents the results obtained from different models, detailing the simultaneous fitting of the time lag and spectral energy distribution (SED) while assessing the strengths and limitations of each approach. Finally, in Section \ref{sect:distance}, we estimate the luminosity distance based on our best-fitting model. The discussion and conclusions are presented in Sections \ref{ss:discussion} and \ref{concl}, respectively.

\section{Spectral shape and continuum time delay data of NGC 5548} \label{sec:spectral}

NGC 5548 ($z$ = 0.017) is one of the most extensively studied AGNs across multiple wavelengths. It was the primary target of the AGN Watch international collaboration \citep{peterson2002}, which began investigating its emission line variability as early as 1987. Over the years, its optical and UV variability has been the subject of numerous studies \citep{peterson1991, korista1995, peterson1992, peterson2002, peterson2004}. Observations from both ground-based and space-based telescopes have provided critical insights into accretion disk dynamics and emission line variability. Notably, optical variability in both the continuum and emission lines was first reported by \citet{deVaucouleurs1972} in the 1970s, while early UV observations commenced in 1988 with the International Ultraviolet Explorer (IUE; \citealt{ulrich1983}). More recently, intensive multi-wavelength monitoring campaigns \citep{edelson_2015ApJ, derosa_2015ApJ, fausnaugh2016, pei2017, horne2021, Lu2022} have significantly deepened our understanding of the source.

These studies have revealed that the measured continuum time delays do not follow a simple power-law trend with wavelength. Instead, they show clear evidence of interaction with an extended BLR, where different emission lines originate. Consequently, these delays are often approximated by mean time delays, though some emission line delay profiles exhibit double-peaked features. This complexity opens up new avenues for testing specific models of the BLR structure, as well as for explaining the observed inter-band continuum time delays.

In this study, we utilize the broadband SED and global parameters from \citet{mehdipour2015}, covering the period from 2013 to 2014. The broadband SED, corrected for internal extinction, starlight contamination, the Balmer continuum, and the Fe II pseudo-continuum, was first presented by \citet{mehdipour2015}. Subsequently, \citet{kubota2018} fitted this spectrum to an AGN spectral model, revealing that only the outermost part of the accretion flow corresponds to a standard disk. Their results further indicated that the optical/UV emission primarily originates from a region dominated by the warm corona (see their Figure 4).

Therefore, to systematically model the observed inter-band continuum time delays, we propose a step-by-step approach. We select three representative setups to illustrate how the spectral fit and time-delay measurements depend on the pre-set parameters and details of the geometry.  

First, we assume a standard accretion disk model (Model A). This choice is motivated by the delay fitting for NGC 5548 done by \citet{kammoun2021}. The source was fitted by \citet{kammoun2021} without assuming an inner hot flow, only with the corona height, black hole mass ($M_{\text{BH}}$), spin, accretion rate, and the X-ray luminosity as free parameters. The color correction was set to 2.4. We modify the approach: we fix $M_{\text{BH}}$ and the Eddington ratio, but we allow for the inner hot flow and the contribution of the BLR. Our color correction is the same, and no warm corona is present. Since we also fit the spectral shape, we include a starlight component in the model.

Next, we consider an alternative scenario (Model B) based on the spectral decomposition proposed by \citet{kubota2018}.  In this case, the warm corona plays an important role in the spectral fit. We adopt the value of the warm corona temperature from the corresponding paper. In order to calculate the time delay caused by reprocessing, we assume that the optical/UV variability arises solely from the reprocessing of X-rays in the outer cold disk. This is motivated by the fact that warm corona with a temperature of approximately $\sim$  $10^6$ K, is too optically thick to absorb and re-emit incident hard X-rays in the UV/X-ray bands,  although it still contributes to the UV continuum. The model is supplemented with the BLR contribution to the delays and the spectrum, and starlight is included.

Finally, we introduce  Model C, a most general spectral decomposition based on Figure 5 of \citet{mehdipour2015}. Now a number of geometrical and physical parameters are optimized to fit the spectral and time delay data. This includes the warm corona temperature and its optical depth, as well as contribution from BLR and starlight.  

The disk parameters used for Models A, B, and C are summarized in Table \ref{tab:kammoun_sample},  with more detailed descriptions provided in Section \ref{ss:models}.

To test these models, we utilize the continuum time delay data from \citet{fausnaugh2016}, obtained during the STORM campaign from December 2013 to August 2014. This campaign integrated optical light curves data from 16 ground-based observatories, across the $B, V, R, I$ filters, as well as the SDSS $-u, g, r, i, z$ filters, coupled with ultraviolet data from the Hubble Space Telescope (HST) and Swift instruments. The inter-band time delays were measured relative to the HST light curve at 1367 \AA, providing a crucial dataset for evaluating the proposed disk reprocessing scenarios.

\section{Method} \label{sec:method}

 Understanding the origin of continuum time delays in AGNs is crucial for probing the structure of the central engine and its surrounding medium. However, fitting these delays solely as a result of accretion disk reprocessing has often led to the so-called disk-size problem, as discussed earlier. Motivated by previous studies \citep{korista2001, netzer2022, Jaiswal2023, beard2025}, we extend this framework by considering the reprocessing of central flux by both the accretion disk and the BLR medium. Given our twofold motivation, we adopt a theoretical modeling approach that accounts for all relevant aspects. This allows us not only to test the BLR model we employ but also to explore the determination of the Hubble constant, as the relatively small number of model parameters enables us to treat the luminosity distance as an unknown parameter. 

 Building on this foundation, our model of an active nucleus comprises multiple key components: a standard accretion disk in the outer parts of the flow, a warm comptonizing corona, and a hot corona in the inner parts of the flow, the BLR region, and the contribution of the host galaxy to the total spectrum. In the following sections, we examine each of these components in detail.

\subsection{Hot and warm corona}

 The hot corona, responsible for hard X-ray emission, was modeled by \citet{mehdipour2015}. In our approach, we adopt the proposed power-law shape and normalization for the observed spectrum when fitting the warm corona. We do not model the hot corona; instead, the power-law component is treated as part of the broadband spectrum that irradiates the BLR clouds, as discussed in the next section. It is also included in the fit to the global spectrum, in the X-ray region. However, we neglect the contribution of hard X-rays to the optical/UV part of the spectrum.

Unlike \citet{mehdipour2015}, we model the warm corona with a different approach. In our framework, the inner and outer radii of the warm corona are treated as parameters. The outer radius of the warm corona coincides with the inner radius of the standard (outer) disk. The Comptonization process in the warm corona is computed using the analytical formulae from \citet{ST80} and is parameterized by the optical depth and electron temperature. Since the warm corona covers the inner disk, we do not assume a single temperature for the soft photons. Instead, we determine the temperature of the underlying cold disk at each radius and adjust it by the Comptonization amplification factor. These calculations are performed iteratively, ensuring that the soft photon temperature spans a continuous range while preserving the total energy from the accretion flow.  The coronal parameters (optical depth and the temperature) are assumed to be radius-independent as we do not have a firm predictions for these parameters from the warm corona physics. This methodology aligns closely with the approach of \citet{kubota2018}. The computational code used for this purpose was originally presented in \citet{czerny2003}. The warm corona is thus fully modeled, with its spectral shape and normalization determined by global fitted parameters, including the optical depth, temperature, the disk radius at which the warm corona develops, and the accretion rate.  This is incorporated in the spectral fitting, but not included separately in BLR reprocessing where the incident radiation is made at the basis of the broad band SED, instead of a separate modeling of the disk and the warm corona. This choice avoids the computational complexity of calculating the reprocessed emission using the {\tt CLOUDY} code \citep{Cloudy23} at every step of the simulation. It is stated explicitly later on.

 Additionally, we do not include irradiation of the warm corona by hard X-rays, as the high temperature and large optical depth of the warm corona would cause the incident radiation to be predominantly scattered rather than absorbed and reprocessed. Consequently, no significant thermalization is expected, although some high-ionization lines, such as iron lines in the X-ray spectrum, may still be present \citep{petrucci2020, ballantyne2024}.

\subsection{Cold disk reprocessing}

 The irradiating flux is thermalized only in the outer, cold regions of the standard accretion disk. To further explore this process, we simulate the lamp-post model to generate the disk’s response function, assuming a pulse of short duration. In this simulation, we exclude corrections for general relativity (GR) and energy-dependent reflection effects, as introduced by \citet{kammoun2023, Kammoun2024, 2024Papoutsis}, to maintain a relatively simple code suitable for data fitting.  As a result, the incident radiation from the corona is assumed to be fully absorbed by the cold disk, leading to a localized increase in temperature.  We briefly comment on the implications and limitations of this simplification in Section \ref{ss:discussion}.

To accurately model the disk reprocessing, we employ a non-uniform radial grid extending from the innermost stable circular orbit ($r_{\rm isco}$) to the outer disk radius ($r_{\rm out}$),  ensuring sufficient resolution at smaller radii.  The radial grid spacing follows the relation $dr = 0.085 \times (\frac{r}{r_{\rm isco}})^{0.85}$, where $r$ is the radial distance from the central source in units of gravitational radius, $r_g$. For each radial position, the azimuthal angle $\phi$ is sampled using the relation $d\phi = \frac{1.57}{N_{\rm div}}$, where $N_{\rm div} = 3800$ ensures sufficient resolution across the entire radial range. The surface element for each $(r, \phi)$ coordinate is defined as $ds = r \cdot dr \cdot d\phi$ in units of $r_g^2$. To facilitate further calculations, we convert $(r, \phi)$ into Cartesian coordinates, assuming a geometrically thin disk with negligible height.

For each $(x, y)$ coordinate, we compute the total time delay, $\tau_{\rm total}(x,y)$, as the sum of two components: the delay from the corona to the disk, $\tau_d(r)$, and the delay from the disk to the plane intersecting the equatorial plane at ($r=r_{out},\phi = 0$), denoted as $\tau_{\rm do}(x,y)$. The delay is influenced by key parameters, including lamp-luminosity ($L_x$), $M_{\rm BH}$, accretion rate ($\dot{M}$), the corona height ($h$), and the inclination angle of the system ($i$). By default, we assume a cold Keplerian disk extending down to the ISCO at $6 \, r_g$, with an outer radius of $10^4 \, r_g$. However, we also consider cases with different inner and outer radii, which can be set through the parameter $r_{\rm in}$.

The non-irradiated disk is characterized by its flux, \( F_{\text{non-irradiated}} \), and temperature, \( T_{\text{disk}} \), as defined in equations~\ref{eq:flux_non_irr} and \ref{eq:temp_non_irr}, respectively. When the contribution from the corona is included, the resulting total flux and temperature are given by equations~\ref{eq:flux_total} and \ref{eq:temp_total}. For a specified delay, we use the combined flux from both the disk emission and irradiation, \( F_{\text{disk+irradiation}} \), as described in equation~\ref{eq:flux_total}. We then convert this total flux into an effective temperature, \( T_{\text{eff}} \), according to equation~\ref{eq:temp_total}. This temperature is subsequently used to compute the blackbody emission.
\begin{equation}
F_{\text{non-irradiated}}(r,t+\tau_d(r)) = \left(\frac{3GM\dot{M}}{8 \pi r^3}\left(1-\sqrt{\frac{r_{in}}{r}}\right)\right)
\label{eq:flux_non_irr}
\end{equation}
\begin{equation}
T_{\rm disk}(r,t +\tau_d(r)) = \left[\left(\frac{3GM\dot{M}}{8 \pi r^3 \sigma_B}\left(1-\sqrt{\frac{r_{in}}{r}}\right)\right)\right]^{\frac{1}{4}}\\
\label{eq:temp_non_irr}
\end{equation}
\begin{equation}
F_{\text{disk+irradiation}}(r,t+\tau_d(r)) = \left(\frac{3GM\dot{M}}{8 \pi r^3}\left(1-\sqrt{\frac{r_{in}}{r}}\right)\right)+\left(\frac{L_x(t)h}{4\pi r^3}\right)
\label{eq:flux_total}
\end{equation}
\begin{equation}
T_{\rm eff}(r,t +\tau_d(r)) = \left[\left(\frac{3GM\dot{M}}{8 \pi r^3 \sigma_B}\left(1-\sqrt{\frac{r_{in}}{r}}\right)\right)+\left(\frac{L_x(t)h}{4\pi r^3 \sigma_B}\right)\right]^{\frac{1}{4}}\\
\\
\label{eq:temp_total}
\end{equation}
Additionally, we account for a color correction factor $f_c$ to the disk temperature, which raises the temperature of the disk atmosphere to $T_c$, where $f_c = 1$ corresponds to no color correction applied, as described below \citep{1995ApJ...445..780S}:
\begin{equation}
T_c = f_c T_{eff}.
\end{equation}

To compute the full spectrum for a specific differential area, we use Planck's formula and store the results in a photon table, represented as a 2D matrix denoted as \textbf{P(t$^\prime$,${\boldsymbol{\mathrm{\lambda}}}$)}. This involves applying the calculated time delay corresponding to the disk position and wavelength. Each element in the photon table represents a unique delay and wavelength $\lambda$, with the $\lambda$ values ranging from 1000 to 10000 \AA~ and selected using a logarithmic scale grid for simulation. We also introduce a color correction as a free parameter of the model.

Next, we generate the response function by sending a very short light pulse of $\Delta_t$ = 0.05 days duration and normalizing the result by the incident bolometric luminosity.  The pulse duration is chosen to balance accuracy and temporal resolution: a very short pulse can introduce significant errors in the temperature enhancement, while a very long pulse tends to smear the signal and reduce time resolution, especially at the shortest wavelengths. The response function for a bare disk is given by the following equation:
\begin{equation}
\label{eq:resp}
    \psi(t,\lambda) = {1 \over {\Delta_t L_x}} {1 \over f_c^4}\int_{S_{disk}} B_{\lambda}(T_c(r(t^{'}-\tau_{do}(x,y))))ds.
\end{equation}
The time delay in the disk has been described previously by \citet{Jaiswal2023}. Our adopted method, while simpler than the approaches presented by \citet{kammoun2021} and \citet{kammoun_analit2021}, which include corrections for GR and disk albedo, is sufficiently accurate for the current purpose of this pilot study.

\subsection{BLR structure and reprocessing}
\label{ss:frado}

To introduce the contribution from the BLR, we first describe its properties. In this study, we do not parameterize the structure of the BLR using arbitrary numbers or functions; instead, we calculate it based on the FRADO model, which was qualitatively proposed by \citet{czhr2011} and has been further developed in several subsequent studies \citep{czerny2017, naddaf2021, naddaf2022, naddaf2023, naddaf_CF2024}. 


The basic idea behind the model is to associate the low-ionization parts of the BLR -- responsible for emission lines, such as H$\beta$, Mg II, and Fe II, with dust-driven massive winds. In stellar environments, such winds are typically denser, possess higher optical depths, and exhibit significantly slower outflow velocities. Similarly, in AGNs, dust-driven winds are expected to be launched from the outer disk, where the effective temperature falls below the dust sublimation threshold. This temperature constraint naturally defines the inner radius of the BLR. The nature of the wind depends on $M_{\rm BH}$ and Eddington ratio: for sufficiently massive black holes and high accretion rates, the wind escapes, contributing to the BLR outflow structure; otherwise, it forms a failed wind, in which the material ultimately falls back onto the disk. Additionally, irradiation from the central parts of the disk plays an increasingly important role in shaping the dynamics and ionization state of the wind as the cloud elevates above the disk plane.

Specifically, we utilize the code from \citet{naddaf2021}, which provides a detailed numerical description of the wavelength-dependent cross-section for dusty particles. According to this model, radiatively dust-driven pressure lifts the clouds from the disk surface, while preserving their angular momentum derived from the Keplerian motion of the disk surface. As a cloud is lifted, it becomes increasingly illuminated by the inner parts of the disk. However, if the cloud becomes too hot, the dust evaporates, allowing the cloud to continue its motion along a ballistic orbit. The global parameters of the model, such as $M_{\rm BH}$, Eddington rate, and metallicity, govern the behavior of the BLR. For lower black hole masses, lower Eddington ratios, and lower metallicities, the clouds form a failed wind, while in the opposite case, a fraction of the clouds may form an escaping wind. Hence, we emphasize that the inner and outer radii of the BLR, along with the statistical distribution of the clouds, are governed by global parameters, such as $M_{\rm BH}$ and the Eddington ratio.

We assume a universal value of the dust sublimation temperature of 1500 K as representative for all grain species and sizes. It was suggested by \citet{barvainis1987} at the basis of the observational data for PG quasars. It is frequently used for zero order approximation, although it is well known that actually the sublimation temperature depends on the chemical composition as well as size  of the dust grains \citep{draine1984,baskin2018}. However, calculating the range of evaporation temperature of each grain is time consuming and not appropriate for a pilot study. In addition, the chemical composition of grains in AGN is not firmly established, and the presence of the graphite with usual grain size distribution is highly unlikely \citep[e.g.][]{czerny2004,gaskell2004}. We further discuss this issue in Section \ref{ss:dust}.

With knowledge of the global parameters of the source, we can determine the 3-D locations of statistically representative clouds within the BLR.  The code calculates 3-D trajectories of clouds launched from the disk. As described in \citet{naddaf2021}, clouds are then located at these trajectories proportionally to time they spent at each part of the trajectory. This information is then combined with the emissivity law for specific emission lines, particularly low-ionization lines such as H$\beta$, Mg II, or Fe II. In our model, we assume that the emissivity weight of each cloud is influenced by its distance from the central disk: clouds closer to the center receive more radiation, resulting in a higher emissivity weight, while clouds farther away receive less radiation, leading to a lower weight, as defined in equation \ref{eq:weight}:
\begin{equation}
\label{eq:weight}
w \propto d^{-2}, 
\end{equation}
where $d$ is the distance of a cloud from the center. This scaling of the emissivity, which inversely depends on the square of the distance, neglects the local efficiency of converting incident flux into  BLR spectrum. However, since our model does not yet predict the local density, this simplification is reasonable at this stage of development. We fix the local density of the cloud at $10^{11}$ cm$^{-3}$, and the column density at $3 \times 10^{23}$ cm$^{-2}$. 

By supplementing the cloud distribution with emissivity, we can determine the emission line profile for any viewing angle relative to the observer, as demonstrated in \citet{naddaf2022}. Next, we focus on calculating the response function $\psi_{BLR}(t)$ for the BLR based on this distribution. This is accomplished by calculating the time delays for each representative cloud as observed, using a method analogous to that applied for disk emission. At this stage, we assume that the emissivity of the clouds is universal, though wavelength-dependent, which is determined through photoionization calculations for a representative cloud.

\subsection{Photoionization calculations using {\tt CLOUDY}}
\label{sect:photoionization}

To determine the BLR spectral shape, we perform photoionization calculations using the code {\tt CLOUDY}, version C23.00 \citep{2023RMxAA..59..327C}. In these computations, we simplify the 3-D cloud distribution by replacing the entire ensemble of clouds with a single representative cloud, positioned at the radius inferred from the time delay. For the analysis at this established distance, we adopt an incident radiation luminosity of $\log L$ (erg/s) = 44, and a BLR distance from the central source of $\log r$ (cm) = 16  \citep{2020ApJ...903..112D}, a constant hydrogen gas density of $\log n_H$ (cm$^{-3}$) = 11, and a column density of $\log N_H$ (cm$^{-2}$) = 23.5, following the prescriptions of \citep{korista2001,2022AN....34310091P}. The adopted distance in our representative case corresponds to the measured time delay of the H$\beta$ line in NGC 5548. The assumed bolometric luminosity is also representative of this source, based on the redshift-inferred distance within the framework of standard cosmology. In this pilot study, we do not explore a wide range of gas densities or column densities in detail, as photoionization models, particularly single-zone models, still do not perfectly represent the full complexity of BLR properties \citep[e.g.][]{netzer2020,Pandey_2025,floris2025}. We assume a metallicity five times higher than solar, as this was favored by comparisons between the FRADO model and quasar spectra \citep{naddaf2021}. Independent studies of quasar metallicity also support super-solar values \citep[e.g.][]{sniegowska2021}, and in the specific case of NGC 5548, high metallicity was advocated by \citet{netzer2020} based on line emissivity analysis. For simplicity, we neglect turbulence in the medium.

The shape of the incident SED is taken from the file "NGC5548.sed" available in the {\tt CLOUDY} database, which corresponds to the SED derived by \citet{mehdipour2015}. Their study estimated the starlight contribution to NGC 5548 using HST observations, providing a well-constrained incident SED that we adopt for our calculations. 

This is clearly an oversimplification, in two aspects. Each cloud receives the radiation of the same shape, only scaled down with the distance as specified in equation~\ref{eq:weight}. This radiation includes both the hard X-ray radiation as well as the warm corona and the disk emission. Apart from the spectral shape, we also do not differentiate here between the arrival time from the hot corona, and warm extended corona. Accounting for such differences would require modeling the BLR response separately to the hard X-ray emission from the hot corona and to the  emission from the warm corona, followed by merging several resulting transfer functions. As this is a pilot study, we have not undertaken that level of complexity at this stage, although it is feasible in principle, given that our spectral decomposition separates the contributions from all three components. For the current analysis, we adopt a fixed SED shape, a simplification that is unlikely to significantly affect our results. The hard X-ray power-law from the hot corona is consistent with SED, and the warm corona is also fitted in such way as to reproduce the spectrum. Slight error on the time delay introduced by the fact that some photons should come directly from the hot corona instead from the disk should not be essential, taking into account much larger size of the BLR than the disk.

The resulting BLR emissivity profile, $\epsilon(\lambda)$, is then incorporated into the final delay model. When we later account for the assumption of an unknown luminosity distance, the irradiating flux is adjusted accordingly to ensure consistency with the observed hard X-ray flux.

Since {\tt CLOUDY} does not compute higher-order Balmer lines, we incorporate a scaled component into the spectrum, following the method of \citet{kovacevic2014}. This modification improves the spectral fit, as demonstrated in \citet{Pandey_2025}, though it only affects the wavelength range between 3646\AA~ and 4000\AA~ in the rest-frame. The shape of this component is taken directly from \citet{Pandey_2025}; however, the emissivity still appears too low to fully capture the gradual decline observed beyond the Balmer edge in the data. More advanced modeling may be required to improve the fit in that region.

When we perform computations of the Model C for fixed luminosity distance the value of the incident flux in {\tt CLOUDY} computations is fixed and corresponds to the observed luminosity and the time delay to the BLR as above. However, later on, in the Section~\ref{sect:distance} when the luminosity distance is treated as unknown, we scale the incident bolometric luminosity to the fitted accretion rate which is a function of the luminosity distance.

\begin{table*}[]
\centering
 \caption{Parameters utilized in modeling the delay and spectral energy distribution (SED) of NGC 5548 for the three Models A, B, and C. Some values were taken from the literature, others were derived as best fit values in the current work, as noted below. }
 \label{tab:kammoun_sample}

\resizebox{18cm}{!}{
\fontsize{4pt}{4pt}\selectfont
\begin{tabular}{lccr} \hline \hline

\\
Parameters & Model A & Model B& Model C \\
 & &  &     \\
(1) & (2) & (3) & (4)
\\ \hline
 & &  &      \\
Black Hole Mass($M_{\rm BH}$) & $ 5 \times 10^{7(ref 1)}$ & $ 5.5 \times 10^{7(ref 4)}$ & $5 \times 10^{7(ref 1)}$\\ 
 Corona Height( h)& $20^{(a)}$ & $43^{(ref4)} $ & {$48.29^{(a)}$}\\ 
 Inclination Angle( i) & $40^{(ref2)}$ & $45^{(ref4)} $ & $40^{(ref2)}$\\ 
 Warm corona Inner Radius & $-$ & $43^{(ref4)}$ &$ 6^{(a)}$\\
 Warm Corona Temperature & $-$ & $1.98\times10^{6(ref4)}$ & $6.58\times10^{6(a)}$\\
 Warm corona optical depth/Photon index & $-$ & $2.28^{(ref4)}$  & $20.26^{(a)}$\\
 Inner Cold Disk Radius( $r_{in}$) & $35^{(a)}$ & $151^{(ref4)}$ & $94.87^{(a)}$\\
 Outer Cold Disk Radius( $r_{out}$) & $10000^{(a)}$&$282^{(ref4)}$&$10000^{(a)}$\\
 Color Correction & $2.4^{(ref2)}$  & $1.0^{(ref4)}$ & $1.0^{(a)}$\\ 
 BLR Contribution($f_{BLR}$) & 12$\%^{(a)}$ & 40$\%^{(a)}$& 30$\%^{(a)}$\\ 
 Lamp Luminosity($L_x$) & $ 1.26 \times 10^{44(a)}$ &$1.247 \times 10^{44(ref4)}$ & $9.68 \times 10^{43(a)}$ \\ 
 Eddington Ratio & $0.02^{(ref3)}$ & $0.027^{(ref4)}$ & $0.017^{(a)}$\\ 
 Starlight & $6.74 \times 10^{-15}$ & $5.39 \times 10^{-15}$& $6.76 \times 10^{-15}$ \\
 \hline
\\
\end{tabular}}
\vspace{0.05cm}
\begin{minipage}{\textwidth}
\footnotesize
\noindent \textbf{Note.} Columns are (1) name of the parameter used in the model, with values given in columns (2), (3) and (4) for Models A, B, and C, respectively. Black hole mass in unit of $M_{\odot}$, corona height in unit of $r_g$, inclination angle in degree, inner disk radius and outer disk radius in $r_g$, lamp luminosity in $erg~s^{-1}$.  Warm corona optical depth or a photon index are alternative (equivalent) parameters of Comptonization models. Starlight normalization is given in erg s$^{-1}$ cm$^{-2}$ \AA$^{-1}$ at 5100 \AA. References for the fixed parameters during fitting: (ref1) \citet{netzer2022}; (ref2) \citet{kammoun2021}; (ref3) \citet{Crenshaw2009}; (ref4) \citet{kubota2018}; (a) fitted parameters. For model C we assume standard $\Lambda$CDM cosmology, with $H_0 = 70$ km s$^{-1}$ Mpc$^{-1}$, $\Omega_m=0.3$.
\end{minipage}
\end{table*}

\subsection{ Starlight contribution}

The starlight contribution in NGC 5548 was estimated by \citet{mehdipour2015}, and we adopt the same level in some of our models. However, in other models, we allow for flexibility in adjusting the starlight level. To represent the starlight profile, we use the template of an Sa galaxy from \citet{kinney1996}, which is available in the Kinney-Calzetti Spectral Atlas\footnote{\url{https://www.stsci.edu/hst/instrumentation/reference-data-for-calibration-and-tools/astronomical-catalogs/the-kinney-calzetti-spetral-atlas}}.

\subsection{Combined time delay and NGC 5548 data fitting}

After analyzing each component of the central regions assumed in our model for NGC 5548, we derive the final response function by incorporating both disk reprocessing and BLR contributions, as expressed in the following equation
\begin{align}
\label{eq:convol}
\psi(\lambda,t) &= (1 - f_{\text{BLR}})\, \psi_d(\lambda,t) \nonumber \\
&\quad + f_{\text{BLR}}\, \epsilon(\lambda) \int_{t_0}^{t_{\text{max}}} \psi_d(\lambda,t - t')\, \psi_{\text{BLR}}(t')\, dt'
\end{align}
where the parameter $f_{BLR}$ defines the BLR fraction contributing to the continuum time delay.

As mentioned in Section~\ref{sect:photoionization}, we do not treat separately photons reaching the BLR directly from the hot corona and/or from the warm corona, and from the disk. Photons from the hot and warm corona are included in BLR  computations as the incident continuum (i.e. they are included in the $\epsilon(\lambda)$), but their arrival time is the same as for the photon disks under consideration. This is done for the efficiency of computations when fitting the data. Replacing the incident continuum adopted for photoionization modeling with three separate spectral components would additionally require repeating photoionization computations at every computational step.

We generate response functions for 100 wavelengths, covering the range from 1000 $\AA$ to 10000 $\AA$. Using these response functions, we then compute the time delay at each wavelength using the following equation:

\begin{equation}
\label{eq:delay}
\tau(\lambda) = \frac{\int t\psi(t,\lambda)dt}{\int 
\psi(t,\lambda)dt}
\end{equation}
To fit the data of NGC 5548, we utilize all the parameters listed in Table\ref{tab:kammoun_sample}, integrating the contributions from both the disk and the BLR as described in equation \ref{eq:convol}.

When searching for the optimal solution, we impose constraints from both the observed SED and the measured time delays. Specifically, we incorporate all 17 available time delay measurements and select 17 representative points, evenly distributed across the entire observed SED range. To systematically evaluate and compare different solutions, we integrate these constraints into a unified criterion:

\begin{equation}
\chi^2=\sum \frac{(O_i-E_i)^2}{(\delta O_i)^2}
\end{equation}
where 
$O_i=$ observed data, 
$\delta O_i=$ is the error in observed data, and 
$E_i=$ value from the model.  We assume the error in the spectral data of 7\%, and the errors of the delay measurements are taken from \citet{fausnaugh2016}.

\section{Results} \label{sec:result}

 In this section we concentrate on testing the FRADO model against the spectral and time-delay data for NGC 5548 and summarize the results obtained from our model fitting.  We fix $M_{\text{BH}}$ and the bolometric luminosity in our analysis. We assume that the distance to the source is known, and corresponds to the source redshift.  The issue of the Hubble constant determination will be addressed separately in Section~\ref{sect:distance}.

\subsection{BLR properties from FRADO}
\label{sect:BLR_response}

\begin{figure}
    \centering
    \includegraphics[scale=0.3]{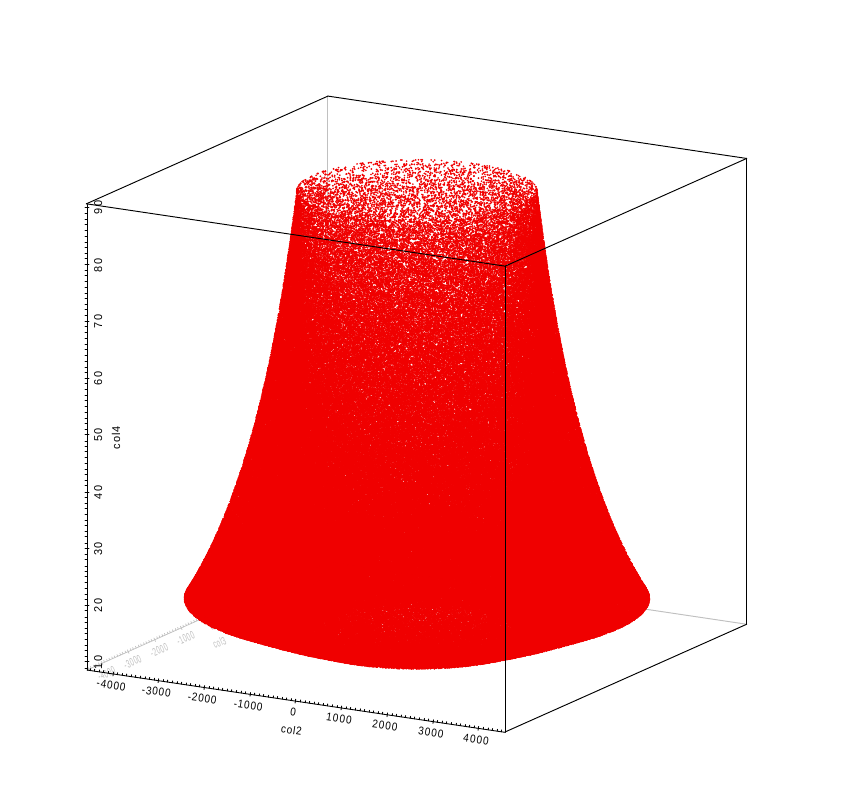}
     \includegraphics[scale=0.55]{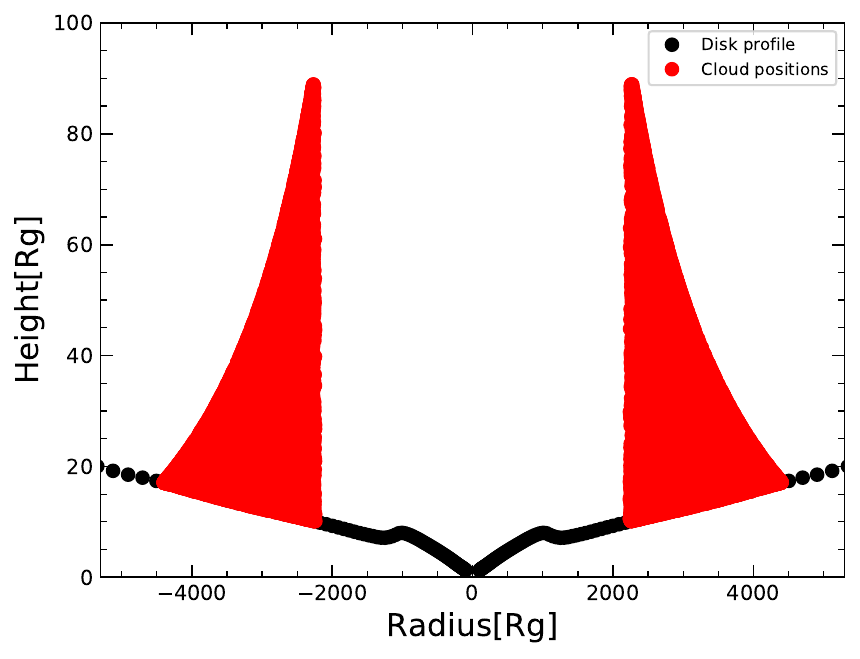}
\caption{Upper panel: 3-D plot of cloud positions from FRADO model  for NGC 5548 with $M_{BH}=5 \times 10^7 M_{\odot}$, $L/L_{Edd} = 0.02$, metallicity Z = 5 in solar units. The axes are in units of $r_g$. Bottom panel: a cross-section of cloud positions for $y>-300 \, r_g$ and $y<300 \, r_g$. The black line indicates the thickness of the Keplerian disk. Clouds form a geometrically thin complex layer above the disk.}
    \label{fig:cloud}
\end{figure}

\begin{figure}
    \centering
    \includegraphics[scale=0.52]{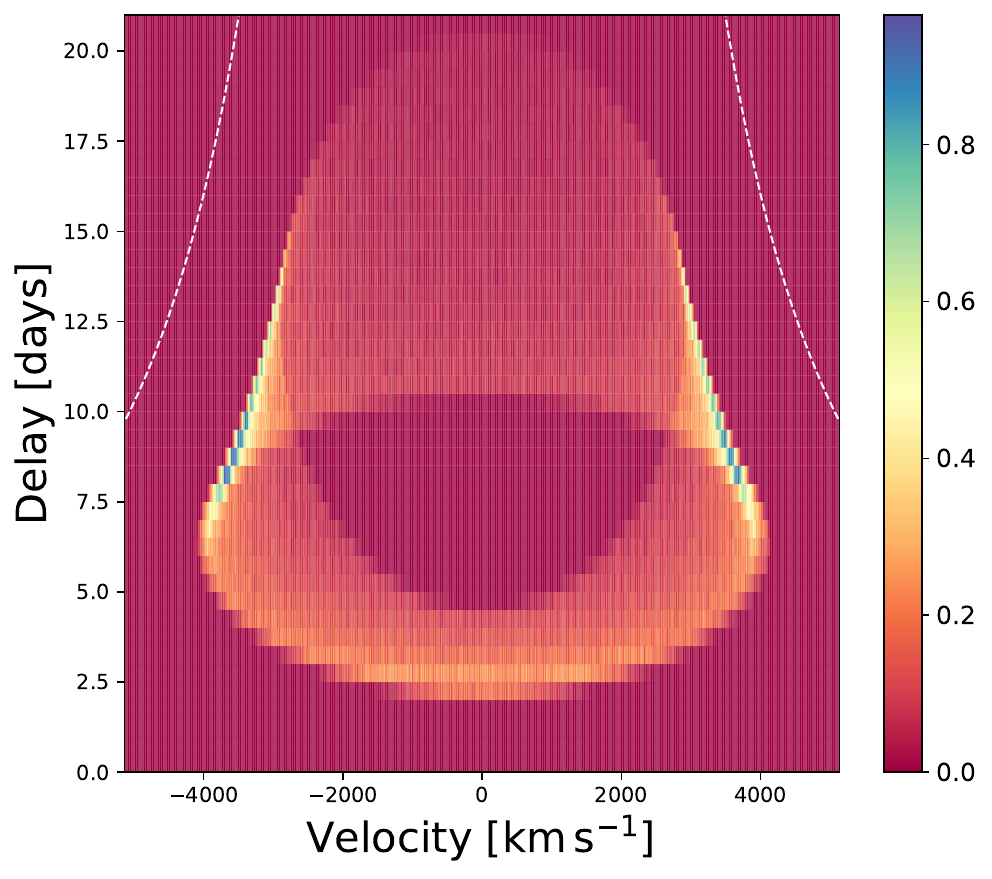}
    \includegraphics[scale=0.54]{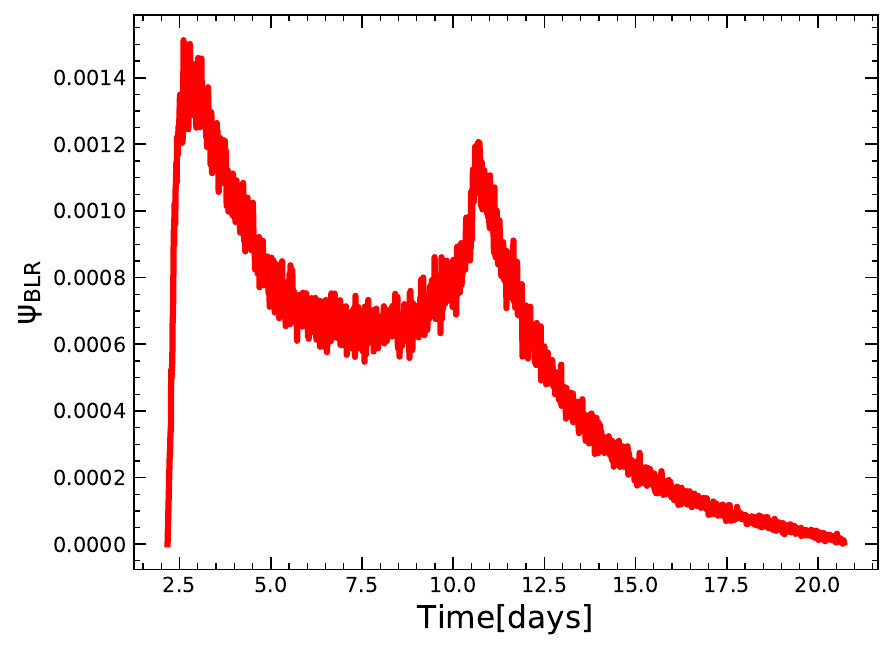}
    \caption{ Upper panel: Velocity–delay map of NGC 5548 constructed from FRADO model with inclination angle $i = 40$ degrees. Color coding represents the number density of BLR clouds in each velocity-delay bin. The white dashed line shows the virial envelope corresponding to Keplerian disk-like rotation for $v^2 \times \tau$ = constant with $M_{\mathrm{BH}} = 5 \times 10^7 \,M_{\odot}$.
    Bottom panel: BLR response function generated using the cloud positions from the model shown in Figure~\ref{fig:cloud}, assuming an inclination angle of $i = 40$ degrees.} 
    \label{fig:blr_response}
\end{figure}

\begin{figure}
    \centering
    \includegraphics[scale=0.5]{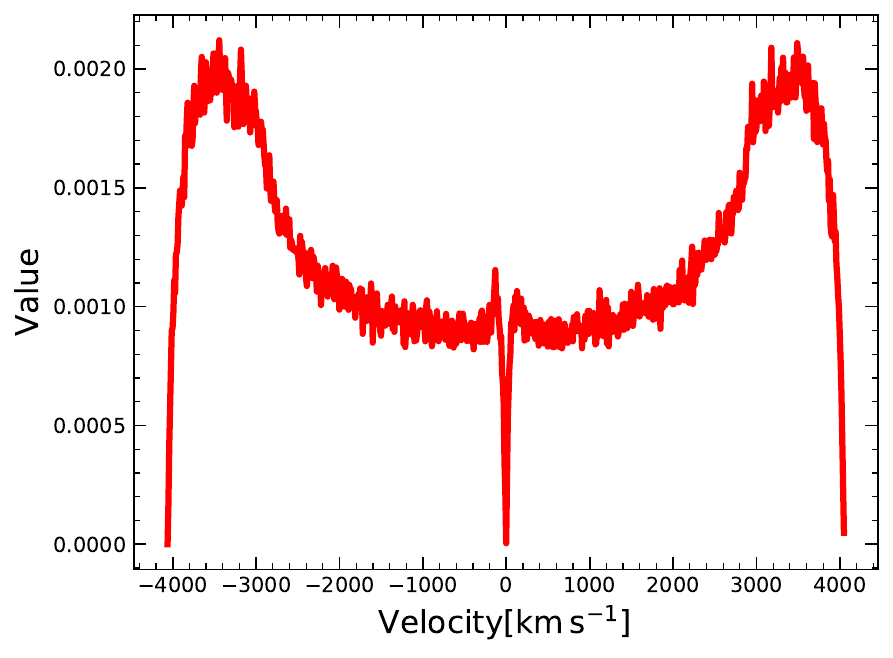}
    \caption{Line profile generated by the FRADO model using the parameters from Figure~\ref{fig:cloud}. 
    }
    \label{fig:line_shape}
\end{figure}

The global model parameters adopted for NGC 5548, combined with the assumption of a dust sublimation temperature of 1500 K, allow us to uniquely determine the structure of the BLR. Notably, the FRADO model is specifically applicable to low-ionization lines such as H$\beta$, Mg II, and Fe II. In this framework, the cloud positions within the BLR are determined by the FRADO model, with parameters relevant to NGC 5548: $M_{\rm BH} = 5 \times 10^7 M_{\odot}$ \citep{netzer2022}, an Eddington ratio of 0.02 \citep{Crenshaw2009}, and an assumed metallicity 5 times solar.

We present the distribution of clouds forming the BLR in NGC 5548 in Figure~\ref{fig:cloud}. The upper panel presents the overall 3-D structure, revealing a complex cone-like configuration where no clouds escape to infinity due to the low Eddington rate. The lower panel provides a cross-section of this distribution, with the black line marking the thickness of the underlying Keplerian disk. This disk structure accounts for the effects of radiation pressure and opacity, as described by \citet{rozanska1999}, both of which play a crucial role in shaping its complex geometry. A notable feature is the abrupt change in disk height, corresponding to the transition from the inner radiation pressure-dominated region to the outer gas-dominated region, which significantly impacts the cooling efficiency of the disk and its overall structure.

In this model, clouds are launched from the disk surface but remain relatively close to it, with the ratio of vertical height ($\boldsymbol{z}$) to radial distance ($r$) staying below 5$\%$. This limited height is a direct consequence of the low Eddington ratio, giving the BLR a geometry that closely resembles a puffed-up disk surface. The $\boldsymbol{z}/r$ ratio  in the model peaks near the inner edge of the BLR, which is defined by the dust sublimation temperature reached at approximately $\sim$ 2260 $r_g$  for the adopted $M_{\text{BH}}$, Eddington rate and metallicity. At this radius, the Keplerian velocity is $\sim 6300$ km s$^{-1}$ for this $M_{\text{BH}}$, implying that the full width at half maximum (FWHM) of the emission lines could reach up to twice this value. This result based on FRADO model is consistent with the H$\beta$ FWHM of $10161 \pm 587$ km s$^{-1}$ reported by \citet{pei2017}. However, contributions from larger radii act to moderate this broadening.

The position of the clouds and their velocity field allow us to construct a 2-D velocity-delay map (see Figure~\ref{fig:blr_response}, upper panel) which we consider as representative for H$\beta$ line, assuming a viewing angle of $40$ degrees, which is likely representative of NGC 5548 \citep{pancoast2014,cappi2016,wildy2021,horne2021}. Our velocity–delay map closely resembles the predictions of a simple flat Keplerian disk model, as the vertical velocities are minimal, no outflows are included, and no selective shielding is assumed. It shows some similarity in the overall shape to the map derived observationally by \cite{horne2021} (see their Figures 5 and 7).

However, the main inconsistency lies in the narrower velocity extent observed in our velocity-delay map, which shows emission up to $\mathrm{\sim \pm 4000 \, km \, s^{-1}}$, while \cite{horne2021} reported observed H$\beta$ emission extending to $\mathrm{\sim \pm 8000-10000 \, km \, s^{-1}}$, although the majority of the response is concentrated within $\sim \pm5000 \, \mathrm{km \, s^{-1}}$. Note that, FRADO inherently assumes that the BLR originates beyond the dust sublimation radius, which is appropriate for low-ionization lines like H$\beta$ and thus naturally predicts narrower velocity distributions, consistent with emission from outer, slower-moving regions of the BLR. While, the broader velocity distributions observed by \cite{horne2021} are shaped by contributions from the inner, dust-free high-ionization zones (dominated by CIV, He II lines) of the BLR, where gas orbits closer to the black hole at higher Keplerian speeds. 

Nevertheless, the exact comparison is difficult since the observational H$\beta$ map in \cite{horne2021} seems to be strongly contaminated by He II emission. Notably, while the red wing of the observed H$\beta$ shows a Keplerian structure, the blue wing appears significantly weaker, suggesting substantial He II contamination. In our model, in its present form, we cannot explain such asymmetry. However, such asymmetries in the H$\beta$ line are well-documented and known to evolve over time in a quasi-periodic fashion \citep{shapovalova2004}. \citet{bon2016} explained such phenomenon as caused by the presence of the secondary black hole, perturbing BLR, by precession of the outer part of the disk, or spiral waves in the disk. All such mechanisms will give asymmetry between blue and red-shifted wing of the flat BLR coexisting with the disk. In the future we could include the warped disk into FRADO model, but this is beyond the scope of the current paper.

Next, we derive the BLR response function by projecting the map onto time axis. The resulting shape, shown in Figure \ref{fig:blr_response}, lower panel, exhibits two distinct peaks: one at shorter time delays, around $\sim$ 2.6 days, corresponding to the clouds on the same side as the observer, and the other peak arising from the clouds on the opposite side of the black hole. The shape of our modeled response function does not perfectly replicate the observed one reported by \citet{horne2021}. In their analysis using the MEMECHO software for H$\beta$ (see Figure \ref{fig:response_comp} for comparison), the projected response exhibits two peaks: a prominent one at approximately 2 days and a second, much smaller peak at around 25 days. While our first peak aligns well with theirs at $\sim$2 days, our second peak appears closer in time, at $\sim$11 days, and is relatively stronger than the second peak reported by \citet{horne2021}. Despite these differences, it is important to emphasize that our model relies solely on global parameters, without introducing arbitrary values such as inner or outer BLR radii. This absence of free parameters enhances the significance of the observed similarity between the modeled and empirical response functions. Additionally, we assess the consistency between the BLR response function recovered from our FRADO model and the H$\beta$ emission-line response function derived from spectroscopic RM data by \citet{horne2021} in Section~\ref{ss:apendix}.

By integrating the BLR response profile derived from the FRADO model, we calculate a mean delay of 8.101 days for a viewing angle of 30 degrees and 8.105 days for a viewing angle of 40 degrees. This delay is measured relative to the X-rays. To compare it with delays at other wavelengths, such as the commonly used 5100 \AA, it is essential to account for the net time delay between the X-rays and the specific wavelength of interest.

The measured time delay of the H$\beta$ line relative to the 5100 \AA~ continuum during the 2014 campaign (January to June) was reported by \citet{pei2017} as $4.17^{+0.36}_{-0.3}$ days, which is shorter than predicted by the radius--luminosity relation. This H$\beta$ delay ranks among the shortest ever recorded for NGC 5548. In contrast, the mean response function from the 1998$-$2001 campaign peaked around 20 days \citep{cackett2006}. \citet{horne2021} later explored the nature of this shorter delay in detail, finding that the response function exhibited a secondary peak at approximately $\sim 25 $ days but began with a high value near to zero time delay. Their observed map also supports the primary response originating from the near side of the BLR, facing the observer.

To compare the predicted and observed time delays, we need to add the delay of the 5100 \AA~ continuum relative to X-rays to the observed data. Since the exact time delay at 5100 \AA~ was not measured by \citet{fausnaugh2016}, we averaged their measurements around this wavelength and included the hard X-ray time delay, also provided in their study. This results in an estimated observed time delay for H$\beta$ relative to X-rays of $6.83 \pm 0.53$ days. Our predicted delay is slightly longer. This discrepancy may stem from the long tail in the delay, which the model accounts for but may not be fully captured in the observations. Alternatively, the delay could be shorter if there is intervening material between the black hole and the BLR, potentially shielding part of the most distant flow. Another possibility is that the dust temperature in the BLR model, as discussed by \citet{2020ApJ...902...76P} and \citet{naddaf2021}, differs from the assumed 1500 K.

Additionally, the model predicts an upper limit for the covering factor of the BLR. If the cloud distribution is not transparent, the region intercepts all radiation emitted within inclinations greater than the aspect ratio $\boldsymbol{z}/r$ of the cloud distribution. As shown in Figure~\ref{fig:cloud}, the model suggests that less than $4\%$ of the radiation can be intercepted by the BLR, as the cloud height is generally small compared to the radius. This is notably lower than the $10\%$ to $30\%$ covering factor typically expected from BLRs \citep[e.g.][]{baldwin1995, korista2019, Panda_cafe2021, 2022AN....34310091P}. In contrast, the yearly variations observed in the line profile of NGC 5548 suggest that the disk can be distorted and/or precessing \citep[e.g.,][see their Figure~2]{shapovalova2009}. Such a departure from the disk model strictly perpendicular to the symmetry axis may affect the irradiation of the launched clouds and this way to modify the BLR emissivity.

The distribution of BLR clouds enables us to roughly estimate the shape of the H$\beta$ line predicted by the model. In this estimation, we neglect the vertical cloud velocities and only consider the projected rotational velocities towards the observer. The resulting line shape, shown in Figure~\ref{fig:line_shape}, reveals a two-peak structure that is even more pronounced than that observed in the data. The FWHM predicted by our model approaches 8000 km s$^{-1}$. In comparison, the measured FWHM of the H$\beta$ line from the rms spectrum was $10161 \pm 587$ km s$^{-1}$ \citep{pei2017}. Notably, the FWHM values have varied over the years, indicating the dynamic nature of the BLR. For instance, \citet{peterson2004} reported an H$\beta$ FWHM of $7202 \pm 392$ km s$^{-1}$, which is in good agreement with our model estimate. These variations suggest that our model captures a plausible state of the BLR, consistent with the historical range of observed line widths.

\subsection{The BLR emissivity}

\begin{figure}
    \centering
    \includegraphics[scale=0.5]{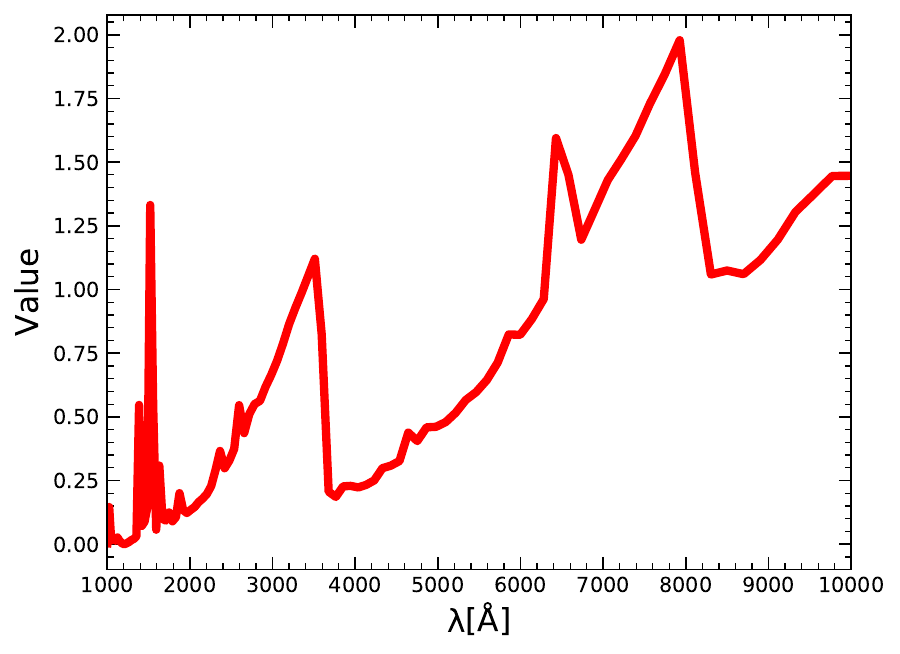}
\caption{ Emissivity profile (ratio of the reprocessed to the incident continuum) of a BLR cloud for the adopted parameters: $\log n_H~[\text{cm}^{-3}] = 11$, $\log L~[\text{erg}~\text{s}^{-1}] = 44$, and the BLR distance of $10^{16}$ cm.
}
    \label{fig:cloudy1}
\end{figure}

To model the emission characteristics of photoionized clouds, we perform {\tt CLOUDY} simulations that account for both line and continuum contributions. During our computations, we consider both line and continuum emissions, incorporating radiation from both inward and outward directions to ensure equal visibility of the illuminated and un-illuminated sides of the clouds. Among the most prominent features that emerge are the Balmer and Paschen edges, as illustrated in Figure~\ref{fig:cloudy1}. The depths of these edges are governed by the specific model parameters adopted \citep[see the most recent study by][]{pandey2023}.


The reprocessing calculations are performed for a single representative cloud, using the full shape of the SED as the incident radiation, as described in Section~\ref{sect:photoionization}. This approach is admittedly a significant simplification, since in reality each cloud is exposed to a different incident spectrum depending on its location and orientation. This spectral variation is properly accounted for in the computation of the radiation force. However, generating comprehensive tables of local emissivity for the full range of incident spectra lies beyond the current scope of this work. Such an extension is essential for future studies to improve the physical realism of the model.

\subsection{Exemplary shape of the response function}

\begin{figure}
    \centering
    \includegraphics[scale=0.5]{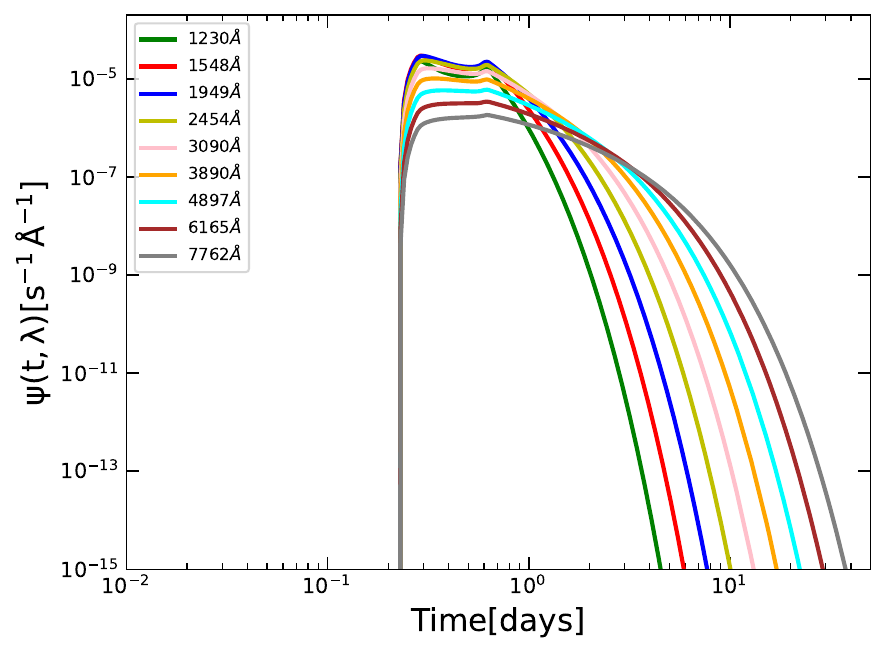}
    \includegraphics[scale=0.5]{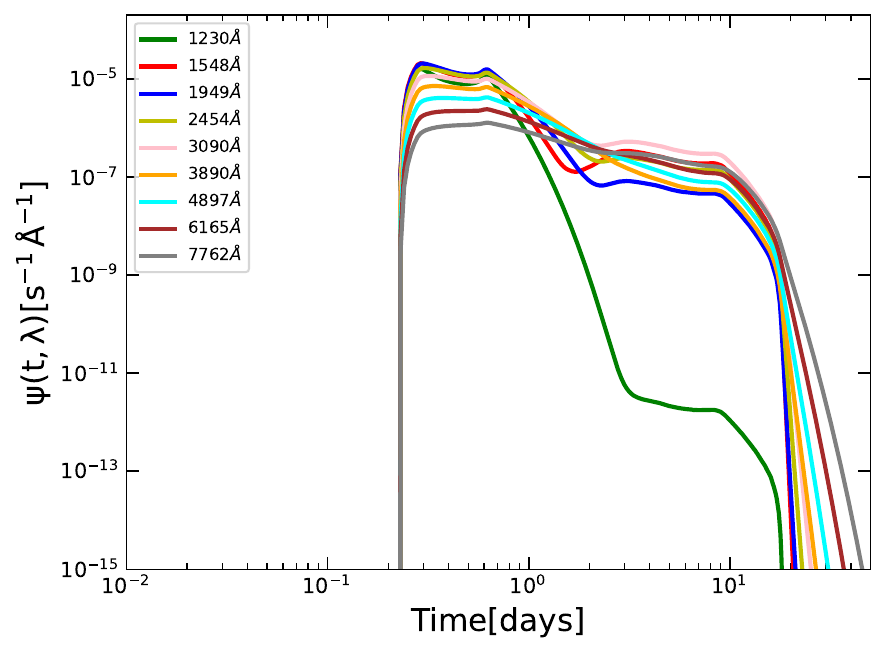}
    \caption{Upper panel: Response functions of the disk at different wavelengths. Bottom panel: Combined response functions of the disk and BLR across the same wavelengths. These representative response functions are obtained for Model C. Parameters used: $M_{\text{BH}}$ = $5.0 \times10^{7}M_{\odot}$, Eddington ratio = 0.015, $L_X = 9.68 \times10^{43}$ erg s$^{-1}$, height: $h = 48.29 r_g$, $R_{in} = 94.87 r_g$, $R_{out} = 10000 r_g$, and viewing angle: $i = 40$ degrees. }
    \label{fig:response}
\end{figure}

After deriving the BLR response function using FRADO model and the emissivity profile from {\tt CLOUDY} for NGC 5548, we compute the disk response function and the combined disk-BLR response function using equations \ref{eq:resp} and \ref{eq:convol}, respectively. These results are presented for nine different wavelengths in the upper and bottom panels of Figure~\ref{fig:response}. The disk response function (upper) displays a single-peaked structure, whereas the combined response function (bottom) exhibits a bimodal shape, as expected, with the first peak corresponding to direct emission from the disk, and the second arises from delayed reprocessing by the BLR. Moreover, the overall time delay associated with the combined response function is significantly longer than that of the disk alone, consistent with the findings of \citet{netzer2022}.

\subsection{Time delays in representative models}
\label{ss:models}

This paper serves as a pilot study, focusing on three distinct global setups rather than exploring the full parameter space. In this section, we assume that the source distance can be determined from its redshift, enabling a unique conversion of bolometric luminosities to fluxes. We define these setups as Model A, Model B, and Model C, which were shortly introduced in Section~\ref{sec:spectral}. Each setup is characterized by different global parameters. In all cases $M_{\rm BH}$ was fixed (in Models A and C, the $M_{\rm BH}$ is derived from observational data \citep{netzer2022}), whereas in Model B, it is adopted from the corresponding model reference  \citep{kubota2018}.  All models are supplemented with the BLR components. A detailed discussion of each model follows in the subsequent sections.

\subsubsection{Model A}

\begin{figure}
    \centering
    \includegraphics[scale=0.5]{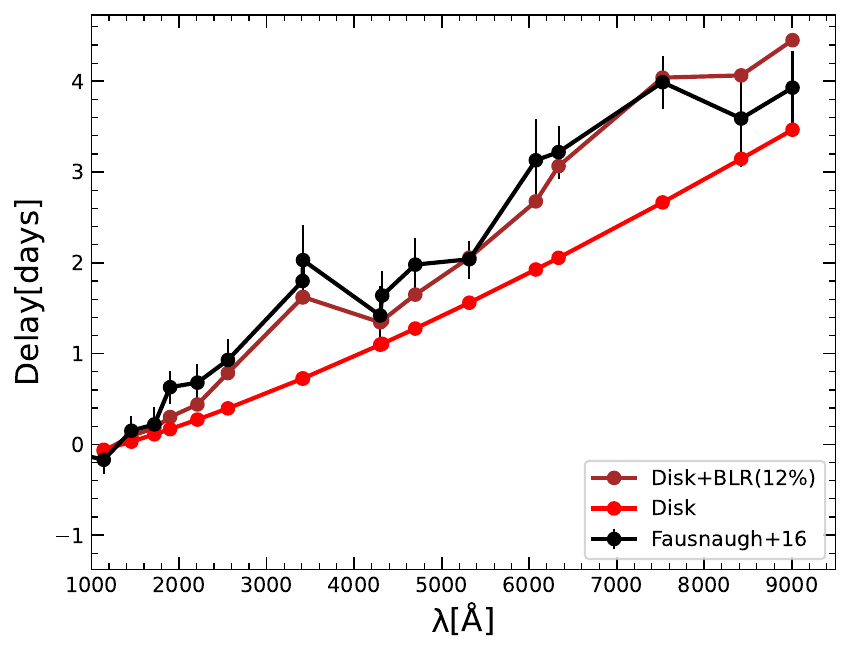}
    \includegraphics[scale=0.5]{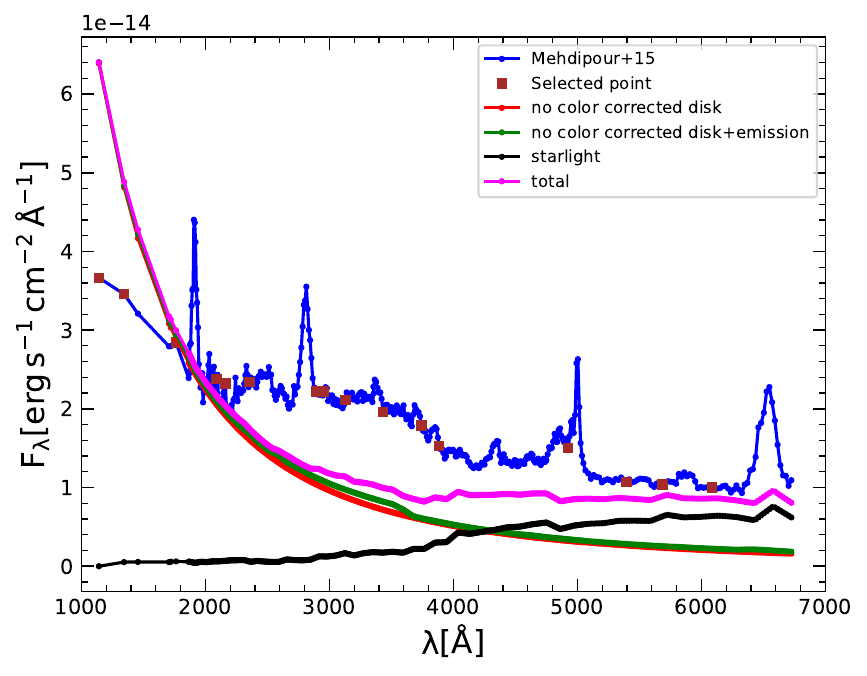}
    \caption{{\bf Model A.} Upper panel:The observed delay from \citet{fausnaugh2016} is represented in black. The delay calculated from the disk response function alone is shown in red, while the delay calculated from the combined response function of the disk and BLR is shown in brown. Bottom panel: The observed SED from \citet{mehdipour2015} is shown in blue. The disk SED is represented in red, while green illustrates the disk SED combined with the emission lines. The starlight contribution is shown in black color. The final SED, incorporating contributions from the disk, star, and emission lines, is shown in purple. The selected points used for estimating the  $\chi^2$ value are represented by brown squares. Parameters: $M_{\rm BH}$ = $5.0 \times10^{7}M_{\odot}$, Eddington ratio = 0.02, $L_X = 1.26 \times {44}$ erg s$^{-1}$, height: $h = 20 r_g$, $R_{in} = 35 r_g$, $R_{out} = 10000 r_g$, color correction$=2.4$, viewing angle: $i = 40$ degrees.}
    \label{fig:modela}
\end{figure}

To account for disk atmosphere effects, Model A employs a standard accretion disk framework with a fixed color correction. It uses well-established literature values for $M_{\rm BH}$ and Eddington ratio, and does not include the warm corona component. With only a few free parameters, the model incorporates the BLR contribution alongside the accretion disk emission to fit the observed continuum time delays and spectral energy distribution.
 We minimize the number of free parameters by relying on the basic parameter values from the literature, including the $M_{\rm BH}$, Eddington rate, color correction, and viewing angle (see Table~\ref{tab:kammoun_sample} for the adopted values). Specifically, the adopted color correction of 2.4 comes from the studies of \cite{kammoun2021, kammoun2023}. The model does not include the warm corona, and the cold disk extends down to the ISCO (6 $r_g$). The BLR model is then directly determined based on the assumed $M_{\rm BH}$, Eddington rate, and a fixed metallicity of 5, but the covering factor is left as a free parameter of the model. The other key parameter is the corona height. The resulting predictions are presented in Figure~\ref{fig:modela}. 

The model reproduces the time delay quite well, the Balmer edge at $\sim 3600$ \AA~ is well reproduced, the region of the Paschen jump is not so well fitted. The covering factor and the corona height optimizing the fit of the time delay are given in Table~\ref{tab:kammoun_sample}.

Below, we outline the key advantages and limitations of this model.

\begin{figure}
    \centering
    \includegraphics[scale=0.5]{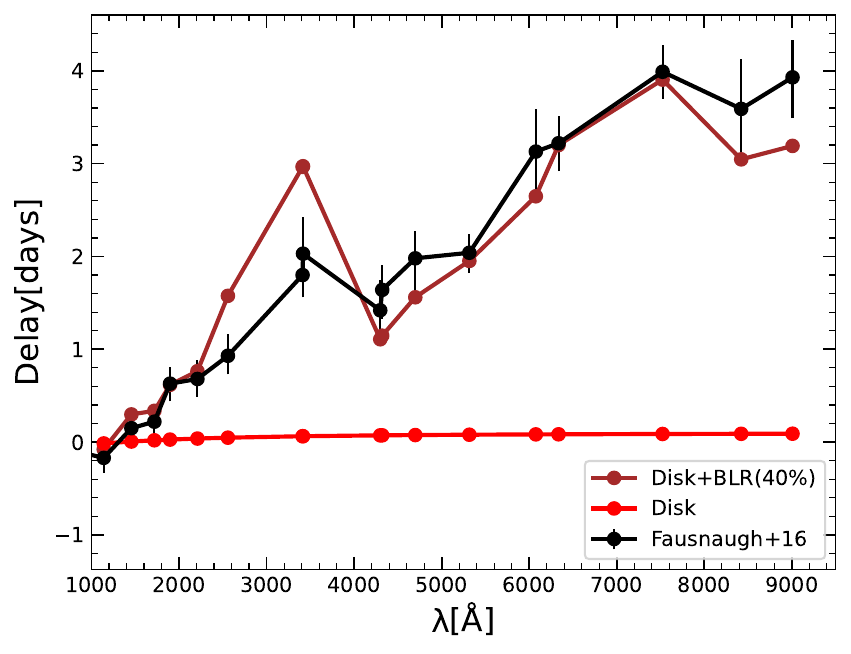}
    \includegraphics[scale=0.5]{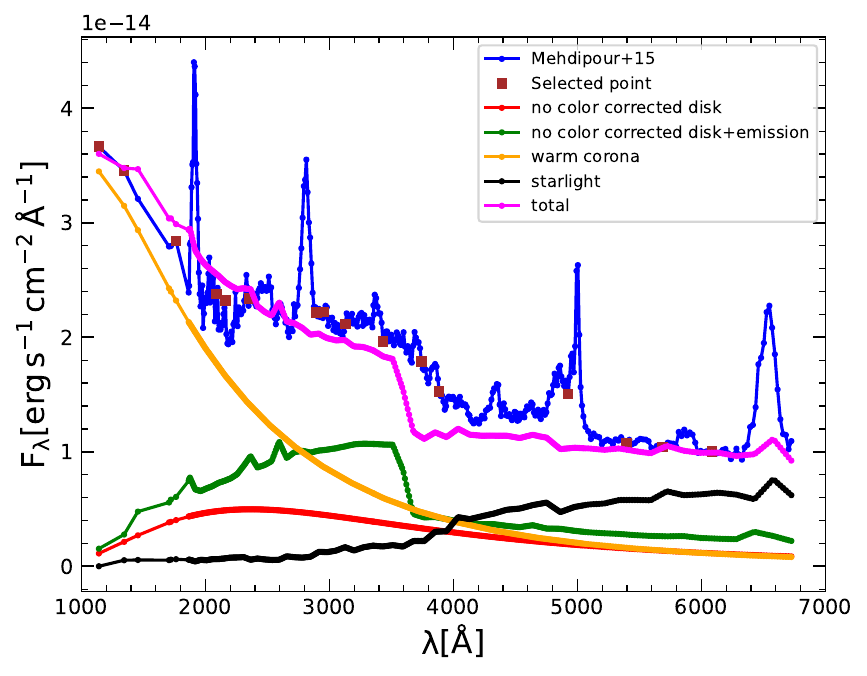}
    \caption{{\bf Model B. }Upper panel: The observed delay from \citet{fausnaugh2016} is represented in black. The delay calculated from the disk response function alone is shown in red, while the delay calculated from the combined response function of the disk and BLR is shown in brown. Bottom panel: The observed SED from \citet{mehdipour2015} is shown in blue. The disk SED is represented in red, while green illustrates the disk SED combined with the emission lines. The starlight contribution is represented in black, while the warm corona is shown in orange. The final SED, incorporating contributions from the disk, star, warm corona, and emission lines, is shown in purple. The selected points used for estimating the $\chi^2$ value are represented by brown squares.  Parameters: $M_{\text{BH}}$ = $5.5 \times10^{7}M_{\odot}$, Eddington ratio = 0.02, $L_X = 1.247 \times {44}$ erg s$^{-1}$, height: $h = 43 r_g$, $R_{in} = 151 r_g$, $R_{out} = 282 r_g$, color correction$=1.0$, viewing angle: $i = 45$ degrees.}
    \label{fig:modelb}
\end{figure}

Model advantages:

\begin{itemize}
\item three free parameters: starlight level, inner disk radius, BLR fraction
\item the H$\beta$ delay is approximately recovered
\item the double peak shape is approximately recovered
\item the time delay is well modeled
\end{itemize}

Model limitations:
\begin{itemize}
\item The optical/UV SED is inconsistent with the observed data
\end{itemize}

In Model A, the discrepancy between the observed and modeled SED is substantial in the blue part, and cannot be resolved by adjusting any model parameters without making the time delay fit worse. The required lamp height lead to a very blue spectrum in the UV due to the color correction applied. The time delay is primarily driven by the disk response, with the BLR contribution appearing only near the Balmer edge$-$contrary to the findings of \citet{netzer2022}. This discrepancy arises because the disk contribution is underestimated due to the fixed color correction being set too high. While this increases the time delay by pushing the fixed-color emission outward, it simultaneously shifts the disk spectrum toward the far-UV. Consequently, we do not include the SED fit from Model A in the $\chi^2$ minimization, as further adjustments would not improve the fit.

The starlight contribution in our decomposition is $6.74 \times 10^{-15}$ erg s$^{-1}$ cm$^{-2}$ \AA$^{-1}$, comparable to reported value of $6.2 \times 10^{-15}$ erg s$^{-1}$ cm$^{-2}$ \AA$^{-1}$ \citep{mehdipour2015}. Furthermore, this model fails to reproduce both soft and hard X-rays, as it includes only a cold disk. As a result, there is an inconsistency between the shape of the incident continuum assumed in the {\tt CLOUDY} modeling and the continuum produced by the model itself.

\subsubsection{Model B}
\label{sect:modelB}

In Model B, we adopt the framework of \citet{kubota2018}, where the innermost part of the accretion disk is replaced by a hot corona. In our setup, the standard accretion disk extends from 151 $r_g$ to 282 $r_g$, and we do not apply any color correction since the disk in this region remains cold. The global system parameters and the inclination angle are consistently taken from \citet{kubota2018}.  The free parameters in the model are the BLR contribution, $f_{BLR}$, and the level of starlight. The corona height is fixed at the value of the inner radius of the warm corona,  i.e., 43 $r_g$.

The results of this model are presented in Figure~\ref{fig:modelb}. As in Model A, we employ the same BLR response function (see Section~\ref{sect:BLR_response}). However, this approach may introduce minor inconsistencies due to slight differences in $M_{\rm BH}$ and accretion rate. While the time delay fit remains broadly satisfactory, the SED is again not well reproduced. The reduced inner disk radius results in a steep decline in the disk spectrum at shorter wavelengths. 

In this configuration, the total time delay is predominantly influenced by the BLR, with only a minimal contribution from the accretion disk. Consequently, the contribution from BLR to the time delay is significantly higher than in Model A. Additionally, the amount of starlight in this case is $5.39 \times 10^{-15}$ erg s$^{-1}$ cm$^{-2}$ \AA$^{-1}$, less than corresponding value in Model A. 

Model advantages:
\begin{itemize}
\item only two free parameters: starlight level, BLR fraction
\item the H$\beta$ delay is approximately recovered
\item the double peak shape is approximately recovered
\item the time delay is reasonably well modeled
\end{itemize}

Model limitations:
\begin{itemize}
\item   Although the high-energy portion of the SED is reasonably well recovered, this model fails to accurately fit the 3000$-$4000 $\AA$ wavelength range, which includes the Balmer jump, and shows some inconsistencies at longer wavelengths in the optical regime. This is related to the limited flexibility due to a small number of free parameters.
\item The modeled time-delay spectrum is predominantly influenced by the BLR, with negligible contribution from the accretion disk, in contrast to observed AGN time-delay spectra that exhibit significant contributions from both the accretion disk and the BLR \citep{2025ApJ...985...30M}.

\end{itemize}

\begin{figure}
    \centering
    \includegraphics[scale=0.5]{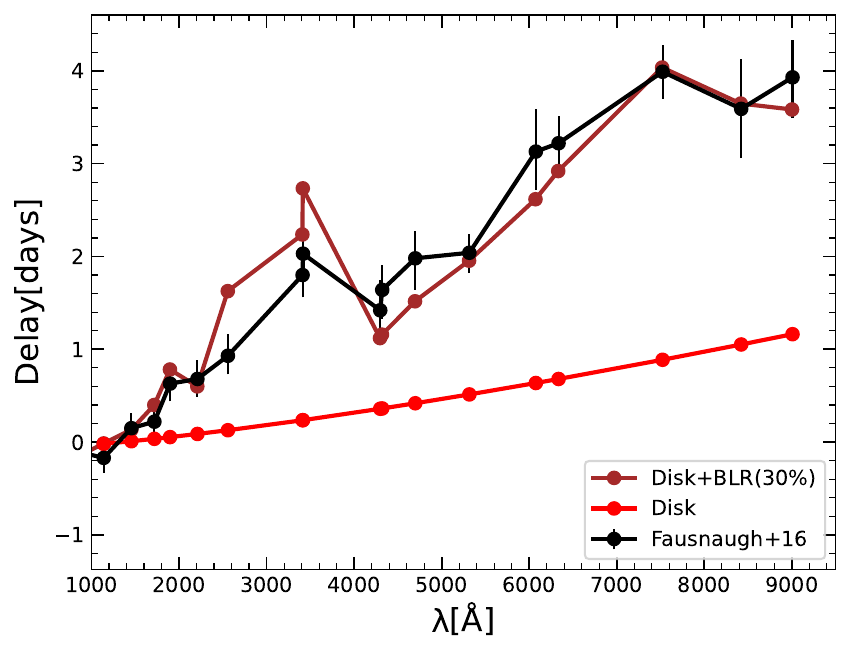}
    \includegraphics[scale=0.5]{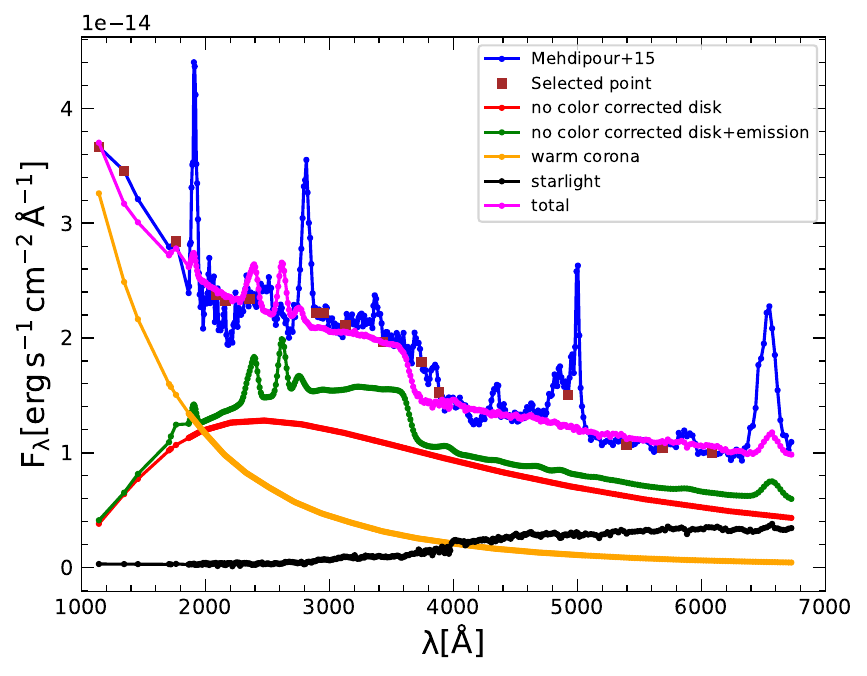}
    \includegraphics[scale=0.42]{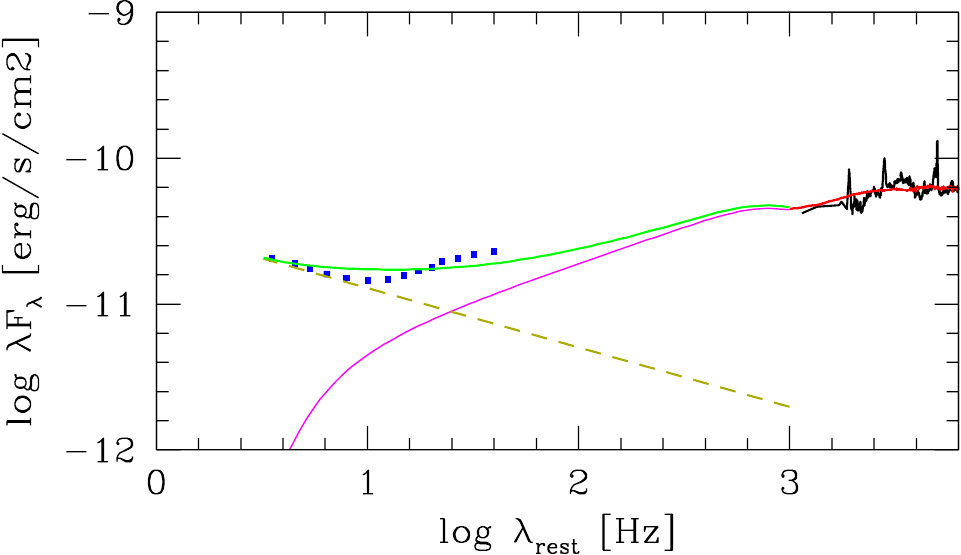}
    \caption{{\bf Model C.} Upper panel: The observed delay from \citet{fausnaugh2016} is represented in black. The delay calculated from the disk response function alone is shown in red, while the delay calculated from the combined response function of the disk and BLR is shown in brown. Middle panel: The observed SED from \citet{mehdipour2015} is shown in blue. The disk SED is represented in red, while green illustrates the disk SED combined with the emission lines.The starlight contribution is represented in black, while the warm corona is shown in orange. The final SED, incorporating contributions from the disk, star, warm corona, and emission lines, is shown in purple. The selected points used for estimating the $\chi^2$ value are represented by brown squares. Parameters used: $M_{\text{BH}}$ = $5.0 \times10^{7}M_{\odot}$, Eddington ratio = 0.015, $L_X = 9.68 \times10^{43}$ erg s$^{-1}$, height: $h = 48.29 r_g$, $R_{in} = 94.87 r_g$, $R_{out} = 10000 r_g$, and viewing angle: $i = 40$ degrees. Bottom panel: X-ray--UV-optical SED fitting. The warm corona is represented by the magenta line, the hard X-ray power law is shown in dashed gold, the green line represents the sum of the two, and the black line displays the optical data.
     }
    \label{fig:modelc}
\end{figure}

\subsubsection{Model C}
\label{sect:modelc_results}
\begin{table}[]
\centering
 \caption{Time delay measurements.}
 \label{tab:delay_table}
\resizebox{8cm}{!}{
\fontsize{3pt}{3pt}\selectfont
\begin{tabular}{lcc} \hline \hline

\\

 Wavelength & $\tau_{centroid}$ & $\tau_{model}$ \\
$~~~~~~(\AA)$ & (days) & (days)\\ 
\\ \hline
$1158$& $-0.17^{+0.16}_{-0.16}$ & ${-0.14}$\\
$1479$& $0.15^{+0.18}_{-0.16}$ & ${0.13}$\\
$1746$& $0.22^{+0.16}_{-0.19}$ & ${0.39}$\\
$1928$& $0.63^{+0.19}_{-0.18}$ & ${0.78}$\\
$2246$& $0.68^{+0.19}_{-0.20}$ & ${0.60}$\\
$2600$& $0.93^{+0.20}_{-0.23}$ & ${1.62}$\\
$3467$& $1.80^{+0.24}_{-0.24}$ & ${2.23}$\\
$3472$& $2.03^{+0.43}_{-0.39}$ & ${2.73}$\\
$4369$& $1.42^{+0.36}_{-0.33}$ & ${1.12}$\\
$4392$& $1.64^{+0.31}_{-0.27}$ & ${1.15}$\\
$4776$& $1.98^{+0.34}_{-0.29}$ & ${1.51}$\\
$5404$& $2.04^{+0.22}_{-0.20}$ & ${1.95}$\\
$6176$& $3.13^{+0.41}_{-0.46}$ & ${2.61}$\\
$6440$& $3.22^{+0.30}_{-0.29}$ & ${2.92}$\\
$7648$& $3.99^{+0.29}_{-0.29}$ & ${4.03}$\\
$8561$& $3.59^{+0.53}_{-0.54}$ & ${3.64}$\\
$9157$& $3.93^{+0.44}_{-0.40}$ & ${3.58}$\\
\hline
\end{tabular}}
\vspace{0.05cm}

\parbox{\linewidth}{
        \vspace{1em} 
        \noindent
        \textbf{Note.} Columns are (1) pivot-wavelength of the filters in the observed-frame, (2) measured inter-band delays from \cite{fausnaugh2016} in the rest-frame, (3) recovered inter-band delays from Model C. The inter-band delays are with respect to HST 1367 {\AA}.}

 \end{table}

This model represents a generalization of Model B. Rather than fixing certain parameters to the values derived by \citet{kubota2018}, we treat them as free parameters and allow the data to constrain them through fitting. First, we allow for the variation of the radii dividing the zones of hot corona, warm corona and outer disk. We assume that the outer radius of the cold disk is large. Then, we vary the warm corona parameters (optical depth, temperature) as well as the Eddington rate, to achieve the best fit to the broad band spectrum (including X-rays), and to time delays. Model parameters are listed in Table~\ref{tab:kammoun_sample}.

The time delays recovered from this model, along with the observed inter-band delays from \citet{fausnaugh2016}, are presented in Table~\ref{tab:delay_table}.  The corresponding fitted SEDs, are displayed in the upper, middle, and bottom panels of Figure~\ref{fig:modelc}, respectively. The Balmer edge is well reproduced in the lag-spectrum. Paschen edge is expected to be less prominent, as shown by \citet{pandey2023} at the basis of CLOUDY simulations. 

Additionally, the broad-band SED is now well represented, with the starlight continuum of $6.76 \times 10^{-15}$ erg s$^{-1}$ cm$^{-2}$ \AA$^{-1}$, which is nearly identical to the value of $6.2 \times 10^{-15}$ erg s$^{-1}$ cm$^{-2}$ \AA$^{-1}$ reported by \citet{mehdipour2015}. 
The middle panel of the figure provides a detailed view of the optical/UV portion of the spectrum, consistent with previous presentations. Furthermore, the bottom panel displays the full X-ray-UV-optical spectral range on a logarithmic scale, allowing for clearer visualization of the broad-band fit. In this representation, both the hard X-ray component and the contribution of the warm corona to the X-ray band are clearly visible. This contrasts with Model A, which lacked a warm corona and therefore failed to reproduce the observed soft X-ray data. In comparison, Model B, by construction from \citet{kubota2018}, is consistent with the X-ray band and thus its X-ray spectral range is not plotted in this figure. However, it is important to note that \citet{kubota2018} did not fit the spectroscopic optical data, and as a result, the discrepancy between their model and the actual optical observations was not apparent in their analysis.

The disk contribution to the time delay in this model falls between the contributions in Model A and Model B. Overall, the time delay is reproduced much more accurately compared to the other two models, providing a better fit to the observed data. The contribution of the corona to the spectrum is dominating the part of the spectrum below $\sim 2000$ \AA.

Model advantages:

\begin{itemize}
\item limited number of parameters (although larger than in Model B)
\item the H$\beta$ delay is approximately recovered
\item the double peak shape is approximately recovered
\item the time delay is well modeled
\item We successfully reproduce the Balmer jump in the observed SED, and our modeled SED is consistent with the observed spectrum.
\end{itemize}

In conclusion, this model offers a significant improvement in both time delay and SED fitting, with a more realistic representation of the BLR contribution to the observed continuum time delays and a better overall alignment with the observed spectral features. 

\section{Distance Estimation from generalized Model C}
\label{sect:distance}

Having considered the final fit from Model C as successful, we now assess whether our model improves the determination of the Hubble constant for this source, as done by \citet{cackett2007}. To begin, we first explore analytical estimates and subsequently evaluate the method that should be employed for actual data fitting.

\subsection{Analytical estimate of $H_0$}

\citet{collier1999} proposed a method for determining $H_0$ based on continuum time delays, under the assumption that the inter-band time delays are a result of a classical accretion disk structure, as given by the following equation

\begin{equation}
H_{0}=89.6 \left( \frac{\lambda}{10^4} \right)^{3/2} \left( \frac{z}{0.001} \right) \left( \frac{\tau}{day} \right)^{-1} \left( \frac{f_{\nu}/cosi}{Jy} \right)^{1/2}\left( \frac{\chi}{4} \right)^{4/3} \left[\frac{\rm km}{\rm s~ Mpc} \right ]
\label{eq:H0_analitic}
\end{equation}
where  $\tau$ is the time delay measured at the wavelength $\lambda$, $z$ is the source redshift, $f_{\nu}$ is the measured flux at the frequency $\nu$ corresponding to the wavelength $\lambda$, and  $\chi$ is the parameter which is used to account for systematic discrepancies due to conversion of annulus temperature to the corresponding wavelength $\lambda$ at a given radius. According to Wien's law, $\chi =4.97$ \citep{netzer2022}; however, under the assumption of a flux-weighted radius, $\chi$ is typically taken to be 2.49 \citep{2024ApJ...973..152E}. In our analysis, considering the AGN accretion disk SED as a multicolored blackbody spectrum, we adopt $\chi =4$. 

We can test the model using this formula, but it is important to use the disk delay rather than the total combined disk plus BLR time delay. The disk delay from Model C is $\tau_{5100} = 1.008$ days at $\lambda = 5100$ {\AA} relative to the X-ray source. For a corona height of $h = 0$, this delay becomes $\tau_{5100} = 0.898$ days (i.e., $1.008 - 0.110$ days), and this value is therefore used in equation \eqref{eq:H0_analitic}. We convert the flux from \cite{Lu2022}, $F_{\lambda}=5.22\times10^{-15}$ erg s$^{-1}$cm$^{-2}$\AA$^{-1}$, to $f_{\nu}$ = $4.590\times10^{-26}$ erg s$^{-1}$cm$^{-2}$Hz$^{-1}$. Assuming an inclination angle of $i=40$ degrees for the source,  we derive $H_0$ of 47.54 km s$^{-1}$Mpc$^{-1}$. This represents a considerable improvement over the value of $H_0 = 15$ km s$^{-1}$ Mpc$^{-1}$ obtained by \citet{cackett2007} for the same source. The difference is largely due to the fact that we are using only the disk delay. If we had used the total time delay (disk + BLR) of 2.00 days, we would have obtained a much lower value of  $H_0 = 21.34$ km s$^{-1}$ Mpc$^{-1}$, closer to the result from \citet{cackett2007}.

However, adopting $\chi = 2.49$ yields a derived $H_0$ of 25.27 km s$^{-1}$ Mpc$^{-1}$ for the disk delay, or 11.34 km s$^{-1}$ Mpc$^{-1}$ when including both the disk+BLR delay. Since there is no universally accepted value of $\chi$, equation \ref{eq:H0_analitic} inherently carries additional uncertainty associated with this parameter.

\subsection{Direct method to measure the distance to the source and $H_0$  from combined spectral and time delay fitting}

The Model C in the version described in Section~\ref{sect:modelc_results} is  based on the known distance  which in turn constraints the accretion rate and the bolometric luminosity  of the source.  Therefore, if we indeed want to measure the distance to the source we cannot rely on quantities that were determined knowing its distance.  We thus must allow the Eddington rate to cover a broad range, reflecting the possibly broad range of the luminosity distance of the source. However, we keep $M_{\rm BH}$ fixed, as it is derived from spectral fitting$-$using the FWHM/line dispersion ($\sigma$) of the H$\beta$ line and the time delay measured from RM, making it independent of the luminosity distance. We also fix the viewing angle.

In this Section, we present a series of computations using a grid of pre-selected luminosity distances. For each luminosity distance, we fit several model parameters by minimizing the $\chi^2$ value. The parameters we vary when fitting the spectrum include the Eddington ratio ($\dot{m}$), the cold disk inner radius ($r_{in}$), the height of the corona ($h$), the optical depth of the warm corona ($\tau_{\mathrm{op-depth}}$), the temperature of the hot corona ($T_e$), the covering factor of the BLR ($f_{BLR}$), and the normalization of the starlight ($A_{star}$). 
 
\begin{figure}
    \centering
    \includegraphics[scale=0.5]{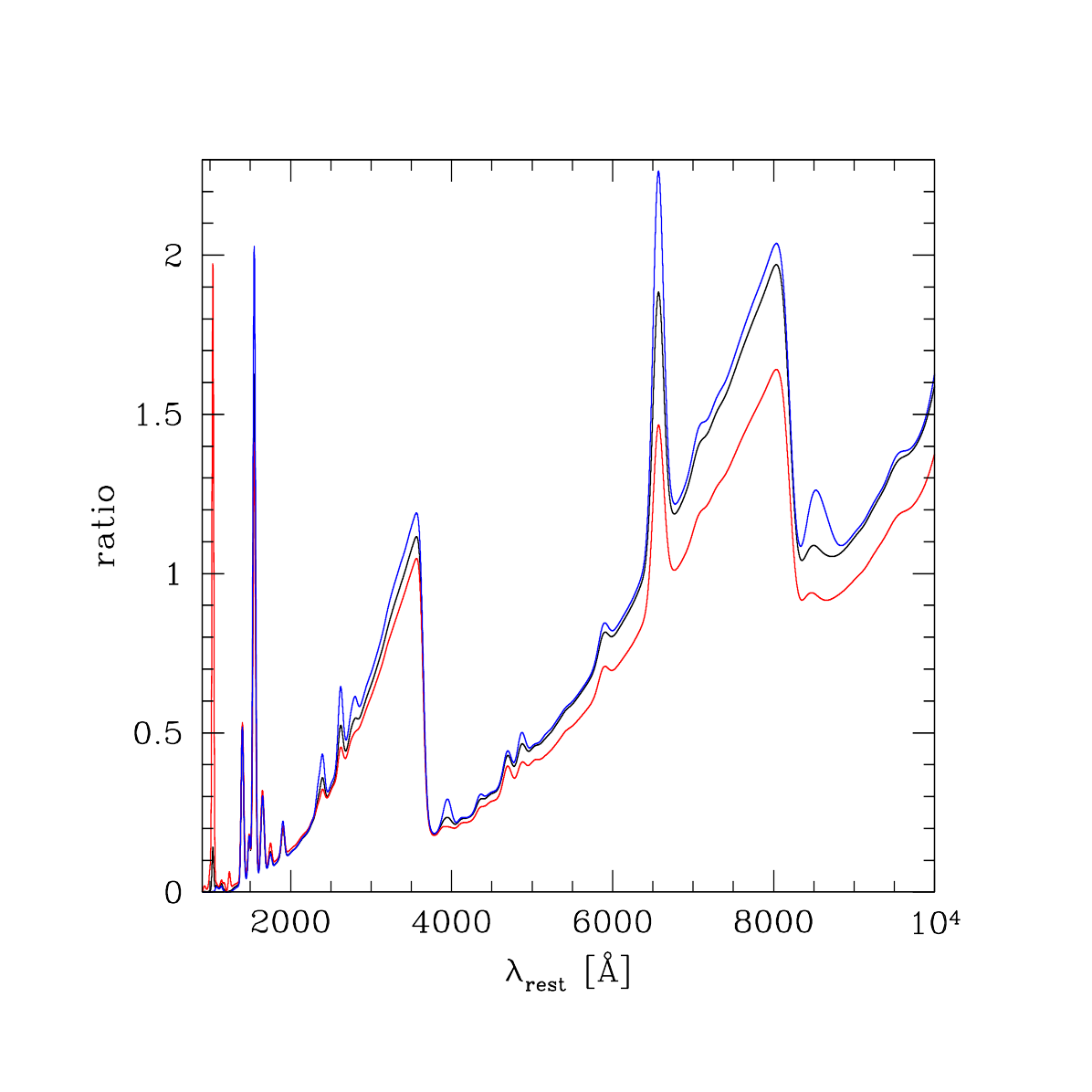}
    \caption{ Examples of  the BLR emission profile for three values of the logarithm of the bolometric luminosity $\log L~[\text{erg}~\text{s}^{-1}] = 44.0$ (black line), 43.8 (blue line), and 44.2 (red line) corresponding to the three values of the luminosity distance, $D_L$ = 70, 56, and 84 Mpc, respectively.}
    \label{fig:CLOUDY_set}
\end{figure}

The incident radiation in the {\tt CLOUDY} computations is determined by the choice of the luminosity distance, as the incident radiation must match the observed hard X-ray flux. Therefore, for each value of the luminosity distance, we use a different spectral shape for the BLR component. Figure~\ref{fig:CLOUDY_set} illustrates exemplary shapes corresponding to different normalizations of the incident flux. It is important to note that the distance to the BLR and the internal parameters of the clouds remain unchanged throughout these computations.

 The dependence of the BLR shape on the incident flux is not monotonic. However, the changes are relatively small, as the range of the considered flux amplitude variation is limited. It is important to emphasize that we do not modify the spectral shape of the incident radiation, as it was derived from observations rather than the model. During spectral fitting, we construct the final model spectrum based on absolute luminosity and compare it to the observed flux, taking into account the adopted distance as

\begin{equation}
F_{\lambda}=\frac{L_{\lambda}}{4\pi D_L^{2}}
\end{equation}
 The value of $\chi^2$ is calculated at several wavelengths, excluding the intense emission lines. Soft X-ray data points, taken from \citet{mehdipour2015}, are also included in the calculation. We assume a $10\%$ error in the spectral fitting process. For the time delay, the $\chi^2$ value is based on the errors complementing the time delay measurements \citep{fausnaugh2016}.

\begin{figure}
    \centering
     \includegraphics[scale=0.5]{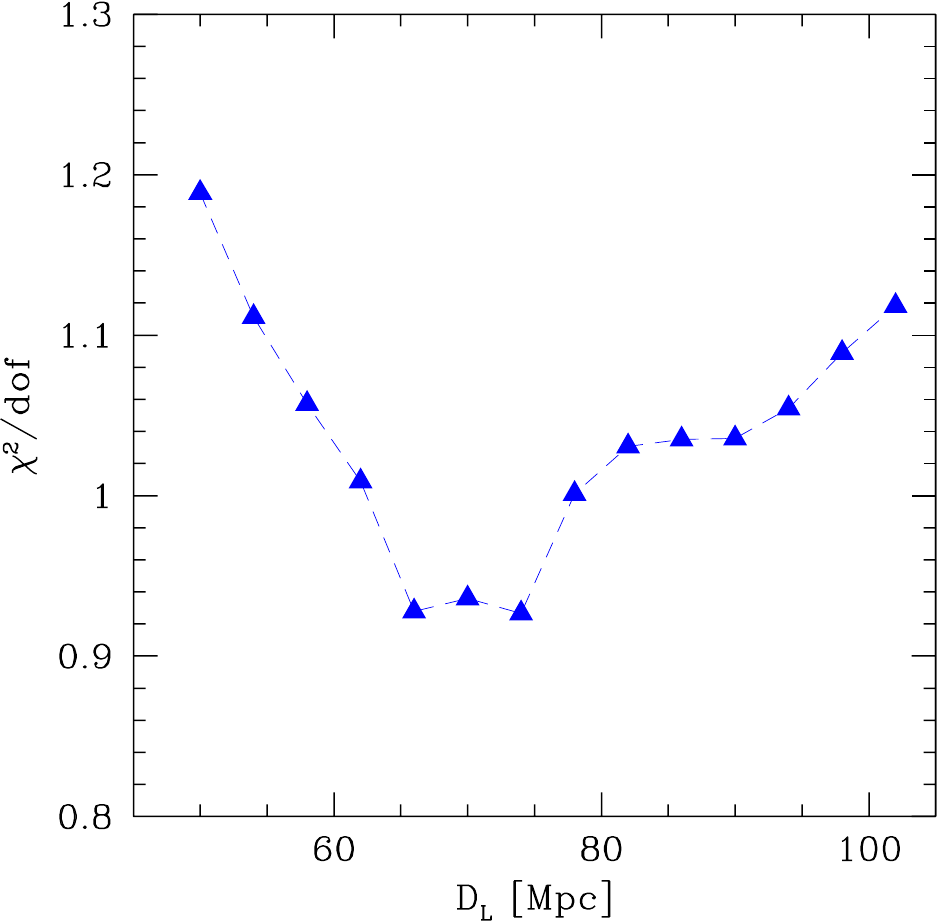}
    \includegraphics[scale=0.5]{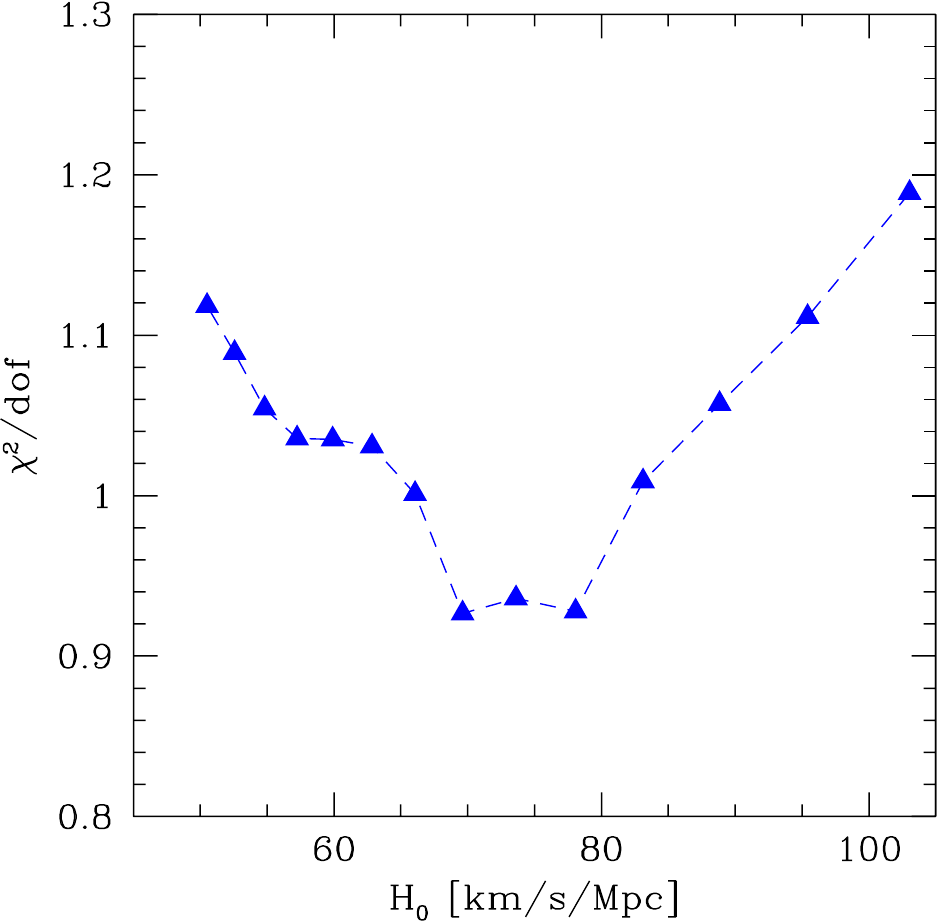}
    \caption{ Upper panel: The sequence of best fits of both the spectrum and the delay as a function of the luminosity distances, marginalized for other model parameters. Bottom panel: The same plot as a function of the resulting Hubble constant.}
    \label{fig:chi2_both}
\end{figure}

 The results of the combined delay and spectra fitting are shown in the upper panel of Figure~\ref{fig:chi2_both} in luminosity distance space. Now, in the $\chi^2$--$D_L$ plane, each point in the sequence corresponds to the same Eddington ratio and other parameters in both the time delay and spectral fitting. The minimum occurs at $D_L = 74$ Mpc in our grid, though the minimum is relatively shallow. The best fits for both the spectrum and the time delay at this grid are also presented in Figure~\ref{fig:modelc}.

The accretion rate predictably increases along the sequence, from 0.008 to 0.038, while the covering factor remain roughly the same, at $\sim 30$ \%. The results are weakly sensitive to the lamp height, which remains close to $\sim 25 r_g$, as well as to the inner radius of the cold disk, which is located at $\sim 90 r_g$ in all fits.

The value of the luminosity distance can be directly translated into the Hubble constant using the expression $H_0 = zc/D_L$, given that the redshift to the source is small.  Thus, the same result, but as a function of the implied $H_0$, is shown in the lower panel of Figure~\ref{fig:chi2_both}. The distance of $D_L = 74$ Mpc corresponds to a Hubble constant of 66.8 km s$^{-1}$ Mpc$^{-1}$. Interpolating between the grid points we find the minimum location at 73.96 Mpc. Since the minimum is shallow, the error is large, leading to our constraint of $D_L$ as $D_L = 74.0^{+2.4}_{-10.0}$ Mpc. This translates to Hubble constant constraints:
$H_0 = 66.9^{+10.5}_{-2.1}\, {\rm km~} {\rm s}^{-1} {\rm Mpc}^{-1}$.

The error estimate presented here is necessarily approximate, as it is based on a qualitative assessment of the effective number of independent degrees of freedom (dof) in the model. Our dataset includes 17 measurements of time delays and we sample the spectrum in the same number of wavelengths. The model involves 7 free parameters, in addition to 4 parameters that are held fixed: $M_{\text{BH}}$, viewing angle, local cloud density, and column density. Each model realization is computed for a fixed luminosity distance. This setup yields 23 dof. We simply determine the 1 sigma error of the reduced $\chi^2$ as $\chi^2/{\text dof} \le \chi^2_{\text{min}}/{\text dof} + 1/23$. We did not perform tests whether all these measurements are statistically independent.

A more robust approach to error estimation would be to use a bootstrap resampling method, which could account for eventual correlations. Unfortunately, our current implementation is not optimized for such computationally intensive analysis, which would require at least 100 realizations to obtain reliable error estimates. Code optimization is underway, as we view this methodology as promising for future studies.

 Note that, in this analysis, we do not vary all the parameters involved in Model C when calculating the spectral and time delay fits, as described above. Therefore, the derived constraint should be interpreted as an indication that the methodology is working, with further progress expected to reduce the error.

The value is well within the error range of the expected values \citep{eleonora_2022,freedman2024}. The measurement error may decrease with improved modeling of the continuum in this source. Additionally, combining measurements from more sources will help further reduce the error. This value is higher than the $47.54 \, {\rm km~} {\rm s}^{-1} {\rm Mpc}^{-1}$ obtained from the analytical formula, which suggests that the proper methodology -- avoiding forced power-law solutions and the extra $\chi$ parameter -- is essential.

To illustrate the necessity of combining delay and spectral fitting, we also present the results obtained when only one type of data, either delays or spectra is used independently.

In experiments where only the time delay is fitted, we vary four parameters ($\dot{m}$, $r_{in}$, $h$, $f_{BLR}$), while the other three ($T_{e}$, $\tau_{\mathrm{op-depth}}$, $A_{star}$) are fixed at the values determined by the spectral best-fit, as these parameters have a negligible impact on the time delay. This strategy is motivated by the fact that spectral fitting is computationally efficient, whereas time delay fitting is considerably more time-consuming. Moreover, when fitting the SED, all  parameters must be optimized simultaneously.

\begin{figure}
    \centering
     \includegraphics[scale=0.5]{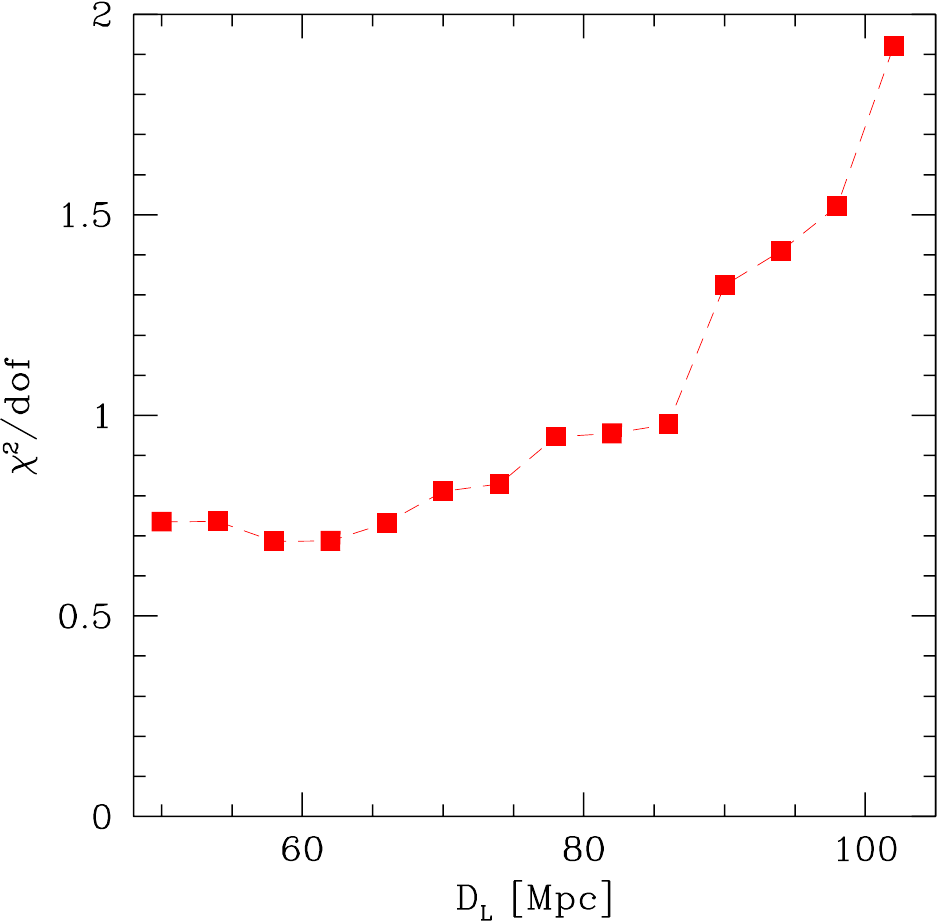}
    \includegraphics[scale=0.5]{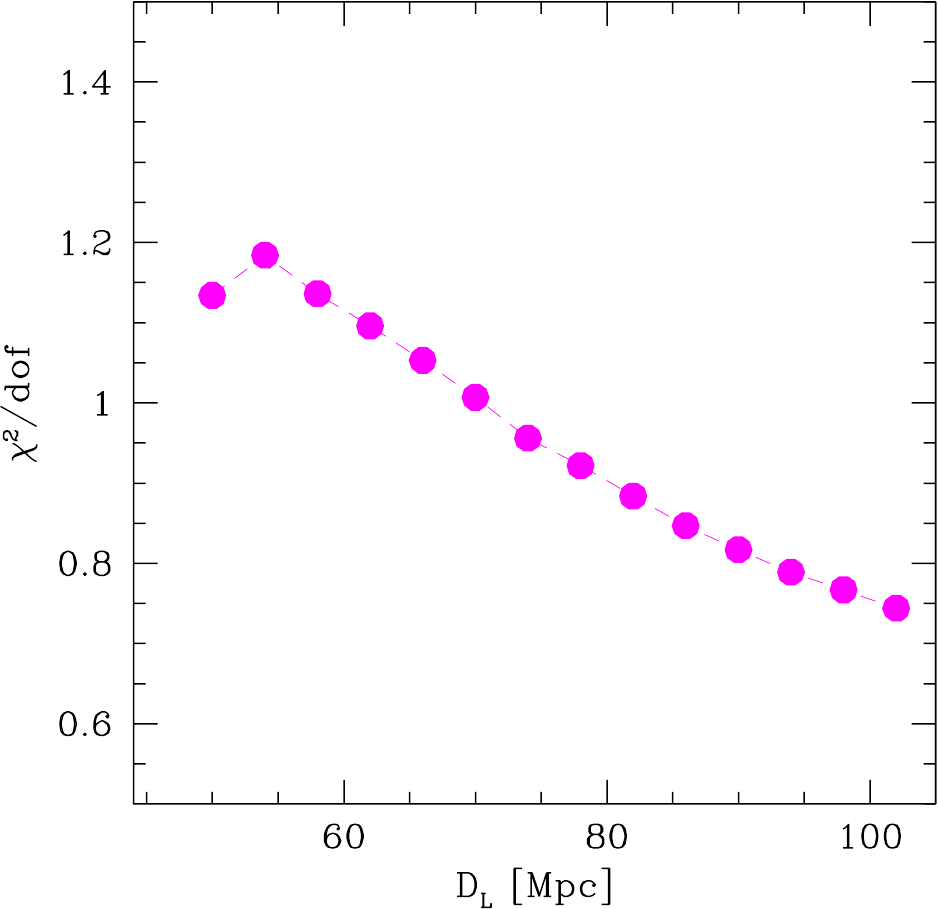}
    \caption{Upper panel: The sequence of best fits of the spectrum as a function of the luminosity distances, marginalized for other model parameters. Bottom panel: The same plot for the time-delay fitting.}
    \label{fig:chi_square}
\end{figure}

The $\chi^2$ values in the luminosity distance space for the SED-only fit are shown in the upper panel of Figure~\ref{fig:chi_square}. The key parameter that changes with $D_L$ is the accretion rate. The formal minimum of $\chi^2$ occurs at $D_L = 58$ Mpc, with the Eddington ratio varying from $\sim 0.008$ at $D_L = 50$ Mpc to $\sim 0.034$ at $D_L = 102$ Mpc. Along with this change in the accretion rate, the corona height increases from 16.5 $r_g$ to 53.0 $r_g$. The covering factor of the BLR, $f_{BLR}$, remains roughly the same, $\sim 0.30$. The stellar contribution also decreases by a factor of 4 as $D_L$ increases. Despite these variations, the optical depth of the hot corona remains roughly unchanged at 20, while the electron temperature increases slightly with  $D_L$, from $6.6 \times 10^6$ K to $8.0 \times 10^6$ K. The reduced $\chi^2$ is of order of 1, suggesting that the adopted spectral errors of 7\% are realistic.

The $\chi^2$ values for the case when we fit only the delay data set in the lag-spectrum, are shown in the lower panel of Figure~\ref{fig:chi_square}. 
The dependence of $\chi^2$ on $D_L$ is relatively weak, exhibiting an almost monotonic decrease in $\chi^2$ values with increasing luminosity distance. This sequence exhibits very different properties compared to the spectral sequence. The Eddington ratio remains constant at 0.03, independent of $D_L$, since there is no direct coupling between the delay, the distance, and the source luminosity. The lamp height is monotonically increasing with $D_L$.
Additionally, the weak variation in the irradiation flux does not lead to a substantial change in the model structure.

 Therefore, the combined fitting of both the SED and the time lag-spectrum is essential. Spectral fitting enforces a change in the best-fit Eddington ratio with $D_L$, which, in turn, affects the delay fitting. A small Eddington ratio implies a smaller source size and shorter time delays  as expected from standard disk.

\section{Discussion}
\label{ss:discussion}

The aim of this paper is two-fold: first, to test the FRADO model for the formation of the BLR, and second, to attempt the estimation of the Hubble constant from such a self-consistent model by determining the redshift-independent distance to the source NGC 5548. While this remains a pilot study, our results demonstrate that the method holds significant potential.

The model itself involves very few parameters. $M_{\rm BH}$ has been fixed, and we have not yet tested its dependence on the results. However, it is worth noting that $M_{\rm BH}$ can be considered independent from cosmology since it is based on line kinematic width and the measured time delay.  The key parameters -- Eddington ratio, BLR contribution, and cold disk inner radius are fitted using the observed mean spectrum and the continuum time delays measured in the continuum--RM campaign by \citet{fausnaugh2016}. The best fit favors a redshift-independent distance of $\sim 74$ Mpc to the source. This distance, when converted to the Hubble constant, implies a value of $H_0$ = $66.9^{+10.5}_{-2.1}$ km s$^{-1}$ Mpc$^{-1}$. Additionally, the model successfully reproduces the H$\beta$ time delay. One hidden parameter in the FRADO model is the dust sublimation temperature, $T_{\rm sub}$, which we adopt as 1500 K. Variations in this temperature would directly influence the favored distance, roughly following the relationship $r \propto (T_{\rm sub}/1500 K)^{-4/3}$.

The multi-wavelength time delays in NGC 5548 have also been modeled in other studies. The original work by \citet{fausnaugh2016}, which provided the delay measurements, employed only power-law parametric modeling. More recently, \citet{kammoun2023} fitted the NGC 5548 time delay using a new model that incorporates relativistic effects, such as the X-ray reflection component and albedo, while also exploring the influence of black hole spin and corona height. Their SED fitting yielded a higher Eddington ratio (0.05) than our solution, along with a higher $M_{\text{BH}}$, resulting in a significantly greater bolometric luminosity for the source compared to our model. However, we cannot analyze this discrepancy in detail, as their best-fit SED was not explicitly presented. While \citet{kammoun2023} strongly advocate for the inclusion of color correction, they do not account for the BLR contribution in their model. Notably, in both of these studies, the distance to NGC 5548 was determined using redshift measurements.

The recent study by \citet{netzer2024} shares several key elements with our work. However, their BLR model is not based on a dynamical framework like ours, requiring additional arbitrary parameters to define the dynamics and geometry. At the same time, their study addresses important factors, such as obscuration, which are not considered in our analysis. Despite successfully fitting both the spectra and time delays for their target, Mrk 817, they did not attempt to interpret their results in terms of a redshift-independent distance or derive an estimate for the Hubble constant.

Constructing a complete model of the BLR remains a significant challenge, with progress being made along several fronts. For instance, \citet{long2025} analyzed the STORM data for NGC 5548 and, by combining a few representative geometries with table models that span a range of cloud ionization properties, concluded that none of the simplest models adequately reproduce the observed data. This highlights the limitations of overly simplified assumptions. Additionally, parametric approaches, such as those implemented in the BELMAG code \citep{rosborough2024,rosborough2025}, exemplify the complexity and richness required to capture the full diversity of BLR structures.

Our approach does not yet incorporate a detailed treatment of the ionization levels of individual clouds, as our objective is more modest and thus we do not aim to reproduce all features of the available data for the NGC 5548 source. Instead, we begin by establishing the dynamics based on the FRADO model, which alone enables us to roughly reproduce key observables such as the H$\beta$ time delay, the transfer function, and the mean spectrum (apart from intensities of all lines), as well as the overall time-delay spectrum.

The geometry adopted in our model naturally leads to double-peaked emission lines. This is consistent with the observed double-peaked structure of the H$\beta$ line, particularly in the  rms profile \citep[e.g.][]{lu2016, shapovalova2009, long2025}. Moreover, other emission lines also exhibit double-peaked profiles \citep{horne2021}, and the velocity-resolved time delays further support this structure \citep{long2025}. Our model recovered transfer function similarly displays a double-peaked profile, consistent with an axially symmetric geometry dominated by Keplerian motion. In contrast, \citet{horne2021} obtained an almost single-peaked transfer function for H$\beta$, which may indicate the presence of cloud emission anisotropy, an effect already considered by \citet{ferland1979}, as well as possible selective shielding of parts of the BLR by outflowing material located within the region.

There remains considerable room for advancement in the coming years, particularly in refining the physical details of BLR modeling. Furthermore, the potential application of continuum time-delay measurements to cosmology provides an additional strong motivation for continued progress in this area.

 Taken together, these comparisons highlight the unique strengths of our approach. Unlike previous studies that either relied on power-law fitting, incorporated relativistic effects without addressing the BLR contribution, or did not pursue an independent distance determination, our method integrates spectral fitting and time delay modeling within a self-consistent framework. By leveraging the FRADO model, we obtain a redshift-independent distance to NGC 5548. While further refinements--such as improved continuum modeling and broader sample studies--are needed to reduce uncertainties, our results demonstrate the potential of this methodology in offering an alternative approach to cosmic distance measurements.

\subsection{Covering factor}

Although we refer to our disk-plus-BLR model as self-consistent, not all of its components are fully self-consistent yet. In principle, for a given $M_{\text{BH}}$ and Eddington ratio, the BLR covering factor should be determined by the model. Some tests of the predicted covering factor have been performed for relatively high Eddington ratios \citep{naddaf2023, naddaf_CF2024}. However, at lower Eddington ratios, the clouds are located too close to the disk, leading to inconsistencies.

The aspect ratio of the cloud distribution, as shown in Figure~\ref{fig:cloud}, implies a covering factor of approximately $5\%$, whereas the value favored by time delay and spectral fitting is $36\%$. While the BLR location is well represented, the vertical motion of the clouds appears insufficient when considering only radiation pressure from dust. This suggests that line driving likely plays a role, but it is not yet included in the FRADO model.

Incorporating line driving, such as by combining FRADO with the QWIND model of \citet{QWIND2010}, is highly complex.   The improved version of the QWIND software \citep{new_QWIND2023} is publicly available, providing a strong foundation for future work. However, it still lacks dust physics, which is crucial for determining the radial structure of the BLR at low luminosities. Future developments integrating these elements will be essential for achieving a fully self-consistent model.


\subsection{Relativistic effects and X-ray albedo}

Currently our model does not account for the effects of GR, which are critically important for phenomena within the inner 10 $r_g$. In these inner regions, spacetime curvature significantly influences both the propagation of photons from the lamppost (a point-like X-ray source) to the accretion disk and their subsequent travel from the disk to the observer. Additionally, the energy dissipation profile in the cold disk depends on the black hole spin \citep[see e.g.,][]{NT1973}.

However, the lamppost geometry itself is a considerable simplification, and in the case of NGC 5548, the cold accretion disk does not extend all the way to the ISCO. As a result, the reprocessing primarily occurs at larger radii, where GR effects are less dominant.

Another important consideration is the disk albedo, which is a complex function of the density in the disk atmosphere and the ionization parameter, both of which are likely to vary with radius. This issue is best addressed in systems that exhibit broad, relativistically smeared X-ray Fe K$\alpha$ lines, where model assumptions can be more directly tested \citep[e.g.,][]{ballantyne2001,dovciak2004,garcia2013}.  Even models of reprocessing by the warm corona were studied recently by \citet{petrucci2020}.

For the case of continuum time delays, GR effects have been specifically studied for NGC 5548 by \citet{kammoun_5548_2024}. However, these models do not include contributions from the BLR and fail to reproduce features such as the Balmer and Paschen edges in the time-delay spectrum.

In the future, it may become possible to integrate these various components into a single model. However, full GR-based computations are computationally expensive and require the inclusion of additional free functions (radial profile of the ionization parameter, disk geometrical height, hard X-ray source geometry) to model the reprocessing accurately. Overall, as demonstrated by \citet{kammoun_analit2021}, the disk response is indeed sensitive to both relativistic effects and albedo. Nonetheless, those models do not incorporate an inner warm corona and therefore cannot be used for direct comparison in our case.

\subsection{Dust sublimation temperature}
\label{ss:dust}

In this work, we adopt a universal dust sublimation temperature of 1500 K, which plays a crucial role in the FRADO model as it sets the onset of BLR cloud launching. This choice directly influences the analytical expression for the Hubble constant derived in equation~\ref{eq:H0_analitic}. Given the temperature profile in the standard accretion disk that underlies this formula, our inferred values for the Hubble constant scale approximately as:

\begin{equation}
H_0 \propto ({T_{sub} \over 1500 K})^{4/3}.
\end{equation}

However, the dust sublimation temperature is not firmly established. While \citet{barvainis1987} originally recommended 1500 K as a representative value, more recent papers recommend 1500 K for standard mixture of silicate/graphite dust \citep[e.g.][]{mor2012}, but for pure graphite, higher values such as 1800 K are proposed \citep[e.g.,][]{mor2012}. A broader, density-dependent range was presented by \citet{baskin2018}, whose Figure~1 shows that at densities around $10^{11}\mathrm{cm}^{-3}$, sublimation temperatures for silicate and graphite grains could reach 1600 K and 2000 K, respectively.

Nonetheless, graphite is unlikely to be present in the nuclear parts of AGN, as the 2175 \AA~ feature is generally not seen in absorption \citep[][]{laor1993,gaskell2004}, which rather suggests the presence of amorphous carbon grains \citep[e.g.][]{czerny2004}. However, there are no reliable estimates for the sublimation temperature of amorphous carbon in AGN environments. Stellar measurements from AGB stars give a temperature of $1170 \pm 60$ K, very close to the sublimation temperature of silicate in these surroundings.

An alternative constraint was proposed by \citet{Vazquez2015}, who argued for dust temperatures as high as 1800 K based on dust time-delay measurements rather than direct spectral analysis. However, their interpretation may be affected by an incomplete subtraction of the accretion disk contribution to the observed dust-delays \citep{2024ApJ...968...59M}.

Given these uncertainties, we adopt 1500 K as a conservative and widely used value, but we acknowledge a plausible error margin of $\pm 100$ K. Based on the scaling relation above, this translates into a $\sim 9\%$ uncertainty in the inferred Hubble constant, representing an important systematic contribution to our results.

The presence of dust, particularly of dust with atypical properties, may additionally imply problems with establishing the intrinsic SED. For example, \citet{gaskell2023} argue that, based on their
extinction curve, the UV flux of NGC 5548 may be underestimated by a factor of seven. Tests of such issues can be performed in the future.

\subsection{Other fixed parameters as sources of systematic errors}

It would be computationally very time consuming to move the other fixed parameters to the parameters fitted to the data, and to study the degeneracy between all these parameters. However, we can easily discuss their expected importance looking at the analytical formula for the Hubble constant of \citet{collier1999} (see equation~\ref{eq:H0_analitic}). Those extra fixed parameters are: $M_{\text{BH}}$, inclination angle, gas density, and column density.  The black hole mass does not enter into equation~\ref{eq:H0_analitic}. The monochromatic flux $f_{\nu}$ is not sensitive on $M_{\text{BH}}$ as long as the optical spectrum does not show strong departure from a power law \citep{SS1973}. This happens only in the far UV. The presence of the warm corona complicates the reasoning. Also the shape of the structure function shows some dependence on $M_{\text{BH}}$ \citep[see e.g.][for most recent study]{Naddaf_2025}, and not just on local monochromatic flux which is actually fitted. The onset of the BLR does not depend on $M_{\text{BH}}$, as it is set by the dust sublimation temperature, but the vertical velocities increase with $M_{\text{BH}}$. Overall, it should not lead to significant effect. The viewing angle is important, as it is present in equation~\ref{eq:H0_analitic}. For the viewing angle of 30 degrees, the Hubble constant would be lower by 6 \%, and for 50 degrees it would be higher by 8 \%. Therefore, this effect is important for any specific source. On the other hand, if the method is later applied to a large sample of objects at different redshifts, the viewing angle effects will average out, as there is no strong evidence for a redshift-dependent trend in the viewing angle or the dusty torus opening angle \citep[e.g.][]{prince2022, ralowski2024}. However, assessing the influence of the local density and column density of the BLR clouds is considerably more challenging, as both parameters impact the ionization state and the effective optical depth of the region. These quantities can be effectively constrained by analyzing the shape of the delay enhancement near the Balmer and Paschen edges. Moreover, any reduction in effective optical depth can be compensated for by adjusting the covering factor, which is treated as a free parameter in the model. We certainly plan to refine and improve the model in future work.


\subsection{FRADO vs LOC models as a method to reproduce the BLR transfer function}
 
 FRADO is a dynamical model which aims to reproduce the physical origin of the clouds, their location and their velocities. The model is not yet fully matured, but there are clear prospects for the model further development. For the dynamics, the addition of the line driving, mentioned above, should be included as its importance is increasing with the height of the clouds.  Next, the emissivity of each clouds should be calculated individually, according to its location. Finally, there are two even more complex problems. One is a good realistic description of the shielding of clouds by themselves, important at low heights from the disk. The second one is determination of the density distribution of material based on thermal instability of the medium, as described already by \citet{krolik1981}. The wind is launched as a continuous medium, with clouds forming out of this wind in short thermal timescale.

 Locally Optimally Emitting Cloud (LOC) models, on the other hand, by their concept (power law distribution of clouds in radius and in density, with the limiting densities and limiting radii, and power law slopes as parameters) are better suited for accurately reproducing the line ratios observed in AGN spectra. The additional parameters are still required if the distribution is flat, concentrated around the accretion disk, instead of a spherical distribution as in the original model of \citet{baldwin1995}. This model is mature, but its parameters must be controlled by the data. However, to derive the BLR response, additional assumptions are required. \citet{lawther2018} combined a spherically symmetric approach with the continuity equation to constrain the radial distribution of clouds, and were thereby able to predict the expected time delays for all emission lines. Additionally, using LOC model, \citet{korista2019} pioneered the study of the disk continuum contamination by BLR continuum. However, it remains to be seen whether LOC models can be applied in the same way as FRADO to simultaneously fit both the spectra and time delays, and thereby enable determination of the Hubble constant. This has not been achieved so far.

\subsection{Other distance measurements to the source}

The heliocentric redshift of NGC 5548 \citep[$z = 0.017175 \pm 2.30 \times 10^{-5}$;][]{deVaucouleurs1991} corresponds to a distance of $79.05 \pm 5.54$ Mpc. It is different when the motion of the source is measured relative to the Cosmic Microwave Background (CMB)\footnote{\url{https://ned.ipac.caltech.edu/}}. The proper motion of the source is thus not negligible, as indicated by the difference between its heliocentric and CMB-based velocities reported by NED ($5149 \pm 7$ km s$^{-1}$ vs. $5359 \pm 16$ km s$^{-1}$).

In addition to redshift-based estimates, several independent distance measurements exist for this source. Two estimates rely on the Tully-Fisher method: an older measurement of 103 Mpc \citep{bottinelli1984} and a more recent one that closely aligns with the redshift-based value ($82.2 \pm 16.4$ Mpc; \citealt{robinson2021}). Meanwhile, the AGN continuum time lag method yielded a significantly higher distance of $341 \pm 62$ Mpc \citep{cackett2007}. In contrast, dust reverberation mapping provided a distance of $92.5 \pm 1.5$ Mpc \citep{yoshii2014}, which, despite its small uncertainty, is not consistent with the redshift-based value.

Applying a velocity correction from the heliocentric frame to the CMB frame shifts the redshift from 0.017175 to 0.017876. This adjustment would, in turn, modify our estimated value of the Hubble constant to $69.6^{+11.0}_{-2.2}$ km s$^{-1}$ Mpc$^{-1}$. However, given the relatively large uncertainties involved, achieving a highly precise correction for the proper motion is not crucial in this context.

\subsection{The Hubble tension}

The recent measurements of the Hubble constant generally highlight a persistent tension between early- and late-Universe determinations \citep[see][for a review]{eleonora_2022}. A key issue in this debate is the determination of measurement uncertainties. For example, studies by \citet{riess2016} and \citet{Planck2016} suggest a significant discrepancy, while the baryon acoustic oscillation (BAO) measurements by \citet{wang2017} are consistent with both within their reported errors. However, the overall tension remains unresolved, as indicated by the most recent results from \citet{riess2024}, \citet{uddin2024}, and \citet{Planck2020}. 

In response to this discrepancy, several independent methods have been proposed to confirm or address the tension \citep[e.g., see][for some of the latest examples]{SNII2022,lensing2024,trefoloni2024,biesiada2024,dainotti2024}. Notably, some local measurements yield lower values of $H_0$ \citep[e.g.,][]{grillo2024}.  The latest findings from the clustering of galaxies, quasars, and Lyman-$\alpha$ forest tracers, based on the first year of observations from the Dark Energy Spectroscopic Instrument (DESI Data Release 1; \citealt{DESI_BAO_2024arXiv240403002D}), support a lower Hubble constant of $68.40 \pm 0.27$ km s$^{-1}$ Mpc$^{-1}$ for a flat $\Lambda$CDM model, with remarkably small uncertainty. Their results, incorporating constraints from Big Bang nucleosynthesis and acoustic angular scale measurements from the CMB, suggest $H_0 = 68.52 \pm 0.62$ km s$^{-1}$ Mpc$^{-1}$, while a joint analysis of CMB anisotropies and CMB lensing from Planck+ACT yields $H_0 = 67.97 \pm 0.38$ km s$^{-1}$ Mpc$^{-1}$. Most recent measurements from the Dark Energy Survey (DES) experiment yields an $H_{0}$ value of  $67.81^{+0.96}_{-0.86}$ km s$^{-1}$ Mpc$^{-1}$ \citep{DES2025}.

Although our method lacks the precision needed to decisively address this tension, it represents a significant improvement over the results obtained by \citet{cackett2007} for this object. This progress is both substantial and encouraging.

Interestingly, distance-ladder-based measurements tend to yield higher values of $H_0$ compared to single-step methods, with \citet{perivolaropoulos2024} reporting an average of $H_0 = 69.0 \pm 0.48$ km s$^{-1}$ Mpc$^{-1}$. Our target source, at a redshift of $z = 0.017175$, lies near the boundary of $z \sim 0.01$, where some studies suggest a possible transition in cosmological behavior \citep[e.g.,][]{gavas2024,verde2024}. However, given the current uncertainties in our results, we cannot yet distinguish between competing interpretations. Future efforts should focus on reducing measurement errors and extending the analysis to AGN at both lower and higher redshifts to further explore this issue.

\subsection{Prospects for continuum reverberation from LSST}

The standard six-color photometry provided by the Legacy Survey of Space and Time (LSST), soon to be available from the Vera Rubin Observatory, may not achieve the same level of precision in time delay measurements as combined SWIFT/ground-based monitoring. However, there remain promising opportunities for detecting continuum time delays, particularly for AGNs with larger black hole masses with high luminosities \citep{pozo2023,pozo2024,2024ApJ...968L..16P}.

For the most massive quasars, however, the contribution of the BLR to the observed time delay may be less significant. This is suggested by results for $PG 2308+098$, a quasar with $M_{\rm BH}$ of $10^{9.6 \pm 0.1} M_{\odot}$, as reported by \citet{kokubo2018}. To better understand this effect, we can also investigate a potential scaling relation between BLR continuum contamination and emission line intensities, which may provide further insight into the role of the BLR in time delay measurements.

\section{Conclusions}
\label{concl}

Our pilot study aims to test the FRADO model for the BLR structure and estimate the Hubble constant based on the mean spectrum data and continuum time delay measurements for NGC 5548. The study has been highly successful, yielding several key outcomes:

\begin{itemize}

\item The time delay of the H$\beta$ line is approximately recovered, even though our model does not allow for direct adjustments of the inner and outer radii of BLR. Instead, these parameters emerge naturally from computations of radiation pressure acting on dust and the dust sublimation temperature.

\item By using the redshift-based distance of NGC 5548 and its known luminosity, we successfully model the overall spectrum, incorporating contributions from both the accretion disk and the BLR. Additionally, we accurately reproduce the time-delay pattern, emphasizing the crucial role played by the Balmer and Paschen edges.

\item Relaxing the assumption of a known redshift-based luminosity, we model the spectrum and time-delay data across a range of Eddington ratios. For each value, we optimize the source distance, ultimately identifying a best-fit Eddington ratio and corresponding distance, which in turn provide an estimate of the Hubble constant.

\item Although the uncertainty in the Hubble constant remains relatively large ($66.9^{+10.5}_{-2.1}$ km s$^{-1}$ Mpc$^{-1}$), our method demonstrates potential for refinement. Further improvements, both for this source and for similar determinations using other AGNs with dense monitoring could lead to significantly more precise measurements.

\end{itemize}

 Our results validate the viability of using AGN continuum time delays and spectral modeling as an independent method for measuring cosmic distances and estimating the Hubble constant. While the current uncertainty is still substantial, the methodology holds promise for future applications. Expanding the sample to include additional AGNs with well-sampled continuum--RM data will be crucial for reducing uncertainties and improving the accuracy of this approach.

\begin{acknowledgements}
We thank the anonymous referee for valuable comments and suggestions. This project has received funding from the European Research Council (ERC) under the European Union’s Horizon 2020 research and innovation program (grant agreement No. [951549]). VKJ acknowledges the OPUS-LAP/GA ˇCR-LA bilateral project (2021/43/I/ST9/01352/OPUS
22 and GF23-04053L). MHN also acknowledges the financial support by the University of Liege under Special Funds for Research, IPD-STEMA Program. SP is supported by the international Gemini Observatory, a program of NSF NOIRLab, which is managed by the Association of Universities for Research in Astronomy (AURA) under a cooperative agreement with the U.S. National Science Foundation, on behalf of the Gemini partnership of Argentina, Brazil, Canada, Chile, the Republic of Korea, and the United States of America. BC acknowledges the support
from COST Action CA21136 - Addressing observational tensions in cosmology with systematics and fundamental physics (CosmoVerse), supported by COST (European Cooperation in Science and Technology). FPN gratefully acknowledges the generous and invaluable support of the Klaus Tschira Foundation. We thank Dr. Missagh Mehdipour for sharing the broadband SED data points with us.
\end{acknowledgements}

\bibliographystyle{aa}
\bibliography{main}

\begin{thebibliography}{152}
\expandafter\ifx\csname natexlab\endcsname\relax\def\natexlab#1{#1}\fi

\bibitem[{{Abdalla} {et~al.}(2022){Abdalla}, {Abell{\'a}n}, {Aboubrahim},
  {Agnello}, {Akarsu}, {Akrami}, {Alestas}, {Aloni}, {Amendola}, {Anchordoqui},
  {Anderson}, {Arendse}, {Asgari}, {Ballardini}, {Barger}, {Basilakos},
  {Batista}, {Battistelli}, {Battye}, {Benetti}, {Benisty}, {Berlin}, {de
  Bernardis}, {Berti}, {Bidenko}, {Birrer}, {Blakeslee}, {Boddy}, {Bom},
  {Bonilla}, {Borghi}, {Bouchet}, {Braglia}, {Buchert}, {Buckley-Geer},
  {Calabrese}, {Caldwell}, {Camarena}, {Capozziello}, {Casertano}, {Chen},
  {Chluba}, {Chen}, {Chen}, {Chudaykin}, {Cicoli}, {Copi}, {Courbin},
  {Cyr-Racine}, {Czerny}, {Dainotti}, {D'Amico}, {Davis}, {de Cruz P{\'e}rez},
  {de Haro}, {Delabrouille}, {Denton}, {Dhawan}, {Dienes}, {Di Valentino},
  {Du}, {Eckert}, {Escamilla-Rivera}, {Fert{\'e}}, {Finelli}, {Fosalba},
  {Freedman}, {Frusciante}, {Gazta{\~n}aga}, {Giar{\`e}}, {Giusarma},
  {G{\'o}mez-Valent}, {Handley}, {Harrison}, {Hart}, {Hazra}, {Heavens},
  {Heinesen}, {Hildebrandt}, {Hill}, {Hogg}, {Holz}, {Hooper}, {Hosseininejad},
  {Huterer}, {Ishak}, {Ivanov}, {Jaffe}, {Jang}, {Jedamzik}, {Jimenez},
  {Joseph}, {Joudaki}, {Kamionkowski}, {Karwal}, {Kazantzidis}, {Keeley},
  {Klasen}, {Komatsu}, {Koopmans}, {Kumar}, {Lamagna}, {Lazkoz}, {Lee},
  {Lesgourgues}, {Levi Said}, {Lewis}, {L'Huillier}, {Lucca}, {Maartens},
  {Macri}, {Marfatia}, {Marra}, {Martins}, {Masi}, {Matarrese}, {Mazumdar},
  {Melchiorri}, {Mena}, {Mersini-Houghton}, {Mertens}, {Milakovi{\'c}},
  {Minami}, {Miranda}, {Moreno-Pulido}, {Moresco}, {Mota}, {Mottola}, {Mozzon},
  {Muir}, {Mukherjee}, {Mukherjee}, {Naselsky}, {Nath}, {Nesseris},
  {Niedermann}, {Notari}, {Nunes}, {{\'O} Colg{\'a}in}, {Owens},
  {{\"O}z{\"u}lker}, {Pace}, {Paliathanasis}, {Palmese}, {Pan}, {Paoletti},
  {Perez Bergliaffa}, {Perivolaropoulos}, {Pesce}, {Pettorino}, {Philcox},
  {Pogosian}, {Poulin}, {Poulot}, {Raveri}, {Reid}, {Renzi}, {Riess}, {Sabla},
  {Salucci}, {Salzano}, {Saridakis}, {Sathyaprakash}, {Schmaltz},
  {Sch{\"o}neberg}, {Scolnic}, {Sen}, {Sehgal}, {Shafieloo}, {Sheikh-Jabbari},
  {Silk}, {Silvestri}, {Skara}, {Sloth}, {Soares-Santos}, {Sol{\`a} Peracaula},
  {Songsheng}, {Soriano}, {Staicova}, {Starkman}, {Szapudi}, {Teixeira},
  {Thomas}, {Treu}, {Trott}, {van de Bruck}, {Vazquez}, {Verde}, {Visinelli},
  {Wang}, {Wang}, {Wang}, {Watkins}, {Watson}, {Webb}, {Weiner}, {Weltman},
  {Witte}, {Wojtak}, {Yadav}, {Yang}, {Zhao}, \&
  {Zumalac{\'a}rregui}}]{eleonora_2022}
{Abdalla}, E., {Abell{\'a}n}, G.~F., {Aboubrahim}, A., {et~al.} 2022, Journal
  of High Energy Astrophysics, 34, 49

\bibitem[{{Baldwin} {et~al.}(1995){Baldwin}, {Ferland}, {Korista}, \&
  {Verner}}]{baldwin1995}
{Baldwin}, J., {Ferland}, G., {Korista}, K., \& {Verner}, D. 1995, \apjl, 455,
  L119

\bibitem[{{Ballantyne} {et~al.}(2001){Ballantyne}, {Ross}, \&
  {Fabian}}]{ballantyne2001}
{Ballantyne}, D.~R., {Ross}, R.~R., \& {Fabian}, A.~C. 2001, \mnras, 327, 10

\bibitem[{{Ballantyne} {et~al.}(2024){Ballantyne}, {Sudhakar}, {Fairfax},
  {Bianchi}, {Czerny}, {De Rosa}, {De Marco}, {Middei}, {Palit}, {Petrucci},
  {R{\'o}{\.z}a{\'n}ska}, \& {Ursini}}]{ballantyne2024}
{Ballantyne}, D.~R., {Sudhakar}, V., {Fairfax}, D., {et~al.} 2024, \mnras, 530,
  1603

\bibitem[{{Bargiacchi} {et~al.}(2025){Bargiacchi}, {Dainotti}, \&
  {Capozziello}}]{dainotti2024}
{Bargiacchi}, G., {Dainotti}, M.~G., \& {Capozziello}, S. 2025, \nar, 100,
  101712

\bibitem[{{Barvainis}(1987)}]{barvainis1987}
{Barvainis}, R. 1987, \apj, 320, 537

\bibitem[{{Baskin} \& {Laor}(2018)}]{baskin2018}
{Baskin}, A. \& {Laor}, A. 2018, \mnras, 474, 1970

\bibitem[{{Beard} {et~al.}(2025){Beard}, {McHardy}, {Horne}, {Cackett},
  {Vincentelli}, {Santisteban}, {Miller}, {Dhillon}, {Knapen}, {Littlefair},
  {Kynoch}, {Breedt}, {Shen}, \& {Gelbord}}]{beard2025}
{Beard}, M.~W.~J., {McHardy}, I.~M., {Horne}, K., {et~al.} 2025, \mnras, 537,
  293

\bibitem[{{Blandford} \& {McKee}(1982)}]{1982ApJ...255..419B}
{Blandford}, R.~D. \& {McKee}, C.~F. 1982, \apj, 255, 419

\bibitem[{{Bon} {et~al.}(2016){Bon}, {Zucker}, {Netzer}, {Marziani}, {Bon},
  {Jovanovi{\'c}}, {Shapovalova}, {Komossa}, {Gaskell}, {Popovi{\'c}},
  {Britzen}, {Chavushyan}, {Burenkov}, {Sergeev}, {La Mura}, {Vald{\'e}s}, \&
  {Stalevski}}]{bon2016}
{Bon}, E., {Zucker}, S., {Netzer}, H., {et~al.} 2016, \apjs, 225, 29

\bibitem[{{Bottinelli} {et~al.}(1984){Bottinelli}, {Gouguenheim}, {Paturel}, \&
  {de Vaucouleurs}}]{bottinelli1984}
{Bottinelli}, L., {Gouguenheim}, L., {Paturel}, G., \& {de Vaucouleurs}, G.
  1984, \aaps, 56, 381

\bibitem[{{Cackett} {et~al.}(2021){Cackett}, {Bentz}, \&
  {Kara}}]{cackett_2021iSci}
{Cackett}, E.~M., {Bentz}, M.~C., \& {Kara}, E. 2021, iScience, 24, 102557

\bibitem[{{Cackett} {et~al.}(2018){Cackett}, {Chiang}, {McHardy}, {Edelson},
  {Goad}, {Horne}, \& {Korista}}]{2018ApJ...857...53C}
{Cackett}, E.~M., {Chiang}, C.-Y., {McHardy}, I., {et~al.} 2018, \apj, 857, 53

\bibitem[{{Cackett} \& {Horne}(2006)}]{cackett2006}
{Cackett}, E.~M. \& {Horne}, K. 2006, \mnras, 365, 1180

\bibitem[{{Cackett} {et~al.}(2007){Cackett}, {Horne}, \&
  {Winkler}}]{cackett2007}
{Cackett}, E.~M., {Horne}, K., \& {Winkler}, H. 2007, \mnras, 380, 669

\bibitem[{{Cao} {et~al.}(2025){Cao}, {Mandal}, {Zaja{\v{c}}ek}, {Czerny}, \&
  {Ratra}}]{2025PhRvD.111h3545C}
{Cao}, S., {Mandal}, A.~K., {Zaja{\v{c}}ek}, M., {Czerny}, B., \& {Ratra}, B.
  2025, \prd, 111, 083545

\bibitem[{{Cao} {et~al.}(2024){Cao}, {Zaja{\v{c}}ek}, {Czerny}, {Panda}, \&
  {Ratra}}]{cao2024}
{Cao}, S., {Zaja{\v{c}}ek}, M., {Czerny}, B., {Panda}, S., \& {Ratra}, B. 2024,
  \mnras, 528, 6444

\bibitem[{{Cao} {et~al.}(2022){Cao}, {Zaja{\v{c}}ek}, {Panda},
  {Mart{\'\i}nez-Aldama}, {Czerny}, \& {Ratra}}]{cao2022}
{Cao}, S., {Zaja{\v{c}}ek}, M., {Panda}, S., {et~al.} 2022, \mnras, 516, 1721

\bibitem[{{Cappi} {et~al.}(2016){Cappi}, {De Marco}, {Ponti}, {Ursini},
  {Petrucci}, {Bianchi}, {Kaastra}, {Kriss}, {Mehdipour}, {Whewell}, {Arav},
  {Behar}, {Boissay}, {Branduardi-Raymont}, {Costantini}, {Ebrero}, {Di Gesu},
  {Harrison}, {Kaspi}, {Matt}, {Paltani}, {Peterson}, {Steenbrugge}, \&
  {Walton}}]{cappi2016}
{Cappi}, M., {De Marco}, B., {Ponti}, G., {et~al.} 2016, \aap, 592, A27

\bibitem[{{Chatzikos} {et~al.}(2023{\natexlab{a}}){Chatzikos}, {Bianchi},
  {Camilloni}, {Chakraborty}, {Gunasekera}, {Guzm{\'a}n}, {Milby}, {Sarkar},
  {Shaw}, {van Hoof}, \& {Ferland}}]{2023RMxAA..59..327C}
{Chatzikos}, M., {Bianchi}, S., {Camilloni}, F., {et~al.} 2023{\natexlab{a}},
  \rmxaa, 59, 327

\bibitem[{{Chatzikos} {et~al.}(2023{\natexlab{b}}){Chatzikos}, {Bianchi},
  {Camilloni}, {Chakraborty}, {Gunasekera}, {Guzm{\'a}n}, {Milby}, {Sarkar},
  {Shaw}, {van Hoof}, \& {Ferland}}]{Cloudy23}
{Chatzikos}, M., {Bianchi}, S., {Camilloni}, F., {et~al.} 2023{\natexlab{b}},
  \rmxaa, 59, 327

\bibitem[{{Clavel} {et~al.}(1991){Clavel}, {Reichert}, {Alloin}, {Crenshaw},
  {Kriss}, {Krolik}, {Malkan}, {Netzer}, {Peterson}, {Wamsteker}, {Altamore},
  {Baribaud}, {Barr}, {Beck}, {Binette}, {Bromage}, {Brosch}, {Diaz},
  {Filippenko}, {Fricke}, {Gaskell}, {Giommi}, {Glass}, {Gondhalekar},
  {Hackney}, {Halpern}, {Hutter}, {Joersaeter}, {Kinney}, {Kollatschny},
  {Koratkar}, {Korista}, {Laor}, {Lasota}, {Leibowitz}, {Maoz}, {Martin},
  {Mazeh}, {Meurs}, {Nair}, {O'Brien}, {Pelat}, {Perez}, {Perola}, {Ptak},
  {Rodriguez-Pascual}, {Rosenblatt}, {Sadun}, {Santos-Lleo}, {Shaw}, {Smith},
  {Stirpe}, {Stoner}, {Sun}, {Ulrich}, {van Groningen}, \&
  {Zheng}}]{clavel1991}
{Clavel}, J., {Reichert}, G.~A., {Alloin}, D., {et~al.} 1991, \apj, 366, 64

\bibitem[{{Collier} {et~al.}(1999){Collier}, {Horne}, {Wanders}, \&
  {Peterson}}]{collier1999}
{Collier}, S., {Horne}, K., {Wanders}, I., \& {Peterson}, B.~M. 1999, \mnras,
  302, L24

\bibitem[{{Crenshaw} {et~al.}(2009){Crenshaw}, {Kraemer}, {Schmitt}, {Kaastra},
  {Arav}, {Gabel}, \& {Korista}}]{Crenshaw2009}
{Crenshaw}, D.~M., {Kraemer}, S.~B., {Schmitt}, H.~R., {et~al.} 2009, \apj,
  698, 281

\bibitem[{{Czerny} {et~al.}(2018){Czerny}, {Beaton}, {Bejger}, {Cackett},
  {Dall'Ora}, {Holanda}, {Jensen}, {Jha}, {Lusso}, {Minezaki}, {Risaliti},
  {Salaris}, {Toonen}, \& {Yoshii}}]{czerny_distances2018}
{Czerny}, B., {Beaton}, R., {Bejger}, M., {et~al.} 2018, \ssr, 214, 32

\bibitem[{{Czerny} \& {Hryniewicz}(2011)}]{czhr2011}
{Czerny}, B. \& {Hryniewicz}, K. 2011, \aap, 525, L8

\bibitem[{{Czerny} {et~al.}(2004){Czerny}, {Li}, {Loska}, \&
  {Szczerba}}]{czerny2004}
{Czerny}, B., {Li}, J., {Loska}, Z., \& {Szczerba}, R. 2004, \mnras, 348, L54

\bibitem[{{Czerny} {et~al.}(2017){Czerny}, {Li}, {Hryniewicz}, {Panda},
  {Wildy}, {Sniegowska}, {Wang}, {Sredzinska}, \& {Karas}}]{czerny2017}
{Czerny}, B., {Li}, Y.-R., {Hryniewicz}, K., {et~al.} 2017, \apj, 846, 154

\bibitem[{{Czerny} {et~al.}(2003){Czerny}, {Niko{\l}ajuk},
  {R{\'o}{\.z}a{\'n}ska}, {Dumont}, {Loska}, \& {Zycki}}]{czerny2003}
{Czerny}, B., {Niko{\l}ajuk}, M., {R{\'o}{\.z}a{\'n}ska}, A., {et~al.} 2003,
  \aap, 412, 317

\bibitem[{{Dalla Bont{\`a}} {et~al.}(2020){Dalla Bont{\`a}}, {Peterson},
  {Bentz}, {Brandt}, {Ciroi}, {De Rosa}, {Fonseca Alvarez}, {Grier}, {Hall},
  {Hern{\'a}ndez Santisteban}, {Ho}, {Homayouni}, {Horne}, {Kochanek}, {Li},
  {Morelli}, {Pizzella}, {Pogge}, {Schneider}, {Shen}, {Trump}, \&
  {Vestergaard}}]{2020ApJ...903..112D}
{Dalla Bont{\`a}}, E., {Peterson}, B.~M., {Bentz}, M.~C., {et~al.} 2020, \apj,
  903, 112

\bibitem[{{de Jaeger} {et~al.}(2022){de Jaeger}, {Galbany}, {Riess}, {Stahl},
  {Shappee}, {Filippenko}, \& {Zheng}}]{SNII2022}
{de Jaeger}, T., {Galbany}, L., {Riess}, A.~G., {et~al.} 2022, \mnras, 514,
  4620

\bibitem[{{De Rosa} {et~al.}(2015){De Rosa}, {Peterson}, {Ely}, {Kriss},
  {Crenshaw}, {Horne}, {Korista}, {Netzer}, {Pogge}, {Ar{\'e}valo}, {Barth},
  {Bentz}, {Brandt}, {Breeveld}, {Brewer}, {Dalla Bont{\`a}}, {De
  Lorenzo-C{\'a}ceres}, {Denney}, {Dietrich}, {Edelson}, {Evans}, {Fausnaugh},
  {Gehrels}, {Gelbord}, {Goad}, {Grier}, {Grupe}, {Hall}, {Kaastra}, {Kelly},
  {Kennea}, {Kochanek}, {Lira}, {Mathur}, {McHardy}, {Nousek}, {Pancoast},
  {Papadakis}, {Pei}, {Schimoia}, {Siegel}, {Starkey}, {Treu}, {Uttley},
  {Vaughan}, {Vestergaard}, {Villforth}, {Yan}, {Young}, \&
  {Zu}}]{derosa_2015ApJ}
{De Rosa}, G., {Peterson}, B.~M., {Ely}, J., {et~al.} 2015, \apj, 806, 128

\bibitem[{{de Vaucouleurs} \& {de Vaucouleurs}(1972)}]{deVaucouleurs1972}
{de Vaucouleurs}, G. \& {de Vaucouleurs}, A. 1972, \aplett, 12, 1

\bibitem[{{de Vaucouleurs} {et~al.}(1991){de Vaucouleurs}, {de Vaucouleurs},
  {Corwin}, {Buta}, {Paturel}, \& {Fouque}}]{deVaucouleurs1991}
{de Vaucouleurs}, G., {de Vaucouleurs}, A., {Corwin}, Herold~G., J., {et~al.}
  1991, {Third Reference Catalogue of Bright Galaxies}

\bibitem[{{DES Collaboration} {et~al.}(2025){DES Collaboration}, {Abbott},
  {Acevedo}, {Adamow}, {Aguena}, {Alarcon}, {Allam}, {Alves},
  {Andrade-Oliveira}, {Annis}, {Armstrong}, {Avila}, {Bacon}, {Bechtol},
  {Blazek}, {Bocquet}, {Brooks}, {Brout}, {Burke}, {Camacho}, {Camilleri},
  {Campailla}, {Carnero Rosell}, {Carr}, {Carretero}, {Castander}, {Cawthon},
  {Chan}, {Chang}, {Chen}, {Conselice}, {Costanzi}, {Crocce}, {da Costa},
  {Pereira}, {Davis}, {De Vicente}, {Deiosso}, {Desai}, {Diehl}, {Dodelson},
  {Doux}, {Drlica-Wagner}, {Elvin-Poole}, {Everett}, {Ferrero}, {Fert{\'e}},
  {Flaugher}, {Frieman}, {Galbany}, {Garc{\'\i}a-Bellido}, {Gatti},
  {Gaztanaga}, {Giannini}, {Gruen}, {Gruendl}, {Gutierrez}, {Hartley},
  {Herner}, {Hinton}, {Hollowood}, {Honscheid}, {Huterer}, {James}, {Jeffrey},
  {Jeltema}, {Kessler}, {Lahav}, {Lee}, {Lee}, {Lidman}, {Lin}, {Lin},
  {Marshall}, {Mena-Fern{\'a}ndez}, {Miquel}, {Muir}, {M{\"o}ller}, {Nichol},
  {Palmese}, {Paterno}, {Percival}, {Pieres}, {Plazas Malag{\'o}n}, {Popovic},
  {Porredon}, {Prat}, {Qu}, {Raveri}, {Rodriguez-Monroy}, {Romer}, {Rykoff},
  {Sako}, {Samuroff}, {Sanchez}, {Sanchez Cid}, {Scolnic}, {Sevilla-Noarbe},
  {Shah}, {Sheldon}, {Smith}, {Suchyta}, {Sullivan}, {Swanson}, {S{\'a}nchez},
  {Tarle}, {Taylor}, {Thomas}, {To}, {Toribio San Cipriano}, {Toy}, {Troxel},
  {Tucker}, {Vikram}, {Vincenzi}, {Walker}, {Weaverdyck}, {Weller}, {Wiseman},
  {Yamamoto}, \& {Yanny}}]{DES2025}
{DES Collaboration}, {Abbott}, T.~M.~C., {Acevedo}, M., {et~al.} 2025, arXiv
  e-prints, arXiv:2503.06712

\bibitem[{{DESI Collaboration} {et~al.}(2024){DESI Collaboration}, {Adame},
  {Aguilar}, {Ahlen}, {Alam}, {Alexander}, {Alvarez}, {Alves}, {Anand},
  {Andrade}, {Armengaud}, {Avila}, {Aviles}, {Awan}, {Bahr-Kalus}, {Bailey},
  {Baltay}, {Bault}, {Behera}, {BenZvi}, {Bera}, {Beutler}, {Bianchi}, {Blake},
  {Blum}, {Brieden}, {Brodzeller}, {Brooks}, {Buckley-Geer}, {Burtin},
  {Calderon}, {Canning}, {Carnero Rosell}, {Cereskaite}, {Cervantes-Cota},
  {Chabanier}, {Chaussidon}, {Chaves-Montero}, {Chen}, {Chen}, {Claybaugh},
  {Cole}, {Cuceu}, {Davis}, {Dawson}, {de la Macorra}, {de Mattia}, {Deiosso},
  {Dey}, {Dey}, {Ding}, {Doel}, {Edelstein}, {Eftekharzadeh}, {Eisenstein},
  {Elliott}, {Fagrelius}, {Fanning}, {Ferraro}, {Ereza}, {Findlay}, {Flaugher},
  {Font-Ribera}, {Forero-S{\'a}nchez}, {Forero-Romero}, {Frenk},
  {Garcia-Quintero}, {Gazta{\~n}aga}, {Gil-Mar{\'\i}n}, {Gontcho},
  {Gonzalez-Morales}, {Gonzalez-Perez}, {Gordon}, {Green}, {Gruen}, {Gsponer},
  {Gutierrez}, {Guy}, {Hadzhiyska}, {Hahn}, {Hanif}, {Herrera-Alcantar},
  {Honscheid}, {Howlett}, {Huterer}, {Ir{\v{s}}i{\v{c}}}, {Ishak}, {Juneau},
  {Kara{\c{c}}ayl{\i}}, {Kehoe}, {Kent}, {Kirkby}, {Kremin}, {Krolewski},
  {Lai}, {Lan}, {Landriau}, {Lang}, {Lasker}, {Le Goff}, {Le Guillou},
  {Leauthaud}, {Levi}, {Li}, {Linder}, {Lodha}, {Magneville}, {Manera},
  {Margala}, {Martini}, {Maus}, {McDonald}, {Medina-Varela}, {Meisner},
  {Mena-Fern{\'a}ndez}, {Miquel}, {Moon}, {Moore}, {Moustakas}, {Mudur},
  {Mueller}, {Mu{\~n}oz-Guti{\'e}rrez}, {Myers}, {Nadathur}, {Napolitano},
  {Neveux}, {Newman}, {Nguyen}, {Nie}, {Niz}, {Noriega}, {Padmanabhan},
  {Paillas}, {Palanque-Delabrouille}, {Pan}, {Penmetsa}, {Percival}, {Pieri},
  {Pinon}, {Poppett}, {Porredon}, {Prada}, {P{\'e}rez-Fern{\'a}ndez},
  {P{\'e}rez-R{\`a}fols}, {Rabinowitz}, {Raichoor}, {Ram{\'\i}rez-P{\'e}rez},
  {Ramirez-Solano}, {Ravoux}, {Rashkovetskyi}, {Rezaie}, {Rich}, {Rocher},
  {Rockosi}, {Roe}, {Rosado-Marin}, {Ross}, {Rossi}, {Ruggeri},
  {Ruhlmann-Kleider}, {Samushia}, {Sanchez}, {Saulder}, {Schlafly}, {Schlegel},
  {Schubnell}, {Seo}, {Shafieloo}, {Sharples}, {Silber}, {Slosar}, {Smith},
  {Sprayberry}, {Tan}, {Tarl{\'e}}, {Taylor}, {Trusov}, {Ure{\~n}a-L{\'o}pez},
  {Vaisakh}, {Valcin}, {Valdes}, {Vargas-Maga{\~n}a}, {Verde}, {Walther},
  {Wang}, {Wang}, {Weaver}, {Weaverdyck}, {Wechsler}, {Weinberg}, {White},
  {Yu}, {Yu}, {Yuan}, {Y{\`e}che}, {Zaborowski}, {Zarrouk}, {Zhang}, {Zhao}, \&
  {Zhao}}]{DESI_BAO_2024arXiv240403002D}
{DESI Collaboration}, {Adame}, A.~G., {Aguilar}, J., {et~al.} 2024, arXiv
  e-prints, arXiv:2404.03002

\bibitem[{{Dov{\v{c}}iak} {et~al.}(2004){Dov{\v{c}}iak}, {Karas}, \&
  {Yaqoob}}]{dovciak2004}
{Dov{\v{c}}iak}, M., {Karas}, V., \& {Yaqoob}, T. 2004, \apjs, 153, 205

\bibitem[{{Draine} \& {Lee}(1984)}]{draine1984}
{Draine}, B.~T. \& {Lee}, H.~M. 1984, \apj, 285, 89

\bibitem[{{Edelson} {et~al.}(2015){Edelson}, {Gelbord}, {Horne}, {McHardy},
  {Peterson}, {Ar{\'e}valo}, {Breeveld}, {De Rosa}, {Evans}, {Goad}, {Kriss},
  {Brandt}, {Gehrels}, {Grupe}, {Kennea}, {Kochanek}, {Nousek}, {Papadakis},
  {Siegel}, {Starkey}, {Uttley}, {Vaughan}, {Young}, {Barth}, {Bentz},
  {Brewer}, {Crenshaw}, {Dalla Bont{\`a}}, {De Lorenzo-C{\'a}ceres}, {Denney},
  {Dietrich}, {Ely}, {Fausnaugh}, {Grier}, {Hall}, {Kaastra}, {Kelly},
  {Korista}, {Lira}, {Mathur}, {Netzer}, {Pancoast}, {Pei}, {Pogge},
  {Schimoia}, {Treu}, {Vestergaard}, {Villforth}, {Yan}, \&
  {Zu}}]{edelson_2015ApJ}
{Edelson}, R., {Gelbord}, J.~M., {Horne}, K., {et~al.} 2015, \apj, 806, 129

\bibitem[{{Edelson} {et~al.}(2024){Edelson}, {Peterson}, {Gelbord}, {Horne},
  {Goad}, {McHardy}, {Vaughan}, \& {Vestergaard}}]{2024ApJ...973..152E}
{Edelson}, R., {Peterson}, B.~M., {Gelbord}, J., {et~al.} 2024, \apj, 973, 152

\bibitem[{{Fausnaugh} {et~al.}(2016){Fausnaugh}, {Denney}, {Barth}, {Bentz},
  {Bottorff}, {Carini}, {Croxall}, {De Rosa}, {Goad}, {Horne}, {Joner},
  {Kaspi}, {Kim}, {Klimanov}, {Kochanek}, {Leonard}, {Netzer}, {Peterson},
  {Schn{\"u}lle}, {Sergeev}, {Vestergaard}, {Zheng}, {Zu}, {Anderson},
  {Ar{\'e}valo}, {Bazhaw}, {Borman}, {Boroson}, {Brandt}, {Breeveld}, {Brewer},
  {Cackett}, {Crenshaw}, {Dalla Bont{\`a}}, {De Lorenzo-C{\'a}ceres},
  {Dietrich}, {Edelson}, {Efimova}, {Ely}, {Evans}, {Filippenko}, {Flatland},
  {Gehrels}, {Geier}, {Gelbord}, {Gonzalez}, {Gorjian}, {Grier}, {Grupe},
  {Hall}, {Hicks}, {Horenstein}, {Hutchison}, {Im}, {Jensen}, {Jones},
  {Kaastra}, {Kelly}, {Kennea}, {Kim}, {Korista}, {Kriss}, {Lee}, {Lira},
  {MacInnis}, {Manne-Nicholas}, {Mathur}, {McHardy}, {Montouri}, {Musso},
  {Nazarov}, {Norris}, {Nousek}, {Okhmat}, {Pancoast}, {Papadakis}, {Parks},
  {Pei}, {Pogge}, {Pott}, {Rafter}, {Rix}, {Saylor}, {Schimoia}, {Siegel},
  {Spencer}, {Starkey}, {Sung}, {Teems}, {Treu}, {Turner}, {Uttley},
  {Villforth}, {Weiss}, {Woo}, {Yan}, \& {Young}}]{fausnaugh2016}
{Fausnaugh}, M.~M., {Denney}, K.~D., {Barth}, A.~J., {et~al.} 2016, \apj, 821,
  56

\bibitem[{{Ferland} {et~al.}(1979){Ferland}, {Netzer}, \&
  {Shields}}]{ferland1979}
{Ferland}, G.~J., {Netzer}, H., \& {Shields}, G.~A. 1979, \apj, 232, 382

\bibitem[{{Floris} {et~al.}(2025){Floris}, {Pandey}, {Czerny}, {Martinez
  Aldama}, {Panda}, {Marziani}, \& {Prince}}]{floris2025}
{Floris}, A., {Pandey}, A., {Czerny}, B., {et~al.} 2025, \aap, 697, A23

\bibitem[{{Freedman} {et~al.}(2024){Freedman}, {Madore}, {Jang}, {Hoyt}, {Lee},
  \& {Owens}}]{freedman2024}
{Freedman}, W.~L., {Madore}, B.~F., {Jang}, I.~S., {et~al.} 2024, arXiv
  e-prints, arXiv:2408.06153

\bibitem[{{Garc{\'\i}a} {et~al.}(2013){Garc{\'\i}a}, {Dauser}, {Reynolds},
  {Kallman}, {McClintock}, {Wilms}, \& {Eikmann}}]{garcia2013}
{Garc{\'\i}a}, J., {Dauser}, T., {Reynolds}, C.~S., {et~al.} 2013, \apj, 768,
  146

\bibitem[{{Gaskell} {et~al.}(2023){Gaskell}, {Anderson}, {Birmingham}, \&
  {Ghosh}}]{gaskell2023}
{Gaskell}, C.~M., {Anderson}, F.~C., {Birmingham}, S.~{\'A}., \& {Ghosh}, S.
  2023, \mnras, 519, 4082

\bibitem[{{Gaskell} {et~al.}(2004){Gaskell}, {Goosmann}, {Antonucci}, \&
  {Whysong}}]{gaskell2004}
{Gaskell}, C.~M., {Goosmann}, R.~W., {Antonucci}, R. R.~J., \& {Whysong}, D.~H.
  2004, \apj, 616, 147

\bibitem[{{Gavas} {et~al.}(2025){Gavas}, {Bagla}, \& {Khandai}}]{gavas2024}
{Gavas}, S., {Bagla}, J.~S., \& {Khandai}, N. 2025, \prd, 111, 043516

\bibitem[{{Grier} {et~al.}(2017){Grier}, {Pancoast}, {Barth}, {Fausnaugh},
  {Brewer}, {Treu}, \& {Peterson}}]{grier2017}
{Grier}, C.~J., {Pancoast}, A., {Barth}, A.~J., {et~al.} 2017, \apj, 849, 146

\bibitem[{{Grillo} {et~al.}(2024){Grillo}, {Pagano}, {Rosati}, \&
  {Suyu}}]{grillo2024}
{Grillo}, C., {Pagano}, L., {Rosati}, P., \& {Suyu}, S.~H. 2024, \aap, 684, L23

\bibitem[{{Guo} {et~al.}(2022){Guo}, {Barth}, \& {Wang}}]{2022ApJ...940...20G}
{Guo}, H., {Barth}, A.~J., \& {Wang}, S. 2022, \apj, 940, 20

\bibitem[{{Homayouni} {et~al.}(2019){Homayouni}, {Trump}, {Grier}, {Shen},
  {Starkey}, {Brandt}, {Fonseca Alvarez}, {Hall}, {Horne}, {Kinemuchi}, {I-Hsiu
  Li}, {McGreer}, {Sun}, {Ho}, \& {Schneider}}]{2019ApJ...880..126H}
{Homayouni}, Y., {Trump}, J.~R., {Grier}, C.~J., {et~al.} 2019, \apj, 880, 126

\bibitem[{{Horne} {et~al.}(2021){Horne}, {De Rosa}, {Peterson}, {Barth}, {Ely},
  {Fausnaugh}, {Kriss}, {Pei}, {Bentz}, {Cackett}, {Edelson}, {Eracleous},
  {Goad}, {Grier}, {Kaastra}, {Kochanek}, {Krongold}, {Mathur}, {Netzer},
  {Proga}, {Tejos}, {Vestergaard}, {Villforth}, {Adams}, {Anderson},
  {Ar{\'e}valo}, {Beatty}, {Bennert}, {Bigley}, {Bisogni}, {Borman}, {Boroson},
  {Bottorff}, {Brandt}, {Breeveld}, {Brotherton}, {Brown}, {Brown}, {Canalizo},
  {Carini}, {Clubb}, {Comerford}, {Corsini}, {Crenshaw}, {Croft}, {Croxall},
  {Dalla Bont{\`a}}, {Deason}, {Dehghanian}, {De Lorenzo-C{\'a}ceres},
  {Denney}, {Dietrich}, {Done}, {Efimova}, {Evans}, {Ferland}, {Filippenko},
  {Flatland}, {Fox}, {Gardner}, {Gates}, {Gehrels}, {Geier}, {Gelbord},
  {Gonzalez}, {Gorjian}, {Greene}, {Grupe}, {Gupta}, {Hall}, {Henderson},
  {Hicks}, {Holmbeck}, {Holoien}, {Hutchison}, {Im}, {Jensen}, {Johnson},
  {Joner}, {Jones}, {Kaspi}, {Kelly}, {Kennea}, {Kim}, {Kim}, {Kim}, {King},
  {Klimanov}, {Korista}, {Lau}, {Lee}, {Leonard}, {Li}, {Lira}, {Lochhaas},
  {Ma}, {MacInnis}, {Malkan}, {Manne-Nicholas}, {Mauerhan}, {McGurk},
  {McHardy}, {Montuori}, {Morelli}, {Mosquera}, {Mudd},
  {M{\"u}ller-S{\'a}nchez}, {Nazarov}, {Norris}, {Nousek}, {Nguyen}, {Ochner},
  {Okhmat}, {Pancoast}, {Papadakis}, {Parks}, {Penny}, {Pizzella}, {Pogge},
  {Poleski}, {Pott}, {Rafter}, {Rix}, {Runnoe}, {Saylor}, {Schimoia},
  {Schn{\"u}lle}, {Scott}, {Sergeev}, {Shappee}, {Shivvers}, {Siegel},
  {Simonian}, {Siviero}, {Skielboe}, {Somers}, {Spencer}, {Starkey}, {Stevens},
  {Sung}, {Tayar}, {Treu}, {Turner}, {Uttley}, {Van Saders}, {Vican},
  {Villanueva}, {Weiss}, {Woo}, {Yan}, {Young}, {Yuk}, {Zheng}, {Zhu}, \&
  {Zu}}]{horne2021}
{Horne}, K., {De Rosa}, G., {Peterson}, B.~M., {et~al.} 2021, \apj, 907, 76

\bibitem[{{Jaiswal} {et~al.}(2023){Jaiswal}, {Prince}, {Panda}, \&
  {Czerny}}]{Jaiswal2023}
{Jaiswal}, V.~K., {Prince}, R., {Panda}, S., \& {Czerny}, B. 2023, \aap, 670,
  A147

\bibitem[{{Kammoun} {et~al.}(2024{\natexlab{a}}){Kammoun}, {Papadakis},
  {Dov{\v{c}}iak}, \& {Panagiotou}}]{Kammoun2024}
{Kammoun}, E., {Papadakis}, I.~E., {Dov{\v{c}}iak}, M., \& {Panagiotou}, C.
  2024{\natexlab{a}}, \aap, 686, A69

\bibitem[{{Kammoun} {et~al.}(2024{\natexlab{b}}){Kammoun}, {Papadakis},
  {Dov{\v{c}}iak}, \& {Panagiotou}}]{kammoun_5548_2024}
{Kammoun}, E., {Papadakis}, I.~E., {Dov{\v{c}}iak}, M., \& {Panagiotou}, C.
  2024{\natexlab{b}}, \aap, 686, A69

\bibitem[{{Kammoun} {et~al.}(2021{\natexlab{a}}){Kammoun}, {Dov{\v{c}}iak},
  {Papadakis}, {Caballero-Garc{\'\i}a}, \& {Karas}}]{kammoun_analit2021}
{Kammoun}, E.~S., {Dov{\v{c}}iak}, M., {Papadakis}, I.~E.,
  {Caballero-Garc{\'\i}a}, M.~D., \& {Karas}, V. 2021{\natexlab{a}}, \apj, 907,
  20

\bibitem[{{Kammoun} {et~al.}(2021{\natexlab{b}}){Kammoun}, {Papadakis}, \&
  {Dov{\v{c}}iak}}]{kammoun2021}
{Kammoun}, E.~S., {Papadakis}, I.~E., \& {Dov{\v{c}}iak}, M.
  2021{\natexlab{b}}, \mnras, 503, 4163

\bibitem[{{Kammoun} {et~al.}(2023){Kammoun}, {Robin}, {Papadakis},
  {Dov{\v{c}}iak}, \& {Panagiotou}}]{kammoun2023}
{Kammoun}, E.~S., {Robin}, L., {Papadakis}, I.~E., {Dov{\v{c}}iak}, M., \&
  {Panagiotou}, C. 2023, \mnras, 526, 138

\bibitem[{{Kinney} {et~al.}(1996){Kinney}, {Calzetti}, {Bohlin}, {McQuade},
  {Storchi-Bergmann}, \& {Schmitt}}]{kinney1996}
{Kinney}, A.~L., {Calzetti}, D., {Bohlin}, R.~C., {et~al.} 1996, \apj, 467, 38

\bibitem[{{Kokubo}(2018)}]{kokubo2018}
{Kokubo}, M. 2018, \pasj, 70, 97

\bibitem[{{Korista} {et~al.}(1995){Korista}, {Alloin}, {Barr}, {Clavel},
  {Cohen}, {Crenshaw}, {Evans}, {Horne}, {Koratkar}, {Kriss}, {Krolik},
  {Malkan}, {Morris}, {Netzer}, {O'Brien}, {Peterson}, {Reichert},
  {Rodriguez-Pascual}, {Wamsteker}, {Anderson}, {Axon}, {Benitez}, {Berlind},
  {Bertram}, {Blackwell}, {Bochkarev}, {Boisson}, {Carini}, {Carrillo},
  {Carone}, {Cheng}, {Christensen}, {Chuvaev}, {Dietrich}, {Dokter},
  {Doroshenko}, {Dultzin-Hacyan}, {England}, {Espey}, {Filippenko}, {Gaskell},
  {Goad}, {Ho}, {Huchra}, {Jiang}, {Kaspi}, {Kollatschny}, {Laor}, {Luminet},
  {MacAlpine}, {MacKenty}, {Malkov}, {Maoz}, {Martin}, {Matheson}, {McCollum},
  {Merkulova}, {Metik}, {Mignoli}, {Miller}, {Pastoriza}, {Pelat}, {Penfold},
  {Perez}, {Perola}, {Persaud}, {Peters}, {Pitts}, {Pogge}, {Pronik}, {Pronik},
  {Ptak}, {Rawley}, {Recondo-Gonzalez}, {Rodriguez-Espinosa}, {Romanishin},
  {Sadun}, {Salamanca}, {Santos-Lleo}, {Sekiguchi}, {Sergeev}, {Shapovalova},
  {Shields}, {Shrader}, {Shull}, {Silbermann}, {Sitko}, {Skillman}, {Smith},
  {Smith}, {Snijders}, {Sparke}, {Stirpe}, {Stoner}, {Sun}, {Thiele}, {Tokarz},
  {Tsvetanov}, {Turnshek}, {Veilleux}, {Wagner}, {Wagner}, {Wanders}, {Wang},
  {Welsh}, {Weymann}, {White}, {Wilkes}, {Wills}, {Winge}, {Wu}, \&
  {Zou}}]{korista1995}
{Korista}, K.~T., {Alloin}, D., {Barr}, P., {et~al.} 1995, \apjs, 97, 285

\bibitem[{{Korista} \& {Goad}(2001)}]{korista2001}
{Korista}, K.~T. \& {Goad}, M.~R. 2001, \apj, 553, 695

\bibitem[{{Korista} \& {Goad}(2019)}]{korista2019}
{Korista}, K.~T. \& {Goad}, M.~R. 2019, \mnras, 489, 5284

\bibitem[{{Kova{\v{c}}evi{\'c}} {et~al.}(2014){Kova{\v{c}}evi{\'c}},
  {Popovi{\'c}}, \& {Kollatschny}}]{kovacevic2014}
{Kova{\v{c}}evi{\'c}}, J., {Popovi{\'c}}, L.~{\v{C}}., \& {Kollatschny}, W.
  2014, Advances in Space Research, 54, 1347

\bibitem[{{Krolik}(1999{\natexlab{a}})}]{krolik1999}
{Krolik}, J.~H. 1999{\natexlab{a}}, {Active galactic nuclei : from the central
  black hole to the galactic environment}

\bibitem[{{Krolik}(1999{\natexlab{b}})}]{1999agnc.book.....K}
{Krolik}, J.~H. 1999{\natexlab{b}}, {Active galactic nuclei : from the central
  black hole to the galactic environment}

\bibitem[{{Krolik} {et~al.}(1991){Krolik}, {Horne}, {Kallman}, {Malkan},
  {Edelson}, \& {Kriss}}]{krolik1991}
{Krolik}, J.~H., {Horne}, K., {Kallman}, T.~R., {et~al.} 1991, \apj, 371, 541

\bibitem[{{Krolik} {et~al.}(1981){Krolik}, {McKee}, \& {Tarter}}]{krolik1981}
{Krolik}, J.~H., {McKee}, C.~F., \& {Tarter}, C.~B. 1981, \apj, 249, 422

\bibitem[{{Kubota} \& {Done}(2018)}]{kubota2018}
{Kubota}, A. \& {Done}, C. 2018, \mnras, 480, 1247

\bibitem[{{Laor} \& {Draine}(1993)}]{laor1993}
{Laor}, A. \& {Draine}, B.~T. 1993, \apj, 402, 441

\bibitem[{{Lawther} {et~al.}(2018){Lawther}, {Goad}, {Korista}, {Ulrich}, \&
  {Vestergaard}}]{lawther2018}
{Lawther}, D., {Goad}, M.~R., {Korista}, K.~T., {Ulrich}, O., \& {Vestergaard},
  M. 2018, \mnras, 481, 533

\bibitem[{{Li} {et~al.}(2025){Li}, {Shangguan}, {Wang}, {Davies}, {Santos},
  {Eisenhauer}, {Songsheng}, {Winkler}, {Aceituno}, {Bai}, {Bai}, {Brotherton},
  {Cao}, {Chen}, {Du}, {Fang}, {Feng}, {Feuchtgruber}, {F{\"o}rster Schreiber},
  {Fu}, {Genzel}, {Gillessen}, {Ho}, {Hu}, {Liu}, {Lutz}, {Ott}, {Petrov},
  {Rabien}, {Shimizu}, {Sturm}, {Tacconi}, {Wang}, {Yao}, {Zhai}, {Zhang},
  {Zhao}, \& {Zhao}}]{2025arXiv250218856L}
{Li}, Y.-R., {Shangguan}, J., {Wang}, J.-M., {et~al.} 2025, arXiv e-prints,
  arXiv:2502.18856

\bibitem[{{Li} {et~al.}(2013){Li}, {Wang}, {Ho}, {Du}, \& {Bai}}]{li2013}
{Li}, Y.-R., {Wang}, J.-M., {Ho}, L.~C., {Du}, P., \& {Bai}, J.-M. 2013, \apj,
  779, 110

\bibitem[{{Li} {et~al.}(2022){Li}, {Wang}, {Songsheng}, {Zhang}, {Du}, {Hu}, \&
  {Xiao}}]{LiYanRong2022}
{Li}, Y.-R., {Wang}, J.-M., {Songsheng}, Y.-Y., {et~al.} 2022, \apj, 927, 58

\bibitem[{{Liu} {et~al.}(2024){Liu}, {Cao}, {Biesiada}, {Zhang}, \&
  {Wang}}]{biesiada2024}
{Liu}, T., {Cao}, S., {Biesiada}, M., {Zhang}, Y., \& {Wang}, J. 2024, \apjl,
  965, L11

\bibitem[{{Long} \& {Dexter}(2025)}]{long2025}
{Long}, K. \& {Dexter}, J. 2025, \apj, 987, 196

\bibitem[{{Lu} {et~al.}(2022){Lu}, {Bai}, {Wang}, {Hu}, {Li}, {Du}, {Xiao},
  {Feng}, {Li}, {Wang}, {Zhang}, \& {Huang}}]{Lu2022}
{Lu}, K.-X., {Bai}, J.-M., {Wang}, J.-M., {et~al.} 2022, \apjs, 263, 10

\bibitem[{{Lu} {et~al.}(2016){Lu}, {Du}, {Hu}, {Li}, {Zhang}, {Wang}, {Huang},
  {Bi}, {Bai}, {Ho}, \& {Wang}}]{lu2016}
{Lu}, K.-X., {Du}, P., {Hu}, C., {et~al.} 2016, \apj, 827, 118

\bibitem[{{Lusso} {et~al.}(2025){Lusso}, {Risaliti}, \& {Nardini}}]{lusso2025}
{Lusso}, E., {Risaliti}, G., \& {Nardini}, E. 2025, arXiv e-prints,
  arXiv:2504.02040

\bibitem[{{Mandal} {et~al.}(2025){Mandal}, {Woo}, \&
  {Wang}}]{2025ApJ...985...30M}
{Mandal}, A.~K., {Woo}, J.-H., \& {Wang}, S. 2025, \apj, 985, 30

\bibitem[{{Mandal} {et~al.}(2024){Mandal}, {Woo}, {Wang}, {Rakshit}, {Cho},
  {Son}, \& {Stalin}}]{2024ApJ...968...59M}
{Mandal}, A.~K., {Woo}, J.-H., {Wang}, S., {et~al.} 2024, \apj, 968, 59

\bibitem[{{Mart{\'\i}nez-Aldama} {et~al.}(2019){Mart{\'\i}nez-Aldama},
  {Czerny}, {Kawka}, {Karas}, {Panda}, {Zaja{\v{c}}ek}, \&
  {{\.Z}ycki}}]{martinezAldama2019}
{Mart{\'\i}nez-Aldama}, M.~L., {Czerny}, B., {Kawka}, D., {et~al.} 2019, \apj,
  883, 170

\bibitem[{{McHardy} {et~al.}(2014){McHardy}, {Cameron}, {Dwelly}, {Connolly},
  {Lira}, {Emmanoulopoulos}, {Gelbord}, {Breedt}, {Arevalo}, \&
  {Uttley}}]{2014MNRAS.444.1469M}
{McHardy}, I.~M., {Cameron}, D.~T., {Dwelly}, T., {et~al.} 2014, \mnras, 444,
  1469

\bibitem[{{Mehdipour} {et~al.}(2015){Mehdipour}, {Kaastra}, {Kriss}, {Cappi},
  {Petrucci}, {Steenbrugge}, {Arav}, {Behar}, {Bianchi}, {Boissay},
  {Branduardi-Raymont}, {Costantini}, {Ebrero}, {Di Gesu}, {Harrison}, {Kaspi},
  {De Marco}, {Matt}, {Paltani}, {Peterson}, {Ponti}, {Pozo Nu{\~n}ez}, {De
  Rosa}, {Ursini}, {de Vries}, {Walton}, \& {Whewell}}]{mehdipour2015}
{Mehdipour}, M., {Kaastra}, J.~S., {Kriss}, G.~A., {et~al.} 2015, \aap, 575,
  A22

\bibitem[{{Mor} \& {Netzer}(2012)}]{mor2012}
{Mor}, R. \& {Netzer}, H. 2012, \mnras, 420, 526

\bibitem[{{Morgan} {et~al.}(2010){Morgan}, {Kochanek}, {Morgan}, \&
  {Falco}}]{2010ApJ...712.1129M}
{Morgan}, C.~W., {Kochanek}, C.~S., {Morgan}, N.~D., \& {Falco}, E.~E. 2010,
  \apj, 712, 1129

\bibitem[{{Mudd} {et~al.}(2018){Mudd}, {Martini}, {Zu}, {Kochanek}, {Peterson},
  {Kessler}, {Davis}, {Hoormann}, {King}, {Lidman}, {Sommer}, {Tucker},
  {Asorey}, {Hinton}, {Glazebrook}, {Kuehn}, {Lewis}, {Macaulay}, {Moeller},
  {O'Neill}, {Zhang}, {Abbott}, {Abdalla}, {Allam}, {Banerji},
  {Benoit-L{\'e}vy}, {Bertin}, {Brooks}, {Carnero Rosell}, {Carollo}, {Carrasco
  Kind}, {Carretero}, {Cunha}, {D'Andrea}, {da Costa}, {Davis}, {Desai},
  {Doel}, {Fosalba}, {Garc{\'\i}a-Bellido}, {Gaztanaga}, {Gerdes}, {Gruen},
  {Gruendl}, {Gschwend}, {Gutierrez}, {Hartley}, {Honscheid}, {James},
  {Kuhlmann}, {Kuropatkin}, {Lima}, {Maia}, {Marshall}, {McMahon}, {Menanteau},
  {Miquel}, {Plazas}, {Romer}, {Sanchez}, {Schindler}, {Schubnell}, {Smith},
  {Smith}, {Soares-Santos}, {Sobreira}, {Suchyta}, {Swanson}, {Tarle},
  {Thomas}, {Tucker}, {Walker}, \& {DES Collaboration}}]{2018ApJ...862..123M}
{Mudd}, D., {Martini}, P., {Zu}, Y., {et~al.} 2018, \apj, 862, 123

\bibitem[{{Naddaf} \& {Czerny}(2022)}]{naddaf2022}
{Naddaf}, M.~H. \& {Czerny}, B. 2022, \aap, 663, A77

\bibitem[{{Naddaf} \& {Czerny}(2024)}]{naddaf_CF2024}
{Naddaf}, M.-H. \& {Czerny}, B. 2024, Universe, 10, 29

\bibitem[{{Naddaf} {et~al.}(2021){Naddaf}, {Czerny}, \&
  {Szczerba}}]{naddaf2021}
{Naddaf}, M.-H., {Czerny}, B., \& {Szczerba}, R. 2021, \apj, 920, 30

\bibitem[{{Naddaf} {et~al.}(2025){Naddaf}, {Martinez-Aldama}, {Hutsemekers},
  {Savic}, \& {Czerny}}]{Naddaf_2025}
{Naddaf}, M.~H., {Martinez-Aldama}, M.~L., {Hutsemekers}, D., {Savic}, D., \&
  {Czerny}, B. 2025, arXiv e-prints, arXiv:2506.01159

\bibitem[{{Naddaf} {et~al.}(2023){Naddaf}, {Martinez-Aldama}, {Marziani},
  {Panda}, {Sniegowska}, \& {Czerny}}]{naddaf2023}
{Naddaf}, M.~H., {Martinez-Aldama}, M.~L., {Marziani}, P., {et~al.} 2023, \aap,
  675, A43

\bibitem[{{Netzer}(2020)}]{netzer2020}
{Netzer}, H. 2020, \mnras, 494, 1611

\bibitem[{{Netzer}(2022)}]{netzer2022}
{Netzer}, H. 2022, \mnras, 509, 2637

\bibitem[{{Netzer} {et~al.}(2024){Netzer}, {Goad}, {Barth}, {Cackett}, {Horne},
  {Hu}, {Kara}, {Korista}, {Kriss}, {Lewin}, {Montano}, {Arav}, {Behar},
  {Brotherton}, {Chelouche}, {De Rosa}, {Dalla Bont{\`a}}, {Dehghanian},
  {Ferland}, {Fian}, {Homayouni}, {Ili{\'c}}, {Kaspi}, {Kova{\v{c}}evi{\'c}},
  {Landt}, {{\v{C}}. Popovi{\'c}}, {Storchi-Bergmann}, {Wang}, \&
  {Zaidouni}}]{netzer2024}
{Netzer}, H., {Goad}, M.~R., {Barth}, A.~J., {et~al.} 2024, \apj, 976, 59

\bibitem[{{Novikov} \& {Thorne}(1973)}]{NT1973}
{Novikov}, I.~D. \& {Thorne}, K.~S. 1973, in Black Holes (Les Astres Occlus),
  ed. C.~{Dewitt} \& B.~S. {Dewitt}, 343--450

\bibitem[{{Pancoast} {et~al.}(2014){Pancoast}, {Brewer}, \&
  {Treu}}]{pancoast2014}
{Pancoast}, A., {Brewer}, B.~J., \& {Treu}, T. 2014, \mnras, 445, 3055

\bibitem[{{Panda}(2021)}]{Panda_cafe2021}
{Panda}, S. 2021, \aap, 650, A154

\bibitem[{{Panda} {et~al.}(2022){Panda}, {Bon}, {Marziani}, \&
  {Bon}}]{2022AN....34310091P}
{Panda}, S., {Bon}, E., {Marziani}, P., \& {Bon}, N. 2022, Astronomische
  Nachrichten, 343, e210091

\bibitem[{{Panda} {et~al.}(2020){Panda}, {Mart{\'\i}nez-Aldama}, {Marinello},
  {Czerny}, {Marziani}, \& {Dultzin}}]{2020ApJ...902...76P}
{Panda}, S., {Mart{\'\i}nez-Aldama}, M.~L., {Marinello}, M., {et~al.} 2020,
  \apj, 902, 76

\bibitem[{{Panda} {et~al.}(2024){Panda}, {Pozo Nu{\~n}ez}, {Ba{\~n}ados}, \&
  {Heidt}}]{2024ApJ...968L..16P}
{Panda}, S., {Pozo Nu{\~n}ez}, F., {Ba{\~n}ados}, E., \& {Heidt}, J. 2024,
  \apjl, 968, L16

\bibitem[{{Pandey} {et~al.}(2023){Pandey}, {Czerny}, {Panda}, {Prince},
  {Jaiswal}, {Martinez-Aldama}, {Zaja{\v{c}}ek}, \&
  {{\'S}niegowska}}]{pandey2023}
{Pandey}, A., {Czerny}, B., {Panda}, S., {et~al.} 2023, \aap, 680, A102

\bibitem[{{Pandey} {et~al.}(2025){Pandey}, {Mart{\'\i}nez-Aldama}, {Czerny},
  {Panda}, {Zaja{\v{c}}ek}, {Wang}, {Li}, \& {Du}}]{Pandey_2025}
{Pandey}, A., {Mart{\'\i}nez-Aldama}, M.~L., {Czerny}, B., {et~al.} 2025,
  \apjs, 277, 36

\bibitem[{{Papoutsis} {et~al.}(2024){Papoutsis}, {Papadakis}, {Panagiotou},
  {Dov{\v{c}}iak}, \& {Kammoun}}]{2024Papoutsis}
{Papoutsis}, M., {Papadakis}, I.~E., {Panagiotou}, C., {Dov{\v{c}}iak}, M., \&
  {Kammoun}, E. 2024, \aap, 691, A60

\bibitem[{{Pei} {et~al.}(2017){Pei}, {Fausnaugh}, {Barth}, {Peterson}, {Bentz},
  {De Rosa}, {Denney}, {Goad}, {Kochanek}, {Korista}, {Kriss}, {Pogge},
  {Bennert}, {Brotherton}, {Clubb}, {Dalla Bont{\`a}}, {Filippenko}, {Greene},
  {Grier}, {Vestergaard}, {Zheng}, {Adams}, {Beatty}, {Bigley}, {Brown},
  {Brown}, {Canalizo}, {Comerford}, {Coker}, {Corsini}, {Croft}, {Croxall},
  {Deason}, {Eracleous}, {Fox}, {Gates}, {Henderson}, {Holmbeck}, {Holoien},
  {Jensen}, {Johnson}, {Kelly}, {Kim}, {King}, {Lau}, {Li}, {Lochhaas}, {Ma},
  {Manne-Nicholas}, {Mauerhan}, {Malkan}, {McGurk}, {Morelli}, {Mosquera},
  {Mudd}, {Muller Sanchez}, {Nguyen}, {Ochner}, {Ou-Yang}, {Pancoast}, {Penny},
  {Pizzella}, {Poleski}, {Runnoe}, {Scott}, {Schimoia}, {Shappee}, {Shivvers},
  {Simonian}, {Siviero}, {Somers}, {Stevens}, {Strauss}, {Tayar}, {Tejos},
  {Treu}, {Van Saders}, {Vican}, {Villanueva}, {Yuk}, {Zakamska}, {Zhu},
  {Anderson}, {Ar{\'e}valo}, {Bazhaw}, {Bisogni}, {Borman}, {Bottorff},
  {Brandt}, {Breeveld}, {Cackett}, {Carini}, {Crenshaw}, {De
  Lorenzo-C{\'a}ceres}, {Dietrich}, {Edelson}, {Efimova}, {Ely}, {Evans},
  {Ferland}, {Flatland}, {Gehrels}, {Geier}, {Gelbord}, {Grupe}, {Gupta},
  {Hall}, {Hicks}, {Horenstein}, {Horne}, {Hutchison}, {Im}, {Joner}, {Jones},
  {Kaastra}, {Kaspi}, {Kelly}, {Kennea}, {Kim}, {Kim}, {Klimanov}, {Lee},
  {Leonard}, {Lira}, {MacInnis}, {Mathur}, {McHardy}, {Montouri}, {Musso},
  {Nazarov}, {Netzer}, {Norris}, {Nousek}, {Okhmat}, {Papadakis}, {Parks},
  {Pott}, {Rafter}, {Rix}, {Saylor}, {Schn{\"u}lle}, {Sergeev}, {Siegel},
  {Skielboe}, {Spencer}, {Starkey}, {Sung}, {Teems}, {Turner}, {Uttley},
  {Villforth}, {Weiss}, {Woo}, {Yan}, {Young}, \& {Zu}}]{pei2017}
{Pei}, L., {Fausnaugh}, M.~M., {Barth}, A.~J., {et~al.} 2017, \apj, 837, 131

\bibitem[{{Perivolaropoulos}(2024)}]{perivolaropoulos2024}
{Perivolaropoulos}, L. 2024, \prd, 110, 123518

\bibitem[{{Peterson}(1993)}]{1993PASP..105..247P}
{Peterson}, B.~M. 1993, \pasp, 105, 247

\bibitem[{{Peterson} {et~al.}(1992){Peterson}, {Alloin}, {Axon}, {Balonek},
  {Bertram}, {Boroson}, {Christensen}, {Clements}, {Dietrich}, {Elvis},
  {Filippenko}, {Gaskell}, {Haswell}, {Huchra}, {Jackson}, {Kollatschny},
  {Korista}, {Lame}, {Leacock}, {Lin}, {Malkan}, {Monk}, {Penston}, {Pogge},
  {Robinson}, {Rosenblatt}, {Shields}, {Smith}, {Stirpe}, {Sun}, {Turner},
  {Wagner}, {Wilkes}, \& {Wills}}]{peterson1992}
{Peterson}, B.~M., {Alloin}, D., {Axon}, D., {et~al.} 1992, \apj, 392, 470

\bibitem[{{Peterson} {et~al.}(1991){Peterson}, {Balonek}, {Barker}, {Bechtold},
  {Bertram}, {Bochkarev}, {Bolte}, {Bond}, {Boroson}, {Carini}, {Carone},
  {Christensen}, {Clements}, {Cochran}, {Cohen}, {Crampton}, {Dietrich},
  {Elvis}, {Ferguson}, {Filippenko}, {Fricke}, {Gaskell}, {Halpern}, {Huchra},
  {Hutchings}, {Kollatschny}, {Koratkar}, {Korista}, {Krolik}, {Lame}, {Laor},
  {Leacock}, {MacAlpine}, {Malkan}, {Maoz}, {Miller}, {Morris}, {Netzer},
  {Oliveira}, {Penfold}, {Penston}, {Perez}, {Pogge}, {Richmond}, {Romanishin},
  {Rosenblatt}, {Saddlemyer}, {Sadun}, {Sawyer}, {Shields}, {Shapovalova},
  {Smith}, {Smith}, {Smith}, {Sun}, {Thiele}, {Turner}, {Veilleux}, {Wagner},
  {Weymann}, {Wilkes}, {Wills}, {Wills}, \& {Younger}}]{peterson1991}
{Peterson}, B.~M., {Balonek}, T.~J., {Barker}, E.~S., {et~al.} 1991, \apj, 368,
  119

\bibitem[{{Peterson} {et~al.}(2002){Peterson}, {Berlind}, {Bertram},
  {Bischoff}, {Bochkarev}, {Borisov}, {Burenkov}, {Calkins}, {Carrasco},
  {Chavushyan}, {Chornock}, {Dietrich}, {Doroshenko}, {Ezhkova}, {Filippenko},
  {Gilbert}, {Huchra}, {Kollatschny}, {Leonard}, {Li}, {Lyuty}, {Malkov},
  {Matheson}, {Merkulova}, {Mikhailov}, {Modjaz}, {Onken}, {Pogge}, {Pronik},
  {Qian}, {Romano}, {Sergeev}, {Sergeeva}, {Shapovalova}, {Spiridonova}, {Tao},
  {Tokarz}, {Valdes}, {Vlasiuk}, {Wagner}, \& {Wilkes}}]{peterson2002}
{Peterson}, B.~M., {Berlind}, P., {Bertram}, R., {et~al.} 2002, \apj, 581, 197

\bibitem[{{Peterson} {et~al.}(2004){Peterson}, {Ferrarese}, {Gilbert}, {Kaspi},
  {Malkan}, {Maoz}, {Merritt}, {Netzer}, {Onken}, {Pogge}, {Vestergaard}, \&
  {Wandel}}]{peterson2004}
{Peterson}, B.~M., {Ferrarese}, L., {Gilbert}, K.~M., {et~al.} 2004, \apj, 613,
  682

\bibitem[{{Petrucci} {et~al.}(2020){Petrucci}, {Gronkiewicz}, {Rozanska},
  {Belmont}, {Bianchi}, {Czerny}, {Matt}, {Malzac}, {Middei}, {De Rosa},
  {Ursini}, \& {Cappi}}]{petrucci2020}
{Petrucci}, P.~O., {Gronkiewicz}, D., {Rozanska}, A., {et~al.} 2020, \aap, 634,
  A85

\bibitem[{{Planck Collaboration} {et~al.}(2016){Planck Collaboration}, {Ade},
  {Aghanim}, {Arnaud}, {Ashdown}, {Aumont}, {Baccigalupi}, {Banday},
  {Barreiro}, {Bartlett}, {Bartolo}, {Battaner}, {Battye}, {Benabed},
  {Beno{\^\i}t}, {Benoit-L{\'e}vy}, {Bernard}, {Bersanelli}, {Bielewicz},
  {Bock}, {Bonaldi}, {Bonavera}, {Bond}, {Borrill}, {Bouchet}, {Boulanger},
  {Bucher}, {Burigana}, {Butler}, {Calabrese}, {Cardoso}, {Catalano},
  {Challinor}, {Chamballu}, {Chary}, {Chiang}, {Chluba}, {Christensen},
  {Church}, {Clements}, {Colombi}, {Colombo}, {Combet}, {Coulais}, {Crill},
  {Curto}, {Cuttaia}, {Danese}, {Davies}, {Davis}, {de Bernardis}, {de Rosa},
  {de Zotti}, {Delabrouille}, {D{\'e}sert}, {Di Valentino}, {Dickinson},
  {Diego}, {Dolag}, {Dole}, {Donzelli}, {Dor{\'e}}, {Douspis}, {Ducout},
  {Dunkley}, {Dupac}, {Efstathiou}, {Elsner}, {En{\ss}lin}, {Eriksen},
  {Farhang}, {Fergusson}, {Finelli}, {Forni}, {Frailis}, {Fraisse},
  {Franceschi}, {Frejsel}, {Galeotta}, {Galli}, {Ganga}, {Gauthier}, {Gerbino},
  {Ghosh}, {Giard}, {Giraud-H{\'e}raud}, {Giusarma}, {Gjerl{\o}w},
  {Gonz{\'a}lez-Nuevo}, {G{\'o}rski}, {Gratton}, {Gregorio}, {Gruppuso},
  {Gudmundsson}, {Hamann}, {Hansen}, {Hanson}, {Harrison}, {Helou},
  {Henrot-Versill{\'e}}, {Hern{\'a}ndez-Monteagudo}, {Herranz}, {Hildebrandt},
  {Hivon}, {Hobson}, {Holmes}, {Hornstrup}, {Hovest}, {Huang}, {Huffenberger},
  {Hurier}, {Jaffe}, {Jaffe}, {Jones}, {Juvela}, {Keih{\"a}nen}, {Keskitalo},
  {Kisner}, {Kneissl}, {Knoche}, {Knox}, {Kunz}, {Kurki-Suonio}, {Lagache},
  {L{\"a}hteenm{\"a}ki}, {Lamarre}, {Lasenby}, {Lattanzi}, {Lawrence}, {Leahy},
  {Leonardi}, {Lesgourgues}, {Levrier}, {Lewis}, {Liguori}, {Lilje},
  {Linden-V{\o}rnle}, {L{\'o}pez-Caniego}, {Lubin}, {Mac{\'\i}as-P{\'e}rez},
  {Maggio}, {Maino}, {Mandolesi}, {Mangilli}, {Marchini}, {Maris}, {Martin},
  {Martinelli}, {Mart{\'\i}nez-Gonz{\'a}lez}, {Masi}, {Matarrese}, {McGehee},
  {Meinhold}, {Melchiorri}, {Melin}, {Mendes}, {Mennella}, {Migliaccio},
  {Millea}, {Mitra}, {Miville-Desch{\^e}nes}, {Moneti}, {Montier}, {Morgante},
  {Mortlock}, {Moss}, {Munshi}, {Murphy}, {Naselsky}, {Nati}, {Natoli},
  {Netterfield}, {N{\o}rgaard-Nielsen}, {Noviello}, {Novikov}, {Novikov},
  {Oxborrow}, {Paci}, {Pagano}, {Pajot}, {Paladini}, {Paoletti}, {Partridge},
  {Pasian}, {Patanchon}, {Pearson}, {Perdereau}, {Perotto}, {Perrotta},
  {Pettorino}, {Piacentini}, {Piat}, {Pierpaoli}, {Pietrobon}, {Plaszczynski},
  {Pointecouteau}, {Polenta}, {Popa}, {Pratt}, {Pr{\'e}zeau}, {Prunet},
  {Puget}, {Rachen}, {Reach}, {Rebolo}, {Reinecke}, {Remazeilles}, {Renault},
  {Renzi}, {Ristorcelli}, {Rocha}, {Rosset}, {Rossetti}, {Roudier},
  {Rouill{\'e} d'Orfeuil}, {Rowan-Robinson}, {Rubi{\~n}o-Mart{\'\i}n},
  {Rusholme}, {Said}, {Salvatelli}, {Salvati}, {Sandri}, {Santos},
  {Savelainen}, {Savini}, {Scott}, {Seiffert}, {Serra}, {Shellard}, {Spencer},
  {Spinelli}, {Stolyarov}, {Stompor}, {Sudiwala}, {Sunyaev}, {Sutton},
  {Suur-Uski}, {Sygnet}, {Tauber}, {Terenzi}, {Toffolatti}, {Tomasi},
  {Tristram}, {Trombetti}, {Tucci}, {Tuovinen}, {T{\"u}rler}, {Umana},
  {Valenziano}, {Valiviita}, {Van Tent}, {Vielva}, {Villa}, {Wade}, {Wandelt},
  {Wehus}, {White}, {White}, {Wilkinson}, {Yvon}, {Zacchei}, \&
  {Zonca}}]{Planck2016}
{Planck Collaboration}, {Ade}, P.~A.~R., {Aghanim}, N., {et~al.} 2016, \aap,
  594, A13

\bibitem[{{Planck Collaboration} {et~al.}(2020){Planck Collaboration},
  {Aghanim}, {Akrami}, {Ashdown}, {Aumont}, {Baccigalupi}, {Ballardini},
  {Banday}, {Barreiro}, {Bartolo}, {Basak}, {Battye}, {Benabed}, {Bernard},
  {Bersanelli}, {Bielewicz}, {Bock}, {Bond}, {Borrill}, {Bouchet}, {Boulanger},
  {Bucher}, {Burigana}, {Butler}, {Calabrese}, {Cardoso}, {Carron},
  {Challinor}, {Chiang}, {Chluba}, {Colombo}, {Combet}, {Contreras}, {Crill},
  {Cuttaia}, {de Bernardis}, {de Zotti}, {Delabrouille}, {Delouis}, {Di
  Valentino}, {Diego}, {Dor{\'e}}, {Douspis}, {Ducout}, {Dupac}, {Dusini},
  {Efstathiou}, {Elsner}, {En{\ss}lin}, {Eriksen}, {Fantaye}, {Farhang},
  {Fergusson}, {Fernandez-Cobos}, {Finelli}, {Forastieri}, {Frailis},
  {Fraisse}, {Franceschi}, {Frolov}, {Galeotta}, {Galli}, {Ganga},
  {G{\'e}nova-Santos}, {Gerbino}, {Ghosh}, {Gonz{\'a}lez-Nuevo}, {G{\'o}rski},
  {Gratton}, {Gruppuso}, {Gudmundsson}, {Hamann}, {Handley}, {Hansen},
  {Herranz}, {Hildebrandt}, {Hivon}, {Huang}, {Jaffe}, {Jones}, {Karakci},
  {Keih{\"a}nen}, {Keskitalo}, {Kiiveri}, {Kim}, {Kisner}, {Knox},
  {Krachmalnicoff}, {Kunz}, {Kurki-Suonio}, {Lagache}, {Lamarre}, {Lasenby},
  {Lattanzi}, {Lawrence}, {Le Jeune}, {Lemos}, {Lesgourgues}, {Levrier},
  {Lewis}, {Liguori}, {Lilje}, {Lilley}, {Lindholm}, {L{\'o}pez-Caniego},
  {Lubin}, {Ma}, {Mac{\'\i}as-P{\'e}rez}, {Maggio}, {Maino}, {Mandolesi},
  {Mangilli}, {Marcos-Caballero}, {Maris}, {Martin}, {Martinelli},
  {Mart{\'\i}nez-Gonz{\'a}lez}, {Matarrese}, {Mauri}, {McEwen}, {Meinhold},
  {Melchiorri}, {Mennella}, {Migliaccio}, {Millea}, {Mitra},
  {Miville-Desch{\^e}nes}, {Molinari}, {Montier}, {Morgante}, {Moss}, {Natoli},
  {N{\o}rgaard-Nielsen}, {Pagano}, {Paoletti}, {Partridge}, {Patanchon},
  {Peiris}, {Perrotta}, {Pettorino}, {Piacentini}, {Polastri}, {Polenta},
  {Puget}, {Rachen}, {Reinecke}, {Remazeilles}, {Renzi}, {Rocha}, {Rosset},
  {Roudier}, {Rubi{\~n}o-Mart{\'\i}n}, {Ruiz-Granados}, {Salvati}, {Sandri},
  {Savelainen}, {Scott}, {Shellard}, {Sirignano}, {Sirri}, {Spencer},
  {Sunyaev}, {Suur-Uski}, {Tauber}, {Tavagnacco}, {Tenti}, {Toffolatti},
  {Tomasi}, {Trombetti}, {Valenziano}, {Valiviita}, {Van Tent}, {Vibert},
  {Vielva}, {Villa}, {Vittorio}, {Wandelt}, {Wehus}, {White}, {White},
  {Zacchei}, \& {Zonca}}]{Planck2020}
{Planck Collaboration}, {Aghanim}, N., {Akrami}, Y., {et~al.} 2020, \aap, 641,
  A6

\bibitem[{{Pozo Nu{\~n}ez} {et~al.}(2023){Pozo Nu{\~n}ez}, {Bruckmann},
  {Deesamutara}, {Czerny}, {Panda}, {Lobban}, {Pietrzy{\'n}ski}, \&
  {Polsterer}}]{pozo2023}
{Pozo Nu{\~n}ez}, F., {Bruckmann}, C., {Deesamutara}, S., {et~al.} 2023,
  \mnras, 522, 2002

\bibitem[{{Pozo Nu{\~n}ez} {et~al.}(2024){Pozo Nu{\~n}ez}, {Czerny}, {Panda},
  {Kovacevic}, {Brandt}, {Horne}, \& {LSST AGN Science
  Collaboration}}]{pozo2024}
{Pozo Nu{\~n}ez}, F., {Czerny}, B., {Panda}, S., {et~al.} 2024, Research Notes
  of the American Astronomical Society, 8, 47

\bibitem[{{Prince} {et~al.}(2025){Prince}, {Hern{\'a}ndez Santisteban},
  {Horne}, {Gelbord}, {McHardy}, {Edelson}, {Onken}, {Donnan}, {Vestergaard},
  {Kaspi}, {Winkler}, {Cackett}, {Landt}, {Barth}, {Treu}, {Valenti}, {Lira},
  {Chelouche}, {Romero Colmenero}, {Goad}, {Gonzalez-Buitrago}, {Kara}, \&
  {Villforth}}]{2025arXiv250606731P}
{Prince}, R., {Hern{\'a}ndez Santisteban}, J.~V., {Horne}, K., {et~al.} 2025,
  arXiv e-prints, arXiv:2506.06731

\bibitem[{{Prince} {et~al.}(2022){Prince}, {Hryniewicz}, {Panda}, {Czerny}, \&
  {Pollo}}]{prince2022}
{Prince}, R., {Hryniewicz}, K., {Panda}, S., {Czerny}, B., \& {Pollo}, A. 2022,
  \apj, 925, 215

\bibitem[{{Quera-Bofarull} {et~al.}(2023){Quera-Bofarull}, {Done}, {Lacey},
  {Nomura}, \& {Ohsuga}}]{new_QWIND2023}
{Quera-Bofarull}, A., {Done}, C., {Lacey}, C.~G., {Nomura}, M., \& {Ohsuga}, K.
  2023, \mnras, 518, 2693

\bibitem[{{Ra{\l}owski} {et~al.}(2024){Ra{\l}owski}, {Hryniewicz}, {Pollo}, \&
  {Stawarz}}]{ralowski2024}
{Ra{\l}owski}, M., {Hryniewicz}, K., {Pollo}, A., \& {Stawarz}, {\L}. 2024,
  \aap, 682, A120

\bibitem[{{Riess} {et~al.}(2016){Riess}, {Macri}, {Hoffmann}, {Scolnic},
  {Casertano}, {Filippenko}, {Tucker}, {Reid}, {Jones}, {Silverman},
  {Chornock}, {Challis}, {Yuan}, {Brown}, \& {Foley}}]{riess2016}
{Riess}, A.~G., {Macri}, L.~M., {Hoffmann}, S.~L., {et~al.} 2016, \apj, 826, 56

\bibitem[{{Riess} {et~al.}(2024){Riess}, {Scolnic}, {Anand}, {Breuval},
  {Casertano}, {Macri}, {Li}, {Yuan}, {Huang}, {Jha}, {Murakami}, {Beaton},
  {Brout}, {Wu}, {Addison}, {Bennett}, {Anderson}, {Filippenko}, \&
  {Carr}}]{riess2024}
{Riess}, A.~G., {Scolnic}, D., {Anand}, G.~S., {et~al.} 2024, \apj, 977, 120

\bibitem[{{Risaliti} \& {Elvis}(2010)}]{QWIND2010}
{Risaliti}, G. \& {Elvis}, M. 2010, \aap, 516, A89

\bibitem[{{Risaliti} \& {Lusso}(2015)}]{risaliti2015}
{Risaliti}, G. \& {Lusso}, E. 2015, \apj, 815, 33

\bibitem[{{Robinson} {et~al.}(2021){Robinson}, {Bentz}, {Courtois}, {Johnson},
  {Crenshaw}, {Meena}, {Polack}, {Silverstein}, \& {Chen}}]{robinson2021}
{Robinson}, J.~H., {Bentz}, M.~C., {Courtois}, H.~M., {et~al.} 2021, \apj, 912,
  160

\bibitem[{{Rokaki} {et~al.}(1993){Rokaki}, {Collin-Souffrin}, \&
  {Magnan}}]{rokaki1993}
{Rokaki}, E., {Collin-Souffrin}, S., \& {Magnan}, C. 1993, \aap, 272, 8

\bibitem[{{Rosborough} {et~al.}(2025){Rosborough}, {Robinson}, {Almeyda},
  {Humphrey}, \& {Noll}}]{rosborough2025}
{Rosborough}, S., {Robinson}, A., {Almeyda}, T., {Humphrey}, D., \& {Noll}, M.
  2025, arXiv e-prints, arXiv:2507.14337

\bibitem[{{Rosborough} {et~al.}(2024){Rosborough}, {Robinson}, {Almeyda}, \&
  {Noll}}]{rosborough2024}
{Rosborough}, S.~A., {Robinson}, A., {Almeyda}, T., \& {Noll}, M. 2024, \apj,
  965, 35

\bibitem[{{R{\'o}{\.z}a{\'n}ska} {et~al.}(1999){R{\'o}{\.z}a{\'n}ska},
  {Czerny}, {{\.Z}ycki}, \& {Pojma{\'n}ski}}]{rozanska1999}
{R{\'o}{\.z}a{\'n}ska}, A., {Czerny}, B., {{\.Z}ycki}, P.~T., \&
  {Pojma{\'n}ski}, G. 1999, \mnras, 305, 481

\bibitem[{{Schmidt}(1963)}]{schmidt1963}
{Schmidt}, M. 1963, \nat, 197, 1040

\bibitem[{{Sergeev} {et~al.}(2005){Sergeev}, {Doroshenko}, {Golubinskiy},
  {Merkulova}, \& {Sergeeva}}]{2005ApJ...622..129S}
{Sergeev}, S.~G., {Doroshenko}, V.~T., {Golubinskiy}, Y.~V., {Merkulova},
  N.~I., \& {Sergeeva}, E.~A. 2005, \apj, 622, 129

\bibitem[{{Seyfert}(1943)}]{Seyfert1943}
{Seyfert}, C.~K. 1943, \apj, 97, 28

\bibitem[{{Shakura} \& {Sunyaev}(1973)}]{SS1973}
{Shakura}, N.~I. \& {Sunyaev}, R.~A. 1973, \aap, 24, 337

\bibitem[{{Shapovalova} {et~al.}(2004){Shapovalova}, {Doroshenko}, {Bochkarev},
  {Burenkov}, {Carrasco}, {Chavushyan}, {Collin}, {Vald{\'e}s}, {Borisov},
  {Dumont}, {Vlasuyk}, {Chilingarian}, {Fioktistova}, \&
  {Martinez}}]{shapovalova2004}
{Shapovalova}, A.~I., {Doroshenko}, V.~T., {Bochkarev}, N.~G., {et~al.} 2004,
  \aap, 422, 925

\bibitem[{{Shapovalova} {et~al.}(2009){Shapovalova}, {Popovi{\'c}},
  {Bochkarev}, {Burenkov}, {Chavushyan}, {Collin}, {Doroshenko}, {Ili{\'c}}, \&
  {Kova{\v{c}}evi{\'c}}}]{shapovalova2009}
{Shapovalova}, A.~I., {Popovi{\'c}}, L.~{\v{C}}., {Bochkarev}, N.~G., {et~al.}
  2009, \nar, 53, 191

\bibitem[{{Shappee} {et~al.}(2014){Shappee}, {Prieto}, {Grupe}, {Kochanek},
  {Stanek}, {De Rosa}, {Mathur}, {Zu}, {Peterson}, {Pogge}, {Komossa}, {Im},
  {Jencson}, {Holoien}, {Basu}, {Beacom}, {Szczygie{\l}}, {Brimacombe},
  {Adams}, {Campillay}, {Choi}, {Contreras}, {Dietrich}, {Dubberley},
  {Elphick}, {Foale}, {Giustini}, {Gonzalez}, {Hawkins}, {Howell}, {Hsiao},
  {Koss}, {Leighly}, {Morrell}, {Mudd}, {Mullins}, {Nugent}, {Parrent},
  {Phillips}, {Pojmanski}, {Rosing}, {Ross}, {Sand}, {Terndrup}, {Valenti},
  {Walker}, \& {Yoon}}]{shappee2014}
{Shappee}, B.~J., {Prieto}, J.~L., {Grupe}, D., {et~al.} 2014, \apj, 788, 48

\bibitem[{{Shimura} \& {Takahara}(1995)}]{1995ApJ...445..780S}
{Shimura}, T. \& {Takahara}, F. 1995, \apj, 445, 780

\bibitem[{{{\'S}niegowska} {et~al.}(2021){{\'S}niegowska}, {Marziani},
  {Czerny}, {Panda}, {Mart{\'\i}nez-Aldama}, {del Olmo}, \&
  {D'Onofrio}}]{sniegowska2021}
{{\'S}niegowska}, M., {Marziani}, P., {Czerny}, B., {et~al.} 2021, \apj, 910,
  115

\bibitem[{{Sunyaev} \& {Titarchuk}(1980)}]{ST80}
{Sunyaev}, R.~A. \& {Titarchuk}, L.~G. 1980, \aap, 86, 121

\bibitem[{{Suyu} {et~al.}(2024){Suyu}, {Goobar}, {Collett}, {More}, \&
  {Vernardos}}]{Suyu2024}
{Suyu}, S.~H., {Goobar}, A., {Collett}, T., {More}, A., \& {Vernardos}, G.
  2024, \ssr, 220, 13

\bibitem[{{Trefoloni} {et~al.}(2024){Trefoloni}, {Lusso}, {Nardini},
  {Risaliti}, {Marconi}, {Bargiacchi}, {Sacchi}, {Pietrini}, \&
  {Signorini}}]{trefoloni2024}
{Trefoloni}, B., {Lusso}, E., {Nardini}, E., {et~al.} 2024, \aap, 689, A109

\bibitem[{{Uddin} {et~al.}(2024){Uddin}, {Burns}, {Phillips}, {Suntzeff},
  {Freedman}, {Brown}, {Morrell}, {Hamuy}, {Krisciunas}, {Wang}, {Hsiao},
  {Goobar}, {Perlmutter}, {Lu}, {Stritzinger}, {Anderson}, {Ashall},
  {Hoeflich}, {Shappee}, {Persson}, {Piro}, {Baron}, {Contreras}, {Galbany},
  {Kumar}, {Shahbandeh}, {Davis}, {Anais}, {Busta}, {Campillay},
  {Castell{\'o}n}, {Corco}, {Diamond}, {Gall}, {Gonzalez}, {Holmbo}, {Roth},
  {Ser{\'o}n}, {Taddia}, {Torres}, {Baltay}, {Folatelli}, {Hadjiyska},
  {Kasliwal}, {Nugent}, {Rabinowitz}, \& {Ryder}}]{uddin2024}
{Uddin}, S.~A., {Burns}, C.~R., {Phillips}, M.~M., {et~al.} 2024, \apj, 970, 72

\bibitem[{{Ulrich} \& {Boisson}(1983)}]{ulrich1983}
{Ulrich}, M.~H. \& {Boisson}, C. 1983, \apj, 267, 515

\bibitem[{{Vazquez} {et~al.}(2015){Vazquez}, {Galianni}, {Richmond},
  {Robinson}, {Axon}, {Horne}, {Almeyda}, {Fausnaugh}, {Peterson}, {Bottorff},
  {Gallimore}, {Eltizur}, {Netzer}, {Storchi-Bergmann}, {Marconi}, {Capetti},
  {Batcheldor}, {Buchanan}, {Stirpe}, {Kishimoto}, {Packham}, {Perez},
  {Tadhunter}, {Upton}, \& {Estrada-Carpenter}}]{Vazquez2015}
{Vazquez}, B., {Galianni}, P., {Richmond}, M., {et~al.} 2015, \apj, 801, 127

\bibitem[{{Verde} {et~al.}(2024){Verde}, {Sch{\"o}neberg}, \&
  {Gil-Mar{\'\i}n}}]{verde2024}
{Verde}, L., {Sch{\"o}neberg}, N., \& {Gil-Mar{\'\i}n}, H. 2024, \araa, 62, 287

\bibitem[{{Wang} {et~al.}(2017){Wang}, {Xu}, \& {Zhao}}]{wang2017}
{Wang}, Y., {Xu}, L., \& {Zhao}, G.-B. 2017, \apj, 849, 84

\bibitem[{{Wildy} {et~al.}(2021){Wildy}, {Landt}, {Ward}, {Czerny}, \&
  {Kynoch}}]{wildy2021}
{Wildy}, C., {Landt}, H., {Ward}, M.~J., {Czerny}, B., \& {Kynoch}, D. 2021,
  \mnras, 500, 2063

\bibitem[{{Wong} {et~al.}(2024){Wong}, {Dux}, {Shajib}, {Suyu}, {Millon},
  {Mozumdar}, {Wells}, {Agnello}, {Birrer}, {Buckley-Geer}, {Courbin},
  {Fassnacht}, {Frieman}, {Galan}, {Lin}, {Marshall}, {Poh}, {Schuldt},
  {Sluse}, \& {Treu}}]{lensing2024}
{Wong}, K.~C., {Dux}, F., {Shajib}, A.~J., {et~al.} 2024, \aap, 689, A168

\bibitem[{{Yoshii} {et~al.}(2014){Yoshii}, {Kobayashi}, {Minezaki}, {Koshida},
  \& {Peterson}}]{yoshii2014}
{Yoshii}, Y., {Kobayashi}, Y., {Minezaki}, T., {Koshida}, S., \& {Peterson},
  B.~A. 2014, \apjl, 784, L11

\bibitem[{{Yu} {et~al.}(2020){Yu}, {Martini}, {Davis}, {Gruendl}, {Hoormann},
  {Kochanek}, {Lidman}, {Mudd}, {Peterson}, {Wester}, {Allam}, {Annis},
  {Asorey}, {Avila}, {Banerji}, {Bertin}, {Brooks}, {Buckley-Geer}, {Calcino},
  {Rosell}, {Carollo}, {Kind}, {Carretero}, {Cunha}, {D'Andrea}, {Costa}, {De
  Vicente}, {Desai}, {Diehl}, {Doel}, {Eifler}, {Flaugher}, {Fosalba},
  {Frieman}, {Garc{\'\i}a-Bellido}, {Gaztanaga}, {Glazebrook}, {Gruen},
  {Gschwend}, {Gutierrez}, {Hartley}, {Hinton}, {Hollowood}, {Honscheid},
  {Hoyle}, {James}, {Kim}, {Krause}, {Kuehn}, {Kuropatkin}, {Lewis}, {Lima},
  {Macaulay}, {Maia}, {Marshall}, {Menanteau}, {Miquel}, {M{\"o}ller},
  {Plazas}, {Romer}, {Sanchez}, {Scarpine}, {Schubnell}, {Serrano}, {Smith},
  {Smith}, {Soares-Santos}, {Sobreira}, {Suchyta}, {Swann}, {Swanson}, {Tarle},
  {Tucker}, {Tucker}, \& {Vikram}}]{2020ApJS..246...16Y}
{Yu}, Z., {Martini}, P., {Davis}, T.~M., {et~al.} 2020, \apjs, 246, 16

\bibitem[{{Zaja{\v{c}}ek} {et~al.}(2024){Zaja{\v{c}}ek}, {Panda}, {Pandey},
  {Prince}, {Rodr{\'\i}guez-Ardila}, {Jaiswal}, {Czerny}, {Hryniewicz},
  {Urbanowicz}, {Trzcionkowski}, {{\'S}niegowska}, {Fa{\l}kowska},
  {Mart{\'\i}nez-Aldama}, \& {Werner}}]{zajacek2024}
{Zaja{\v{c}}ek}, M., {Panda}, S., {Pandey}, A., {et~al.} 2024, \aap, 683, A140

\end{thebibliography}

\begin{appendix}
\label{ss:apendix}
\section{$H_0$ based on observationally determined response function}

\begin{figure}
   \centering
  \includegraphics[scale=0.5]{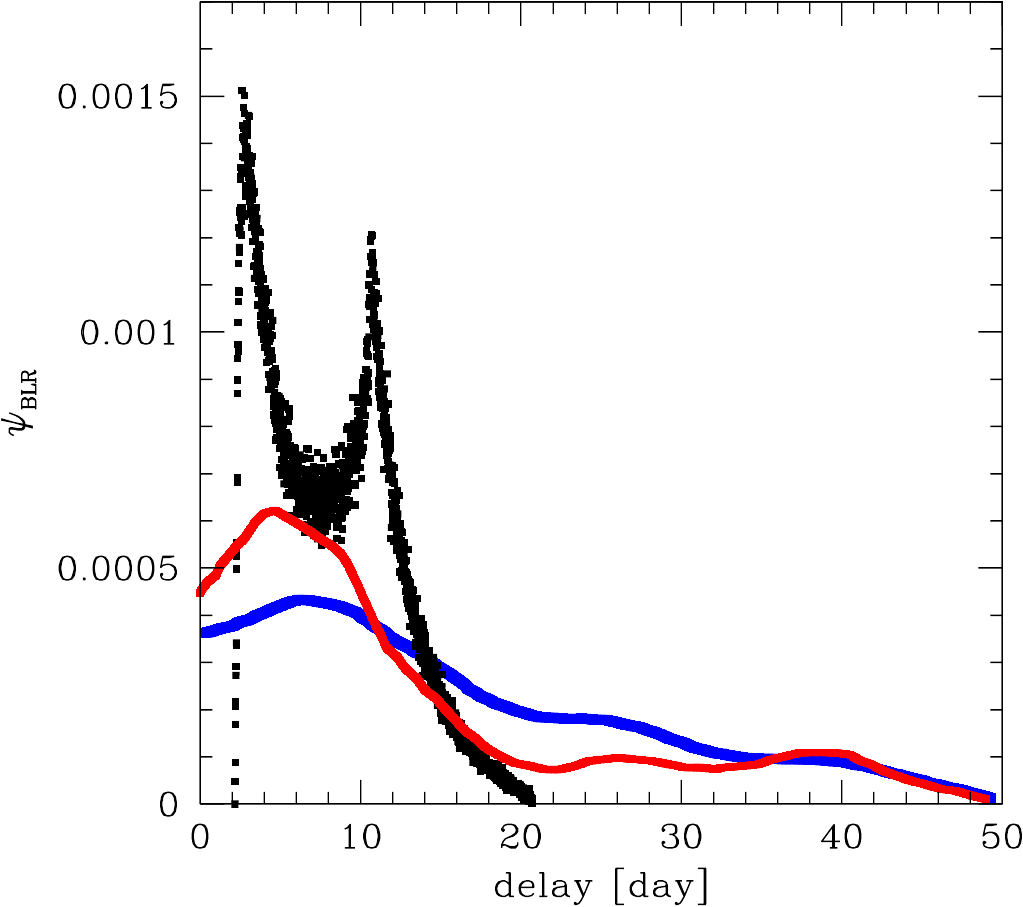}
  \includegraphics[scale=0.5]{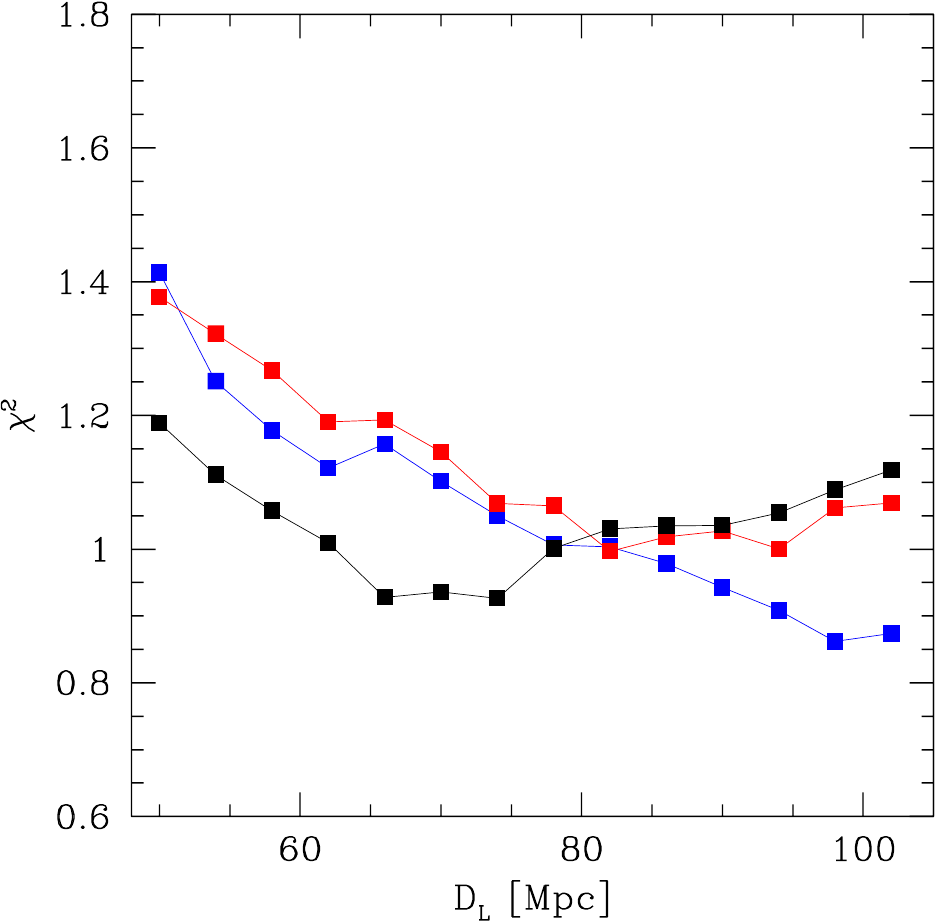}
  \caption{Upper panel: the response function from the FRADO model (black), total H$\beta$ response (blue) presented in Figure~4 of \citet{horne2021}, and the response at 4863 \AA~  (red) shown in Figure~9 of the same paper. Lower panel: The $\chi^2$ values as a function of luminosity distance for the three response functions are shown in their respective colors.} 
  \label{fig:response_comp}
\end{figure}

We employ an alternative method to validate the results obtained from the FRADO model. In rare cases, NGC 5548 being one notable example, the transfer function for H$\beta$ was directly determined from the data using densely sampled spectroscopic RM. For this purpose, we use the results presented  by \citet{horne2021}. These observations were conducted in 2014, roughly during the same period with the photometric time-delay measurements. The study provides response functions for various emission lines, but we focus on those corresponding to the H$\beta$ line, as they are most representative of the regions in the BLR where the Balmer and Paschen continua originate and play a dominant role in shaping the observed time delays. Specifically, we consider two versions of the H$\beta$ response functions presented by \citet{horne2021}: one that encompasses the full H$\beta$ line and another centered narrowly around 4863~\AA. A comparison between our model-derived BLR response function and the two response functions from \citet{horne2021} is shown in the upper panel of Figure~\ref{fig:response_comp}.

Both observed response functions lack a double-peak structure, which, however, is present in the case of the CIV line \citep[see Figure~ 4 of][]{horne2021}. More importantly, these response functions exhibit a significantly longer tail, with this extended tail being notably stronger in the full-H$\beta$ response profile than in the narrower 4863 \AA~ case. This long tail is also clearly visible in the 4830 \AA~ bin shown in Figure~9 of \citet{horne2021}, suggesting that the extended delay may arise from contamination by Fe II emission. On average, the optical and UV Fe II delays tend to be longer than those of typical low-ionization lines such as Mg II and H$\beta$ \citep{zajacek2024}.

Having an observationally derived transfer function significantly reduces our dependence on the FRADO model summarized in Section \ref{ss:frado}. However, we still require {\tt CLOUDY} computations to determine the wavelength-dependent efficiency of re-emission by the BLR, as well as a disk/corona model.

Therefore, we perform fitting of both the mean spectrum and the wavelength-dependent time delay as before, for a set of values of the luminosity distance. For each luminosity distance $D_L$, the model parameters are refitted accordingly.

In the case of the entire H$\beta$ transfer function, we are unable to tightly constrain the luminosity distance (blue data points in the bottom panel of ~\ref{fig:response_comp}); the monotonic trend continues up to 98 Mpc, with marginal rise at 102 Mpc. We did not perform computations beyond this distance. However, for the transfer function centered around 4863 \AA~, a broad minimum is observed from 82 Mpc till 94 Mpc (red data points), which is shifted compared to the 74 Mpc value obtained using the theoretical FRADO model for the transfer function (black data points). Interestingly, the overall fit is the best for the FRADO model. The resulting constraint on the Hubble constant $H_0$ from the 4863 \AA~ bin is $60.3^{+2.0}_{-9.1}$ km s$^{-1}$ Mpc$^{-1}$, generally lower than the FRADO-derived range but only marginally outside the 1 sigma error value due to relatively large uncertainty, further supporting the applicability of the FRADO model in such analyses.

\end{appendix}

\end{document}